\documentclass[aps,pre,showpacs,floatfix]{revtex4}

\usepackage{amssymb,amsmath}
\usepackage[dvips]{graphicx}

\newcommand{\bd}[1]{\mbox{\boldmath$#1$}}

\begin{document}

\title{Noise-induced Turbulence in Nonlocally Coupled Oscillators}

\author{Yoji Kawamura}\email{kawamura@ton.scphys.kyoto-u.ac.jp}
\author{Hiroya Nakao}
\affiliation{Department of Physics, Graduate School of Sciences,
  Kyoto University, Kyoto 606-8502, Japan}

\author{Yoshiki Kuramoto}
\affiliation{Department of Mathematics, Graduate School of Sciences,
  Hokkaido University, Sapporo 060-0810, Japan}

\date{December 11, 2006}


\begin{abstract}
  We demonstrate that nonlocally coupled limit-cycle oscillators
  subject to spatiotemporally white Gaussian noise can exhibit
  a noise-induced transition to turbulent states.
  After illustrating noise-induced turbulent states with numerical
  simulations using two representative models of limit-cycle oscillators,
  we develop a theory that clarifies the effective dynamical
  instabilities leading to the turbulent behavior using a hierarchy
  of dynamical reduction methods.
  We determine the parameter region where the system can exhibit
  noise-induced turbulent states, which is successfully confirmed by
  extensive numerical simulations at each level of the reduction.
\end{abstract}

\maketitle

\section{Introduction} \label{sec:1}

When a dynamical system is driven by external noises, its
effective dynamics generally changes~\cite{ref:horsthemke84,
ref:ojalvo99,ref:anishchenko01,ref:matsumoto83,ref:shibata99}.
We usually expect that the system tends to be more random
and statistically uniform, because the noise destroys the
spatiotemporal structures of the system.
For example, Matsumoto and Tsuda~\cite{ref:matsumoto83} found that
external noise can destabilize chaos and produces ordered behavior
in the Belousov-Zhabotinsky map, and called this phenomenon
``noise-induced order''.
Shibata, Chawanya, and Kaneko~\cite{ref:shibata99} studied
the effect of microscopic external noise on the collective
motion of a globally coupled map in fully desynchronized states.
They demonstrated that the collective motion is successively
simplified with the increase of external noise intensity, while
without the external noise a macroscopic variable shows
high-dimensional chaos distinguishable from random motions.

In this paper, we give a counter-example to this intuition.
Specifically, we demonstrate that nonlocally coupled limit-cycle
oscillators can undergo a noise-induced transition from uniform states
to turbulent states through effective dynamical instabilities, with the
induced turbulent fluctuations far larger than the intensity of the
driving noise.

Our starting point is a general equation describing nonlocally coupled
limit-cycle oscillators subject to spatiotemporally white Gaussian noise.
We first present numerical examples using two representative models
of limit-cycle oscillators, the FitzHugh-Nagumo model and the
Stuart-Landau model, in order to illustrate that weak external noise
can actually cause turbulence in such systems.
To theoretically investigate this noise-induced turbulent state, we
simplify our original equation to a Langevin phase equation by means
of the phase reduction method for limit-cycle oscillators, utilizing
the fact that the external noise intensity and the coupling strength
between the oscillators are sufficiently weak.
The resulting equation describes a system of nonlocally coupled noisy
phase oscillators.
We then derive an equivalent nonlinear Fokker-Planck equation from
this Langevin equation by adopting the mean-field theory, which holds
exactly for our nonlocally coupled oscillators.
Our nonlinear Fokker-Planck equation has a constant solution
corresponding to all the oscillators being in a completely
desynchronized state.
It is linearly stable when the external noise is sufficiently strong,
but, as the noise intensity is decreased, undergoes a Hopf bifurcation
at a certain noise intensity, giving rise to limit-cycle oscillations
of the phase distribution.
In the vicinity of this bifurcation point, we can derive a complex
Ginzburg-Landau equation from the nonlinear Fokker-Planck
equation by applying the center-manifold reduction method, which
governs the small-amplitude deviation of the probability density
from the constant solution.
It is well known that the spatially uniform oscillation of the complex
Ginzburg-Landau equation becomes unstable and spatiotemporal chaos
develops when the Benjamin-Feir instability condition is satisfied.
Therefore, we expect that the Langevin phase equation and the
corresponding nonlinear Fokker-Planck equation also exhibit
spatiotemporal chaos under suitable conditions.
By direct numerical simulations, we will confirm that the amplitude
turbulence typical of the complex Ginzburg-Landau equation actually arises.
In addition, we also confirm that the phase turbulence arises near
the Benjamin-Feir criticality, which is also a hallmark of the
complex Ginzburg-Landau equation.
We then examine the situation far from the Hopf bifurcation.
In our system, another smaller critical noise intensity is expected to
exist, above which the turbulence arises from the spatially uniform
oscillation via a long-wave phase instability.
To confirm this, we derive a Kuramoto-Sivashinsky-type equation by
applying the phase reduction method to the spatially uniform
oscillating solution of the nonlinear Fokker-Planck equation.
Our calculation shows that the phase diffusion coefficient changes its
sign from positive to negative as the noise intensity is increased
from zero, which implies the destabilization of the spatially uniform
oscillating solution.
By a systematic numerical calculation of the phase diffusion coefficient,
we determine the parameter region where our system exhibits a
noise-induced turbulent state.
From the mathematical analysis of the hierarchy of reduced equations
and extensive numerical simulations of the dynamical equation at each
level of the reduction, we will conclude that the appearance of the
turbulence can be considered as a noise-induced transition phenomenon.


The organization of this paper is as follows.
In Sec.~\ref{sec:2}, we introduce our model and present numerical
results for the FitzHugh-Nagumo and Stuart-Landau oscillators.
In Sec.~\ref{sec:3}, we derive Langevin and Fokker-Planck equations
from the original equation by the phase reduction method.
In Sec.~\ref{sec:4}, we derive a complex Ginzburg-Landau equation by a
center-manifold reduction of the Fokker-Planck equation near its Hopf
bifurcation point.
In Sec.~\ref{sec:5}, we reduce the Fokker-Planck equation to a
Kuramoto-Sivashinsky-type equation near the destabilization point of
the spatially uniform oscillation, and draw the complete phase diagram
of the noise-induced turbulence.
Concluding remarks will be given in the final section.

\section{Noise-induced turbulence in nonlocally coupled limit-cycle
  oscillators} \label{sec:2}

In this section, we introduce a general model of nonlocally coupled
noisy oscillators, and numerically demonstrate that the model can
exhibit noise-induced turbulent states using two representative models
of limit-cycle oscillators.

\subsection{General model}

We consider a system of nonlocally coupled limit-cycle oscillators
in one-dimensional space subject to spatiotemporal noise.
The general form of the model is given by
\begin{equation}
  \partial_t\bd{X}\left(x,t\right) = \bd{F}\left(\bd{X}\left(x,t\right)\right)
  +\hat{K}\int_{-\infty}^{\infty}dx'\,G\left(x-x'\right)
  \bd{X}\left(x',t\right) + \sqrt{\sigma}\,\bd{\eta}\left(x,t\right).
\label{eq:1}
\end{equation}
Here, $\bd{X}(x,t)$ represents the state of a local limit-cycle
oscillator at location $x$ and time $t$.
The first term on the right-hand side describes the dynamics of
each oscillator.
In the absence of the coupling and the noise, it is simply given by
$\dot{\bd{X}}=\bd{F}(\bd{X})$, which is assumed to have a single
stable limit-cycle solution.
The second term describes the nonlocal coupling among the oscillators,
where $\hat{K}$ and $G(x)$ represent respectively the coupling matrix
and the nonlocal coupling function.
Throughout this paper, we use a simple exponential function
\begin{equation}
  G\left(x\right) = \frac{1}{2} \exp\left(-\left|x\right|\right),
\label{eq:2}
\end{equation}
which is normalized to unity in the whole space domain.
The last term of Eq.~(\ref{eq:1}) represents the external noise applied to
each oscillator, whose intensity is controlled by the parameter $\sigma$.
$\bd{\eta}(x,t)$ is a spatiotemporally white Gaussian noise with
zero mean specified by
\begin{equation}
  \left\langle\eta_j\left(x,t\right)\right\rangle = 0,\quad
  \left\langle\eta_j\left(x,t\right)\eta_{j'}\left(x',t'\right)\right\rangle
  = 2\delta_{j,j'}\delta\left(x-x'\right)\delta\left(t-t'\right),
\label{eq:3}
\end{equation}
where the subscripts $j$ and $j'$ denote the vector components of
the noise.

Equation~(\ref{eq:1}) can naturally be derived, for example, in the
following situation.
Let us consider a set of equations~\cite{ref:kuramoto06,ref:kuramoto95,
  ref:kuramoto98,ref:kuramoto00,ref:tanaka03,ref:shima04}
\begin{align}
  \partial_t \bd{X}\left(x,t\right) &= \bd{F}\left(\bd{X}\left(x,t\right)\right)
  + \hat{K} \bd{S}\left(x,t\right) + \sqrt{\sigma}\,\bd{\eta}\left(x,t\right), \label{eq:4} \\
  \tau \partial_t\bd{S} \left(x,t\right) &= -\bd{S}\left(x,t\right)
  + \partial_x^2 \bd{S}\left(x,t\right) + \bd{X}\left(x,t\right), \label{eq:5}
\end{align}
which describe a system of mutually interacting oscillatory elements,
where the coupling between the elements is mediated by some substance
that diffuses and decays much faster than the dynamics of the individual
oscillators, e.g. slime molds interacting through signal molecules.
By considering the $\tau\to 0$ limit and eliminating the fast dynamics
of $\bd{S}$ adiabatically, we obtain Eq.~(\ref{eq:1}) with the kernel
given by Eq.~(\ref{eq:2}) from Eqs.~(\ref{eq:4})-(\ref{eq:5}).

\subsection{FitzHugh-Nagumo oscillators}

As the first example, we consider the case that the local oscillator
in Eq.~(\ref{eq:1}) is given by the FitzHugh-Nagumo (FN) oscillator.
The model is explicitly given by
\begin{align}
  \partial_t X(x,t) &= F_X(X,Y) + K_X S_X(x,t) +\sqrt{\sigma}\, \eta_X(x,t), \label{eq:6} \\
  \partial_t Y(x,t) &= F_Y(X,Y) + K_Y S_X(x,t) +\sqrt{\sigma}\, \eta_Y(x,t), \label{eq:7}
\end{align}
where the dynamics of each oscillator is described by
\begin{equation}
  F_X(X,Y) = \epsilon^{-1}\left(X-X^3-Y\right),\quad F_Y(X,Y) = aX+b,
\label{eq:8}
\end{equation}
and the nonlocal coupling term $S_X(x,t)$ is given by
\begin{equation}
  S_X\left(x,t\right)=\int_{-\infty}^{\infty}dx'\,
  G\left(x-x'\right)X\left(x',t\right).
\label{eq:9}
\end{equation}
We consider the case that the oscillators are coupled only through the
variable $X$.
We fix the parameters of the oscillators as $\epsilon=0.5$, $a=1.0$,
and $b=0.3$, with which the oscillators are well in the
self-oscillatory regime.

\begin{figure}
\centering
\includegraphics[height=6cm]{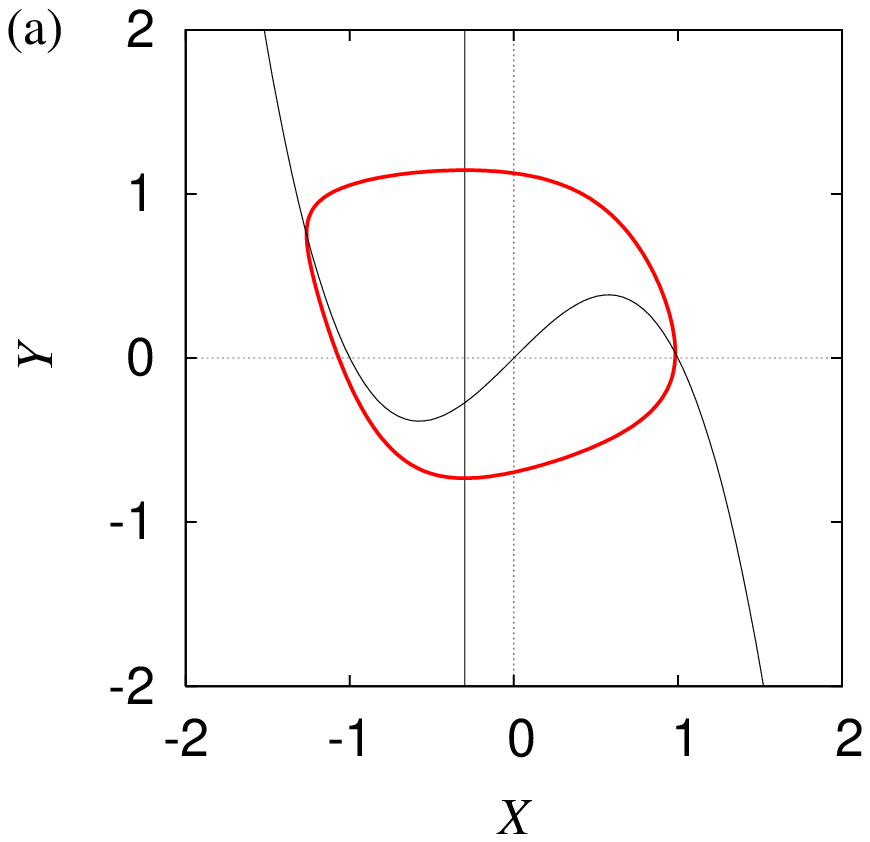}
\includegraphics[height=6cm]{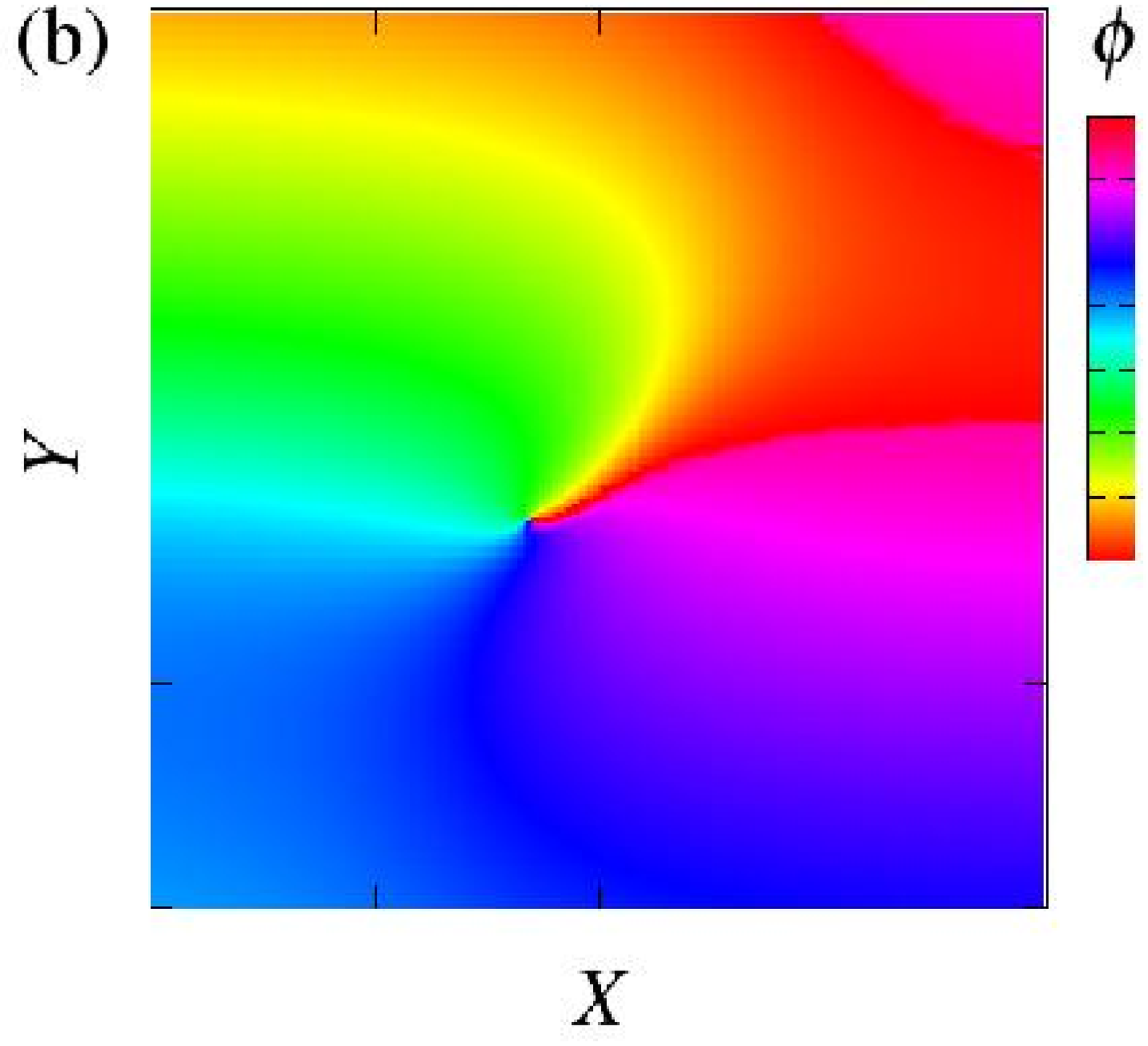}
\caption{(Color online) FitzHugh-Nagumo oscillator with parameters
  $\epsilon=0.5$, $a=1.0$, and $b=0.3$.
  (a): Limit-cycle orbit and nullclines in the $X$-$Y$ plane.
  (b): Generalized phase $\phi(X,Y)\in [0,2\pi]$, which is globally
  defined such that $\dot{\phi}=\omega$ holds identically.}
\label{fig:1}
\end{figure}
Figure~\ref{fig:1}(a) shows the limit-cycle orbit of an individual
FitzHugh-Nagumo oscillator in the $X$-$Y$ plane together with its
nullclines, in the absence of the coupling and the noise.
Throughout this paper, we use a generalized phase variable to describe
the oscillator (see the next section).
Figure~\ref{fig:1}(b) shows the generalized phase variable of the
FitzHugh-Nagumo oscillator defined on the $X$-$Y$ plane, which maps
the state variable $(X,Y)$ of the oscillator to a single real phase
variable $\phi\in [0,2\pi]$.

We performed direct numerical simulations of
Eqs.~(\ref{eq:6})-(\ref{eq:9}), where the oscillator fields $X(x,t)$
and $Y(x,t)$ are discretized using $N=2^{10}$ oscillators with the
spatial grid size of $\Delta x = 0.1$. See Appendix~\ref{sec:C} for
the details of the numerical methods.
The values of the coupling strength $K_X$ and $K_Y$ are chosen
appropriately in such a way that the spatially uniform oscillation
is stable and the system is non-turbulent in the absence of noise.

In order to observe the average deterministic dynamics of the system,
it is desirable to filter statistical fluctuations due to the noise.
We thus introduce a space-time dependent complex order parameter with
modulus $R(x,t)$ and phase $\Theta(x,t)$, calculated from the
oscillator phase field $\phi(x,t)$ through
\begin{equation}
  R\left(x,t\right)\exp\Bigl(i\Theta\left(x,t\right)\Bigr)
  = \int_{-\infty}^{\infty}dx'\,G\left(x-x'\right)
  \exp\Bigl(i\phi\left(x',t\right)\Bigr).
\label{eq:10}
\end{equation}
This order parameter represents a spatial average of the complex phase
factor $\exp(i\phi)$ of the local oscillators over the coupling range.

\begin{figure}
\centering
\includegraphics[height=4cm]{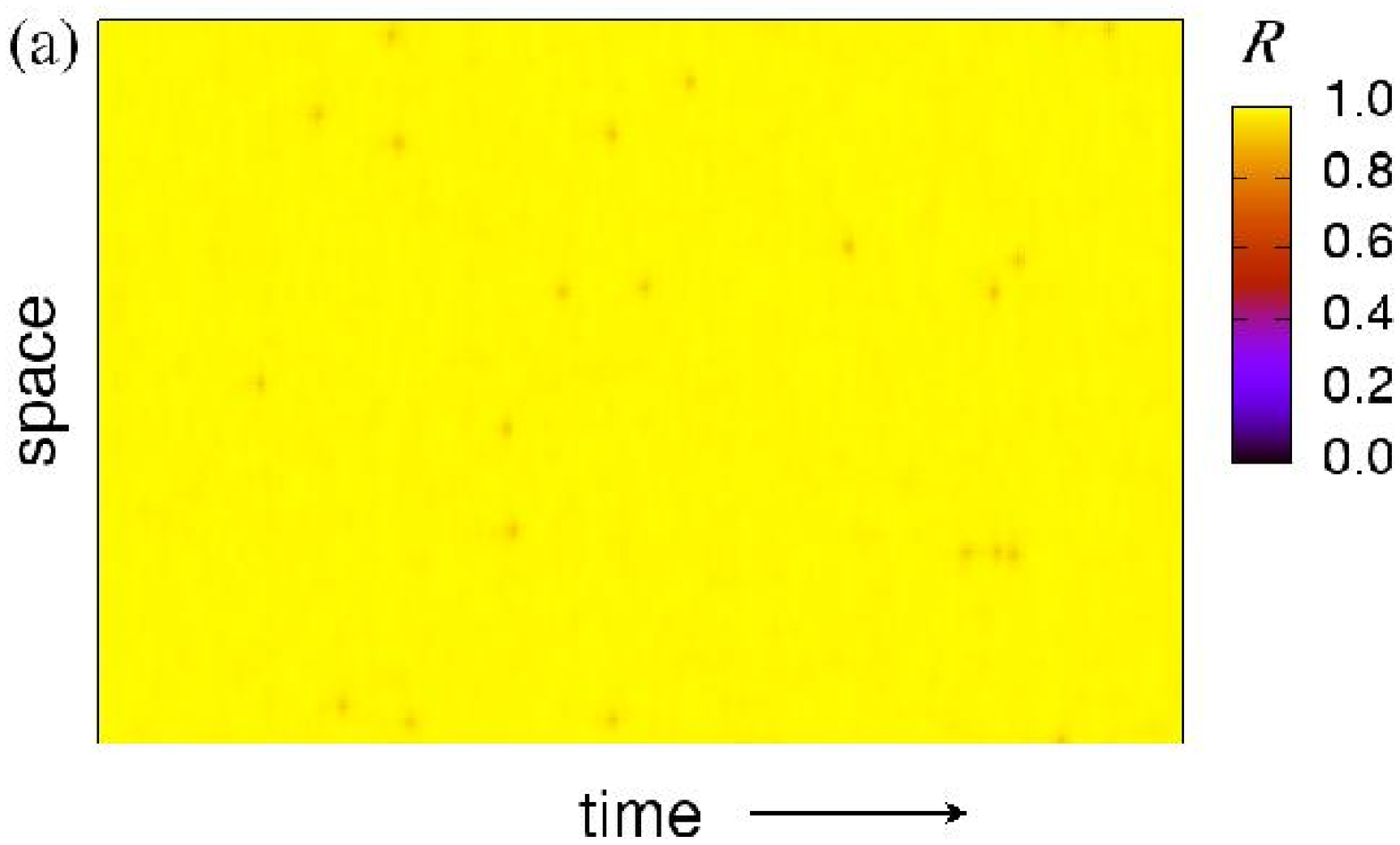}
\includegraphics[height=4cm]{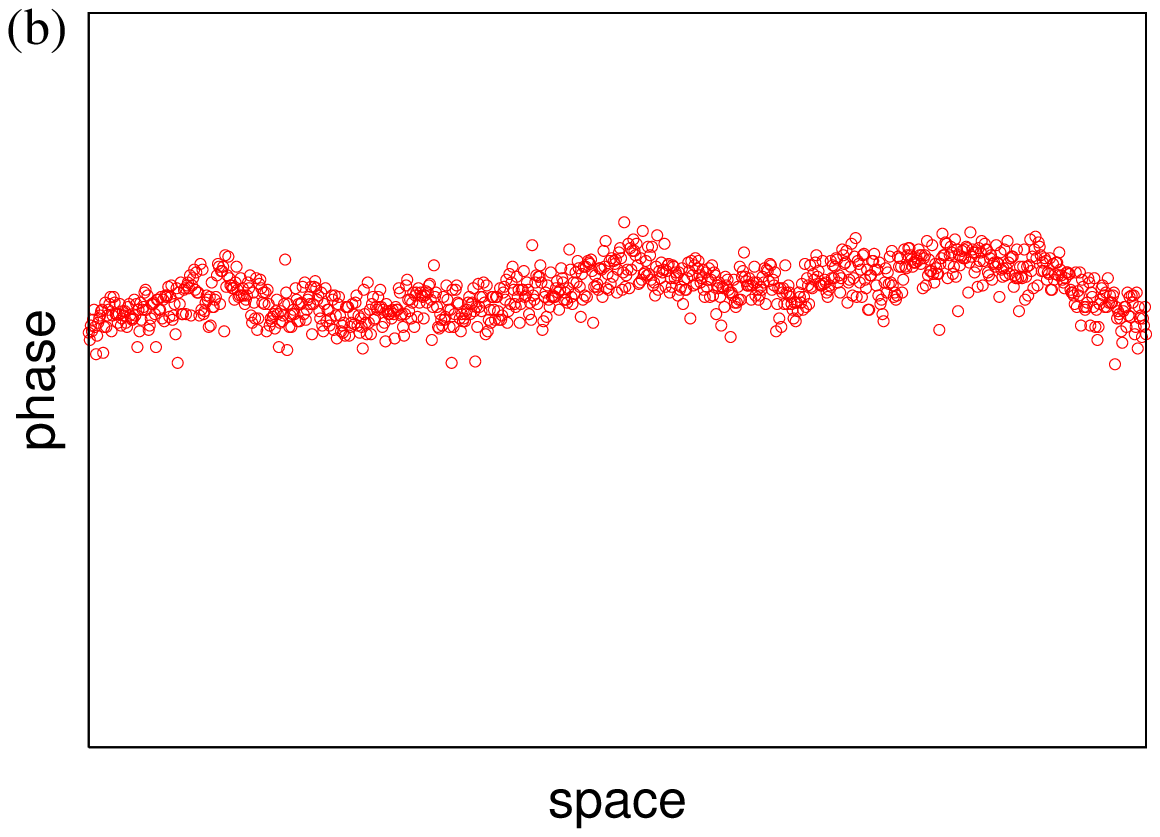}
\includegraphics[height=4cm]{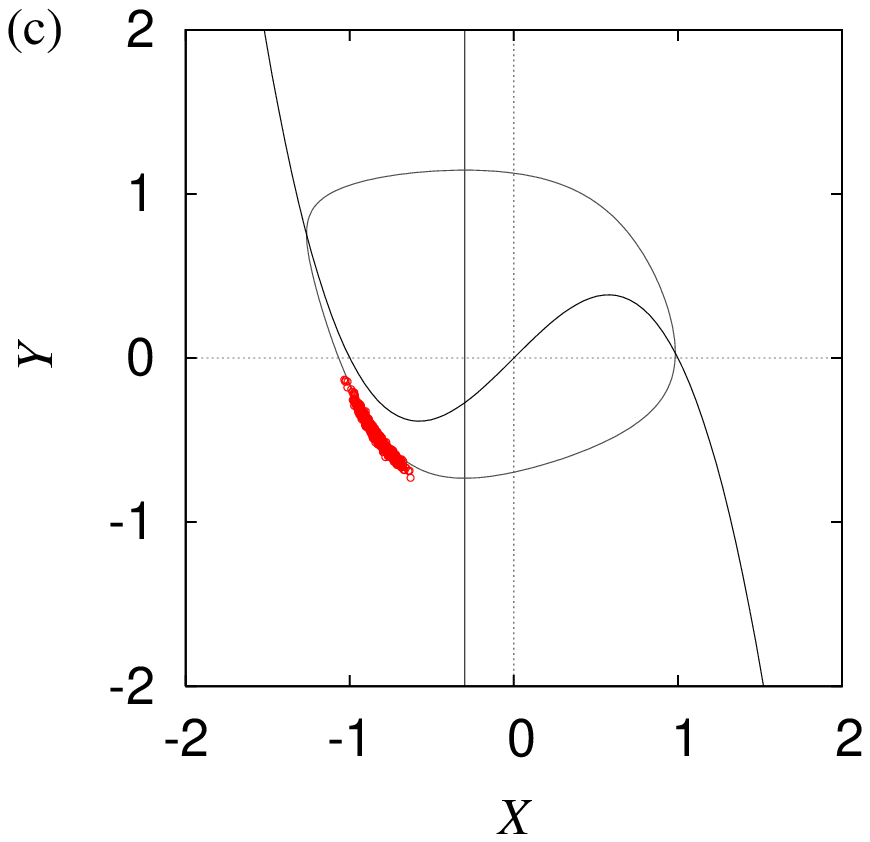}
\includegraphics[height=4cm]{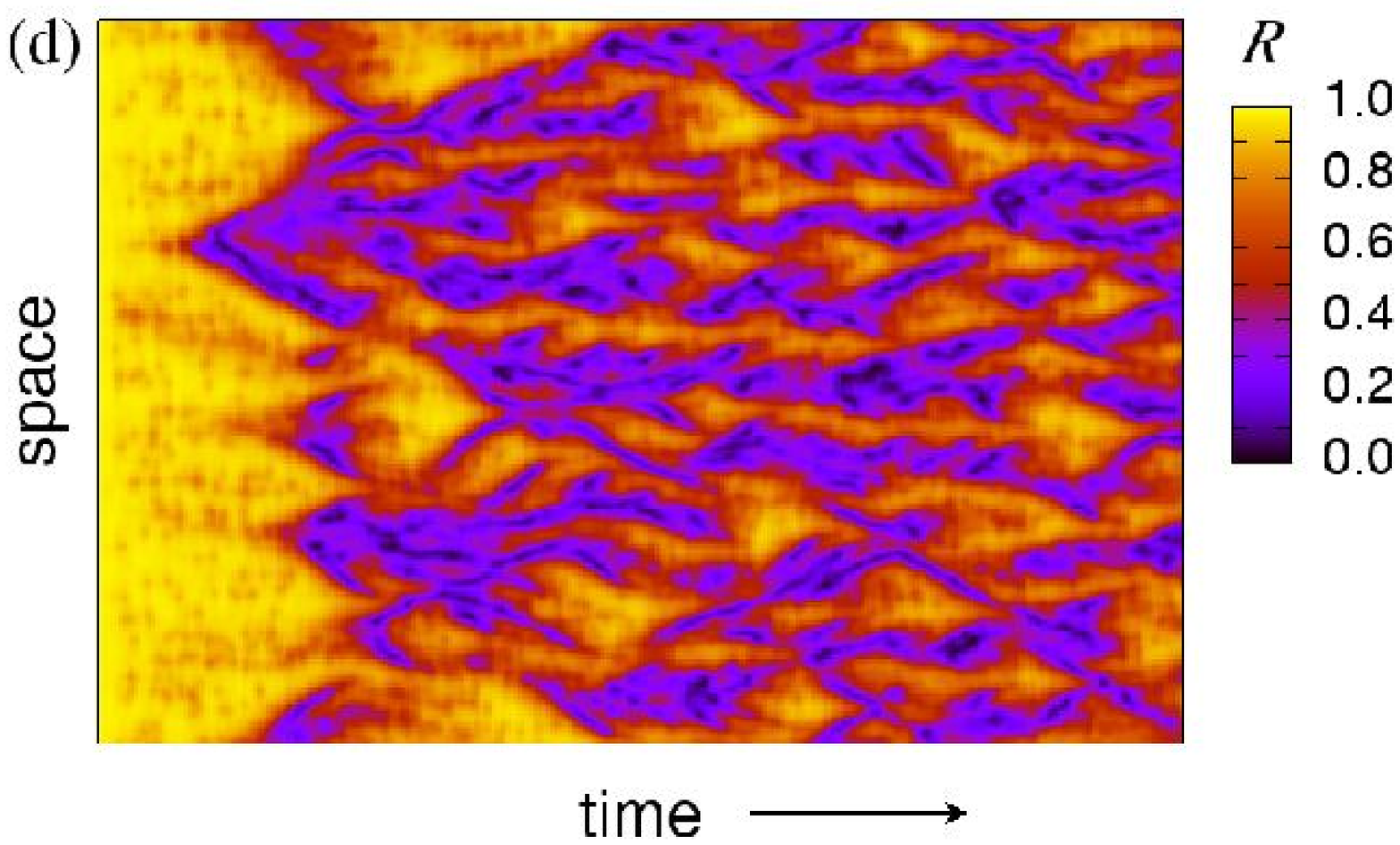}
\includegraphics[height=4cm]{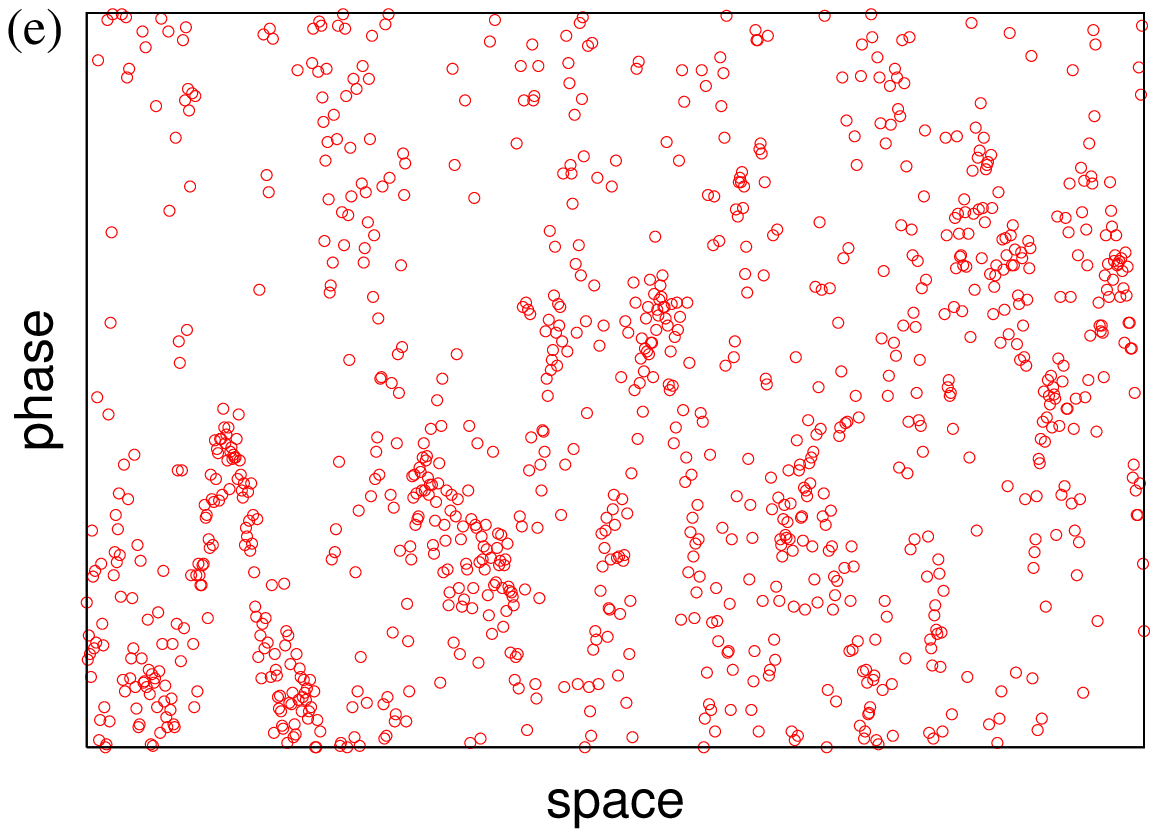}
\includegraphics[height=4cm]{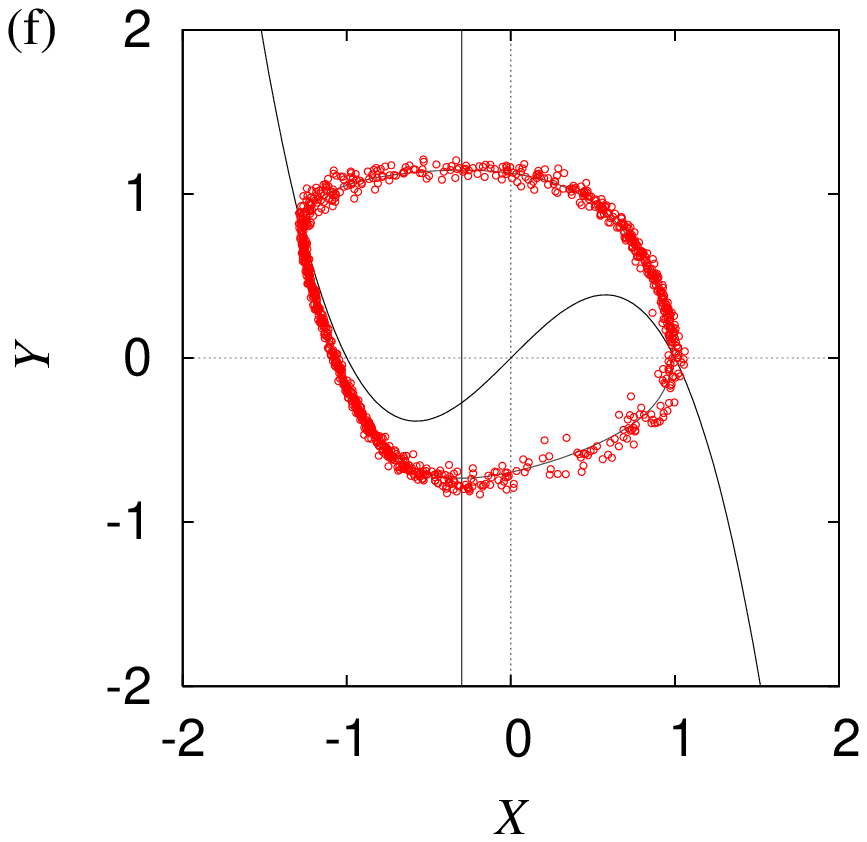}
\includegraphics[height=4cm]{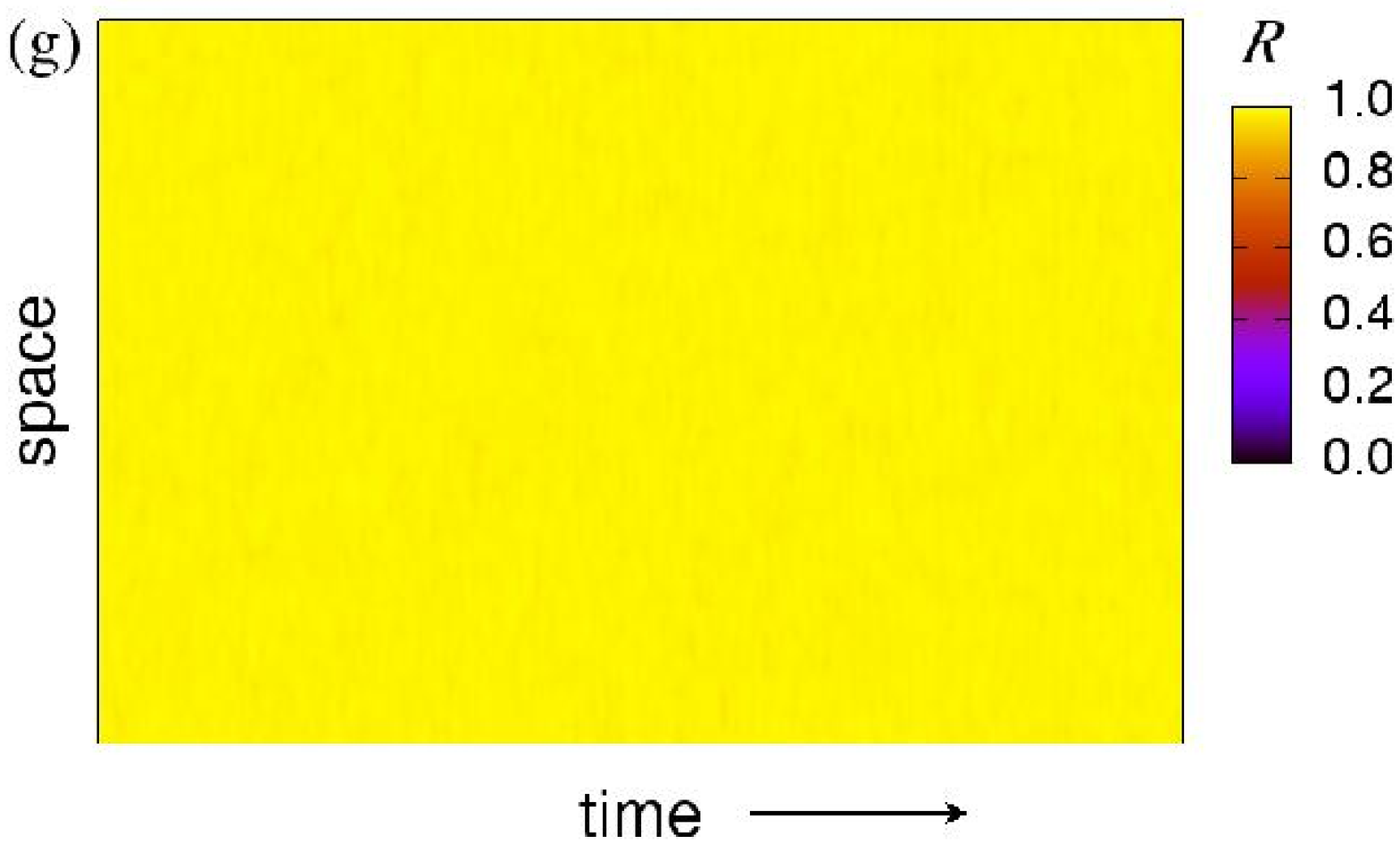}
\includegraphics[height=4cm]{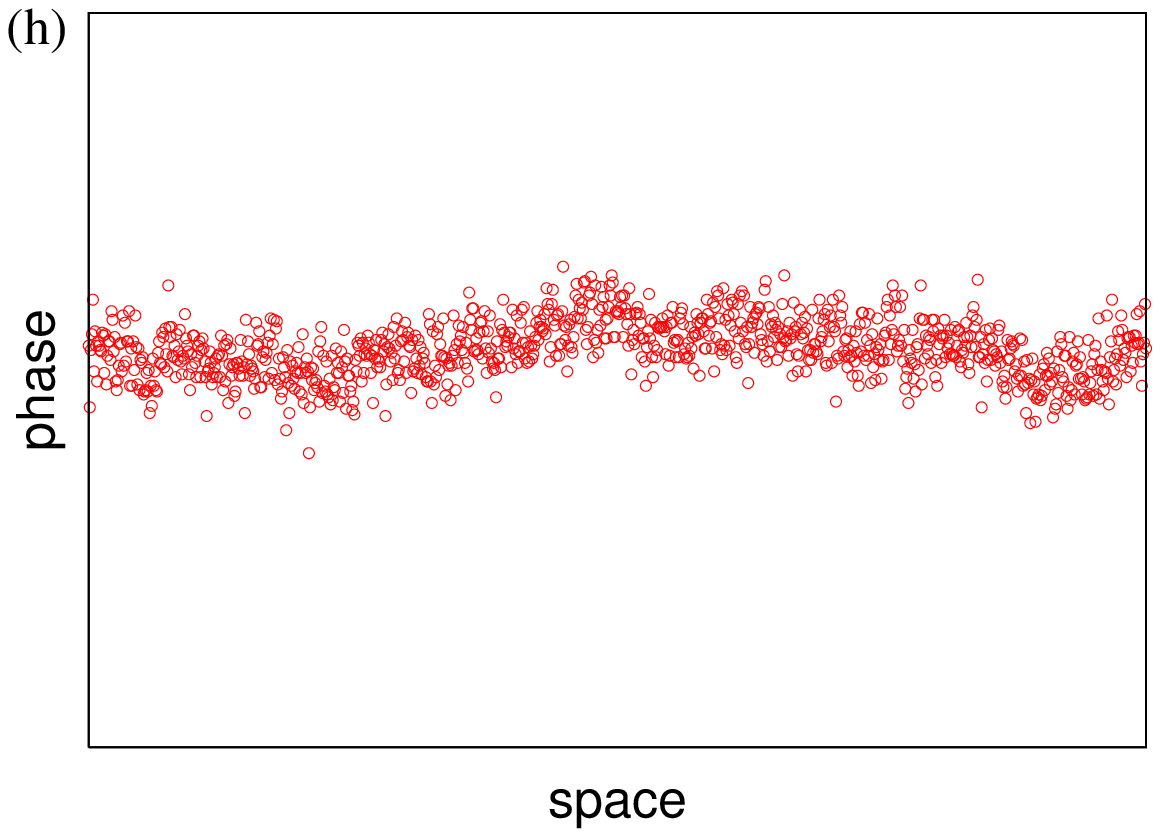}
\includegraphics[height=4cm]{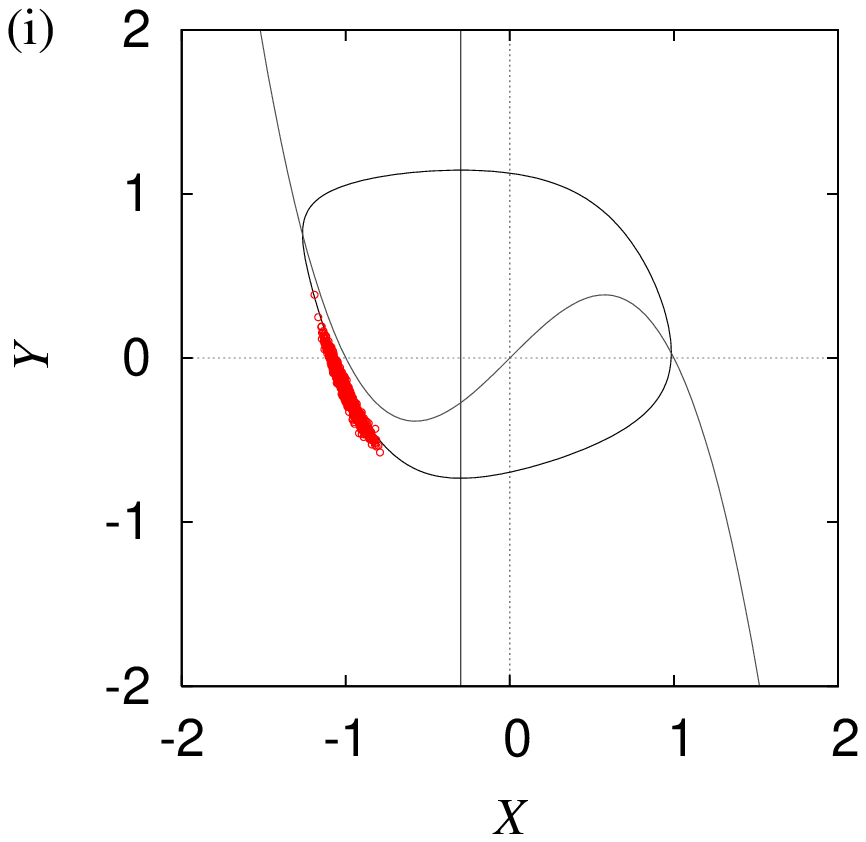}
\caption{(Color online) Numerical simulations of nonlocally coupled
  noisy FitzHugh-Nagumo oscillators for three representative cases.
  (a), (b), and (c): Case FN-I ($K_X=0.02$, $K_Y=0.08$, and $\sigma=0.0002$);
  (d), (e), and (f): Case FN-II ($K_X=0.02$, $K_Y=0.08$, and $\sigma=0.0006$);
  (g), (h), and (i): Case FN-III ($K_X=0.08$, $K_Y=0.02$, and $\sigma=0.0006$).
  Other parameters are fixed as $\epsilon=0.5$, $a=1.0$, and $b=0.3$.
  The left panels (a), (d), and (g) show spatiotemporal patterns of
  the order parameter modulus; the middle panels (b), (e), and (h)
  show instantaneous spatial profile of the local oscillator phase;
  the right panels (c), (f), and (i) show corresponding phase
  portraits in the $X$-$Y$ plane.
  The number of oscillators is $N=2^{10}$ and the separation between
  neighboring oscillators is $\varDelta x=0.1$.}
\label{fig:2}
\end{figure}
Results for three representative cases, denoted by FN-I, FN-II, and
FN-III, are illustrated in Fig.~\ref{fig:2} in rows.
The left panels display the spatiotemporal evolution of the modulus
$R(x,t)$ of the order parameter, the middle panels display how the
phases of the individual oscillators are distributed in space at a
given time, and the right panels display the snapshots of the state
variables of the oscillators on the $X$-$Y$ plane.

To see what happens when we increase the noise intensity from zero, we
first compare the cases FN-I (top row) and FN-II (middle row).
The coupling parameters used in FN-I and FN-II are the same, but the
noise intensity used in FN-II is three times stronger than that used
in FN-I.
In the weak noise case, FN-I, the modulus of the order parameter (left
panel) is almost uniform in space and also constant in time.
The phases of the individual oscillators (middle and right panels)
somewhat fluctuate due to the noise, but the amplitude of the
fluctuation seems to be much smaller than that we expect for turbulent
fluctuations due to a dynamical instability of the system.
In the case FN-II, the noise is three times stronger.
Now the modulus of the order parameter exhibits quite irregular
spatiotemporal behavior, and the amplitude of the phase fluctuations
is much larger than that in the case FN-I, covering the whole range
from $0$ to $2\pi$.
Note also that the amplitude of the phase fluctuations is far larger
than the applied noise intensity, which indicates that they are
produced by a noise-induced dynamical instability of the system.

Only with these results, it might still be suspected that what looks
like turbulence in the case FN-II might actually not be true
turbulence, and the large fluctuations of the order parameter might
simply be the result of increasing the noise intensity by three times.
To clear up this suspicion, we compare the cases FN-II (middle row) and
FN-III (bottom row).
Now the noise intensities are the same, but the coupling parameters
are slightly different.
As clearly be seen, the order parameter observed in the case FN-III is
almost uniform in space and constant in time at the same noise level
as that used in the case FN-II.
Therefore, the violent order parameter fluctuations in the case FN-II
should not simply be statistical ones due to the noise.

To distinguish the two types of order parameter fluctuations more
clearly, we systematically vary the total number $N$ of the
oscillators in each case.
Since we fix the system length, we are controlling the oscillator
number density, or equivalently, the number of oscillators sitting
within the coupling range.
If the fluctuation of the order parameter is simply of statistical
origin coming from the finiteness of $N$, the variance should decrease
as $N^{-1}$ due to the central limit theorem.
In contrast, if the order parameter fluctuation is due to the
dynamical instability of the system, the same quantity should remain
constant even if $N$ is varied.

\begin{figure}
\centering
\includegraphics[height=6cm]{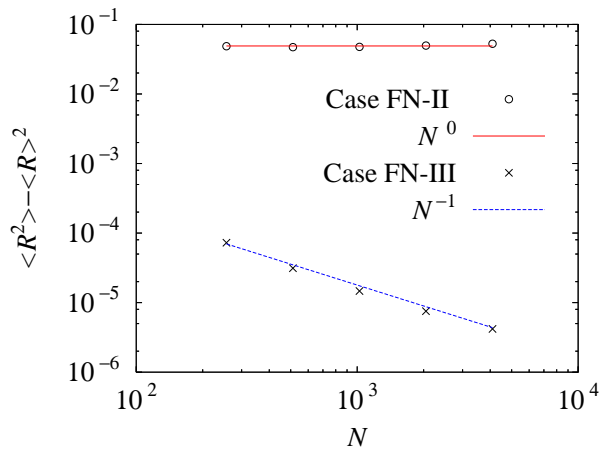}
\caption{(Color online) Order parameter fluctuation $\langle
  R^2\rangle-\langle R\rangle^2$ vs. oscillator number $N$ for the
  Case FN-II (open circles) and the Case FN-III (crosses).
  $\langle\cdot\rangle$ represents the space-time average.
  The solid and dashed lines have slopes 0 and -1, respectively.
  The system size is fixed as $L=N\varDelta x=102.4$.}
\label{fig:3}
\end{figure}
The results are shown in Fig.~\ref{fig:3}, where the variance of each
order parameter is plotted as a function of $N$ in double-logarithmic
scales for the cases FN-II and FN-III.
We see that there is a vast difference in the amplitude of
fluctuations, and, as expected, its dependence on $N$ is clearly
different between the two cases; the fluctuation amplitude is almost
constant in the case FN-II, whereas it decreases in inverse proportion
to $N$ in the case FN-III.

Thus, we conclude that the phase fluctuations observed in the case
FN-II is not merely finite-sample statistical fluctuations, but they
are actually turbulent fluctuations generated by the effective
dynamical instability of the system induced by the weak external noise.

\subsection{Stuart-Landau oscillators}

Our second example is a system of Stuart-Landau (SL) oscillators with
nonlocal coupling, which is described by
\begin{equation}
  \partial_t W(x,t) = \left(1+i\omega_0\right)W
  -\left(1+i\beta\right)\left|W\right|^2 W
  +K S_W(x,t)+\sqrt{\sigma}\,\eta_W(x,t),
\label{eq:11}
\end{equation}
and
\begin{equation}
  S_W\left(x,t\right) = \int_{-\infty}^{\infty}dx'\,
  G\left(x-x'\right) W\left(x',t\right).
\label{eq:12}
\end{equation}
The Stuart-Landau oscillator is the simplest limit-cycle oscillator
derived as a normal form of the supercritical Hopf
bifurcation~\cite{ref:kuramoto84}.
Each oscillator state is now described by a complex amplitude, $W$.
Correspondingly, the coupling term $S_W$ and the noise $\eta_W$
are also complex variables.
We take the parameters $\omega$ and $K$ as $\omega_0=\beta+1.0$ and
$K=0.05$, and control the remaining parameters $\beta$ and $\sigma$.
As in the previous FitzHugh-Nagumo case, the values of the parameter
$\beta$ is chosen in such a way that the spatially uniform oscillation
is stable and the system is non-turbulent in the absence of noise.

We carried out the numerical analysis of this model completely in
parallel with the previous case of the FitzHugh-Nagumo oscillators
(see Appendix~\ref{sec:C} for the numerical methods).
For the Stuart-Landau oscillator, mapping from the complex amplitude
$W$ to the generalized phase $\phi$ can be analytically given 
as~\cite{ref:kuramoto84,ref:pikovsky01}
\begin{equation}
  \phi = \arg W - \beta \ln\left|W\right|.
\label{eq:13}
\end{equation}
Using this definition, we calculate the order parameter of the system
given by Eq.~(\ref{eq:10}).

\begin{figure}
\centering
\includegraphics[height=4cm]{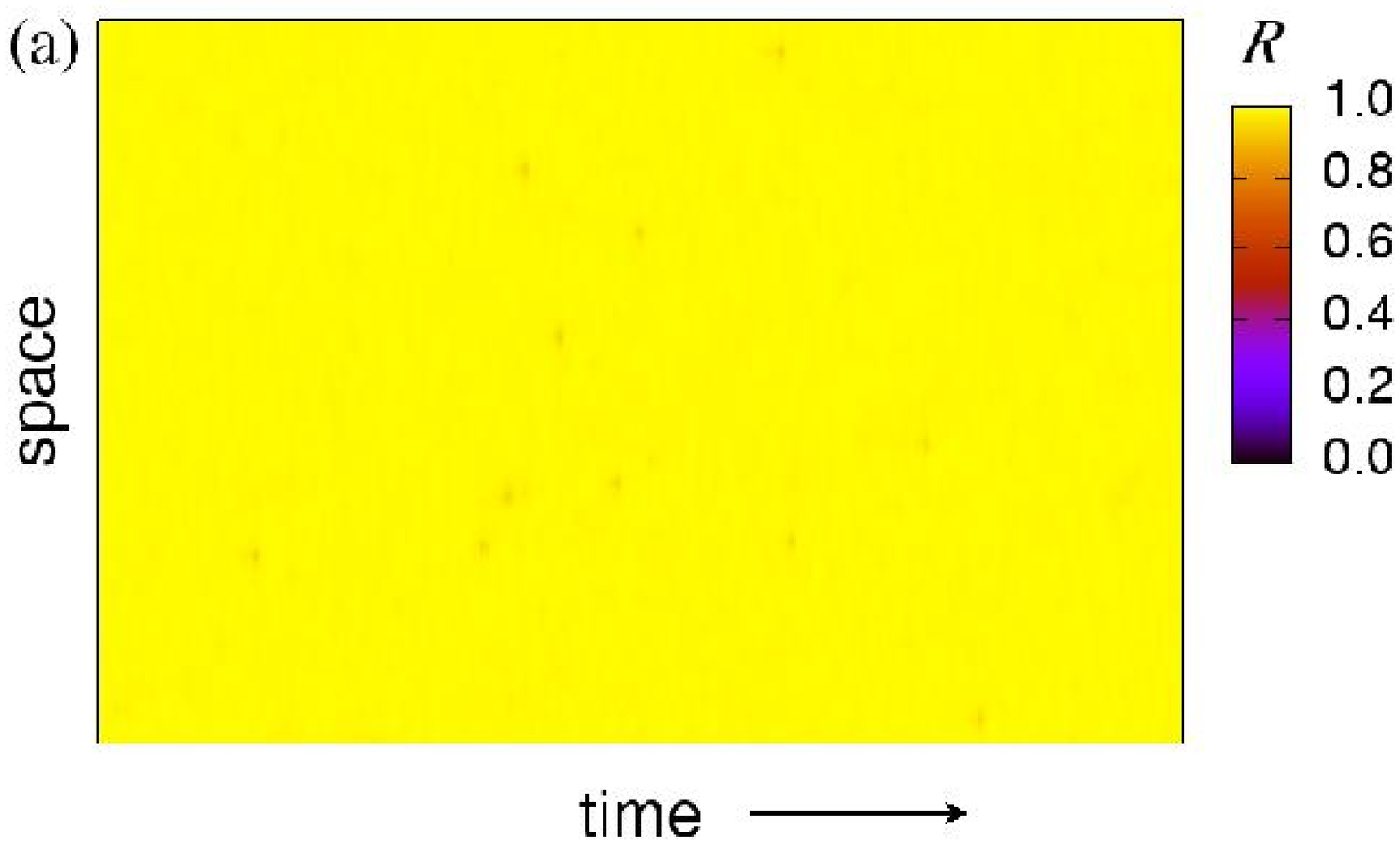}
\includegraphics[height=4cm]{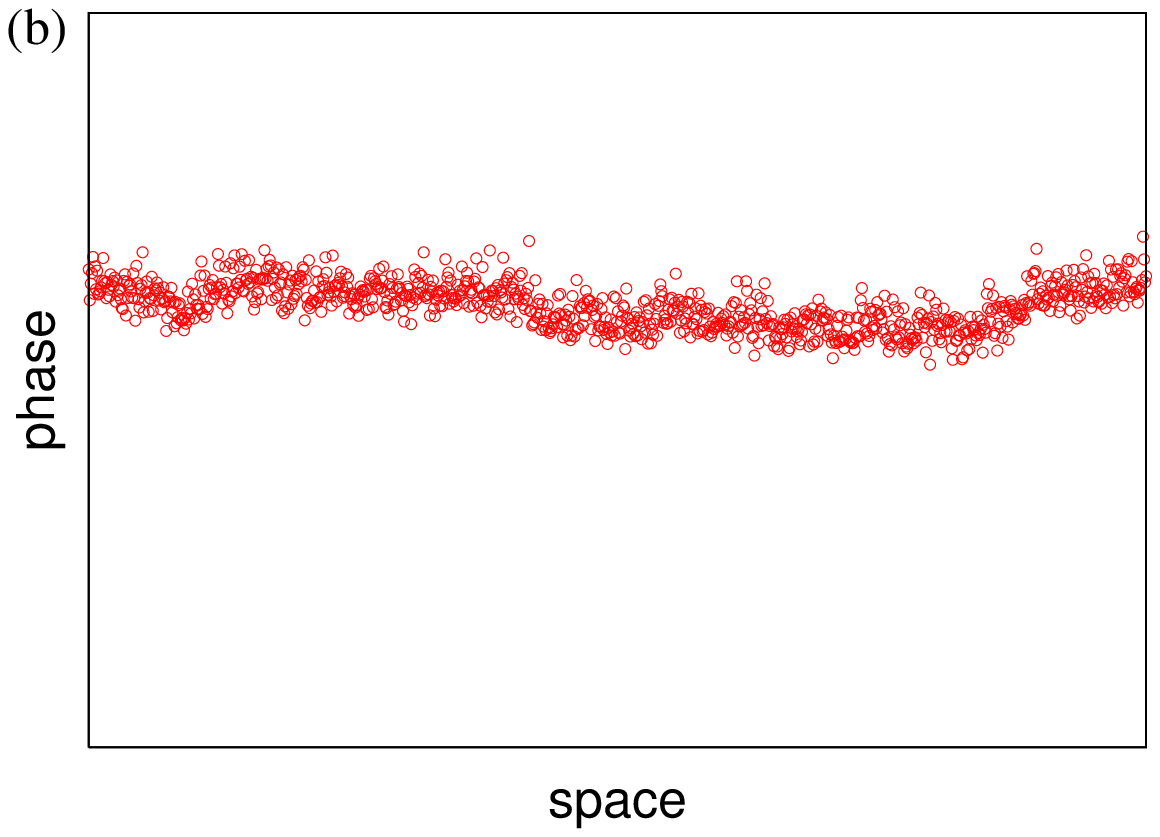}
\includegraphics[height=4cm]{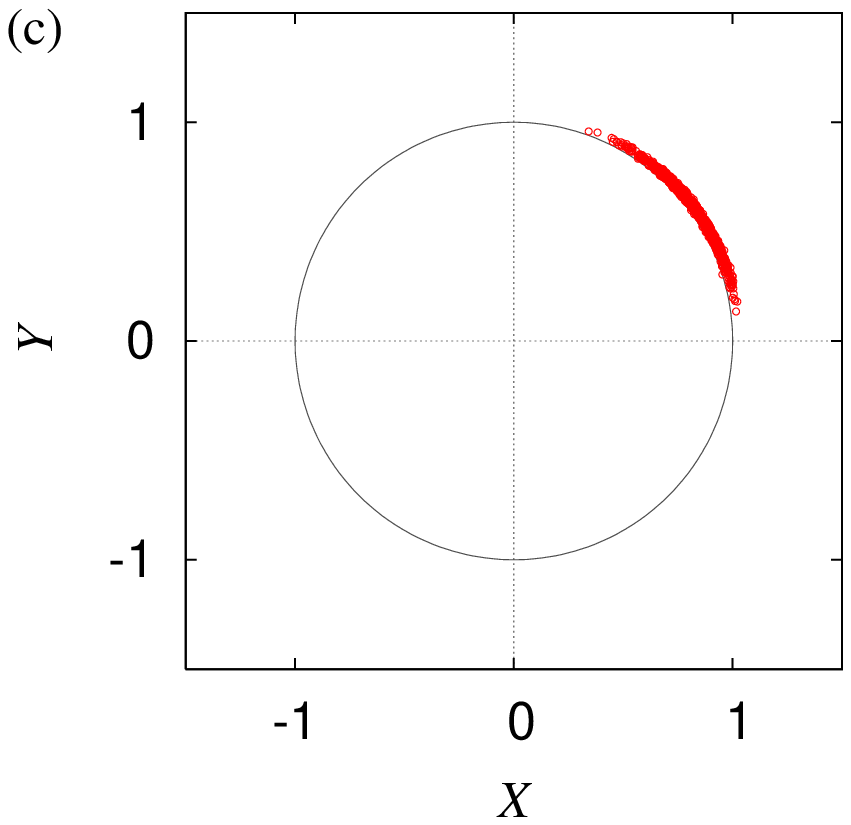}
\includegraphics[height=4cm]{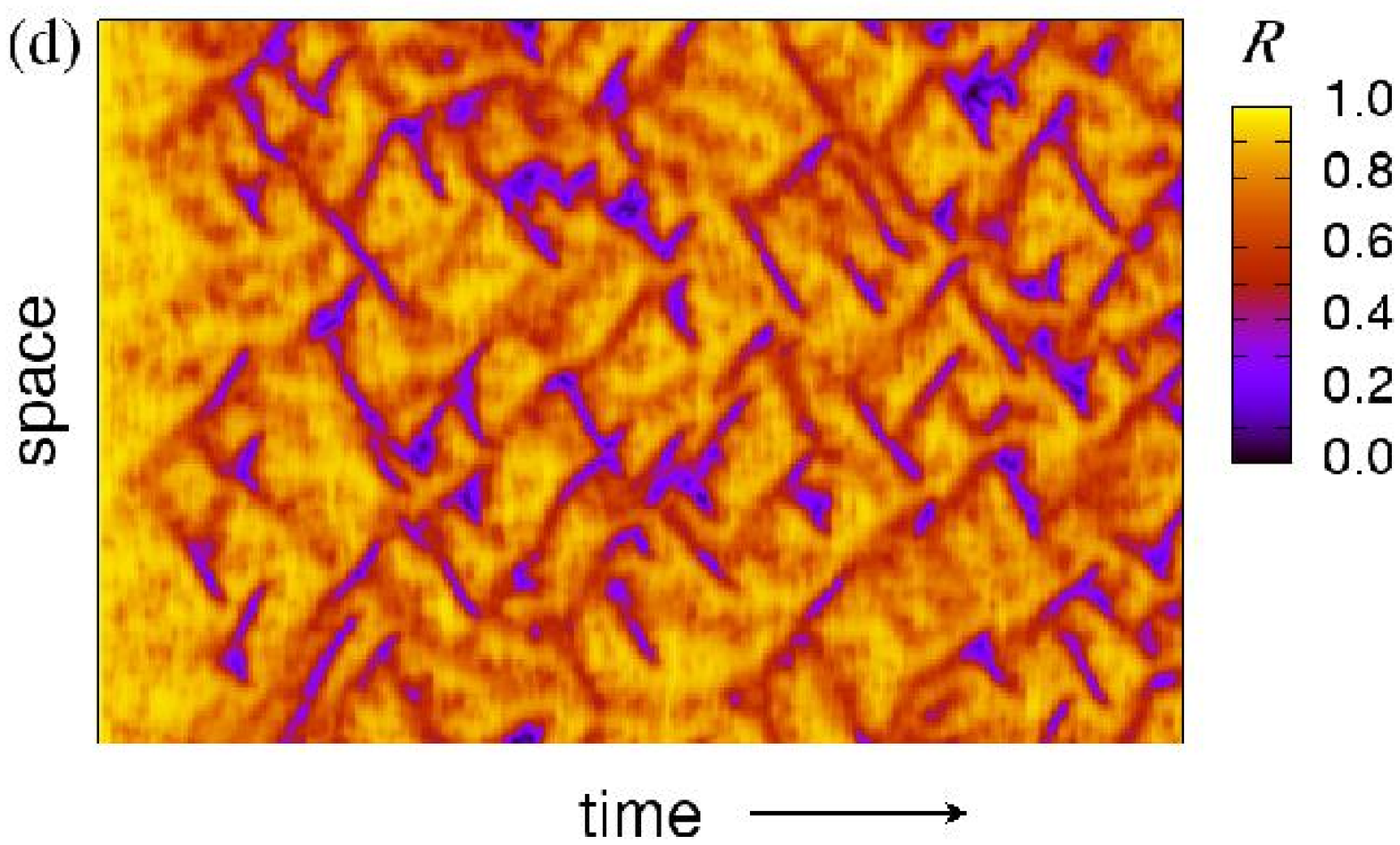}
\includegraphics[height=4cm]{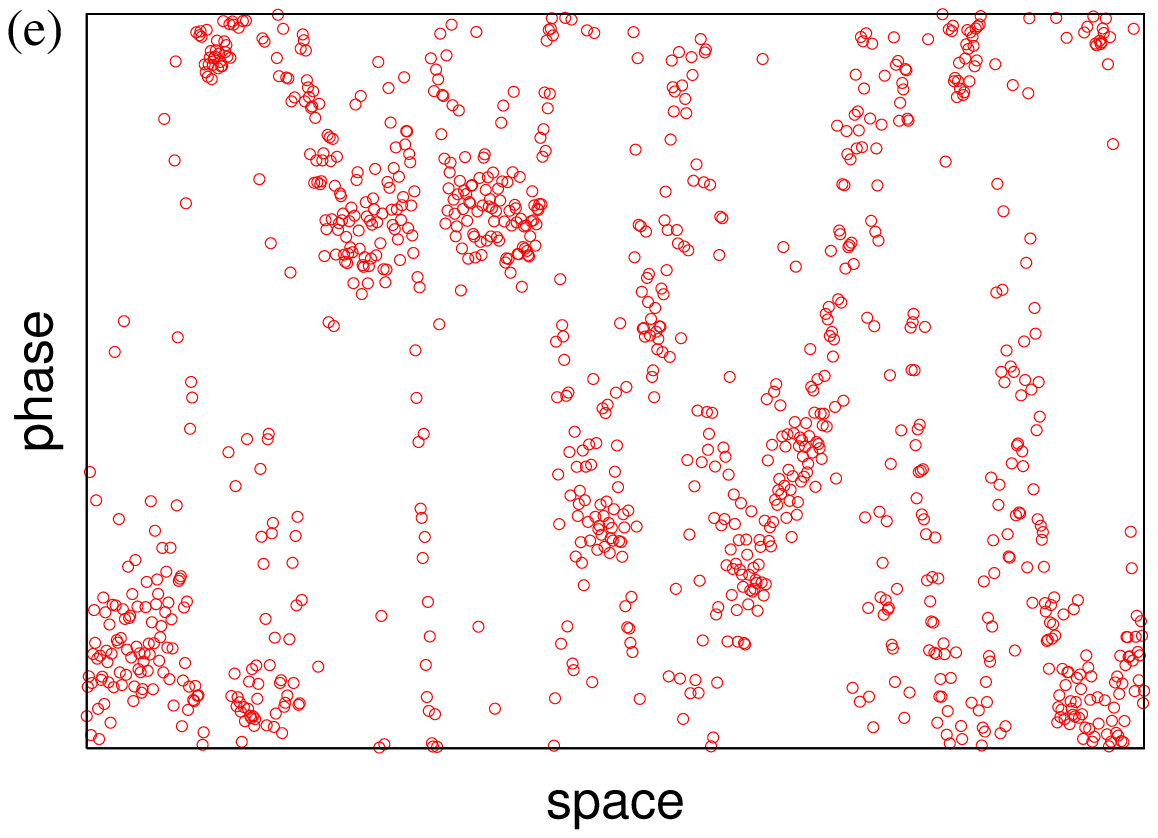}
\includegraphics[height=4cm]{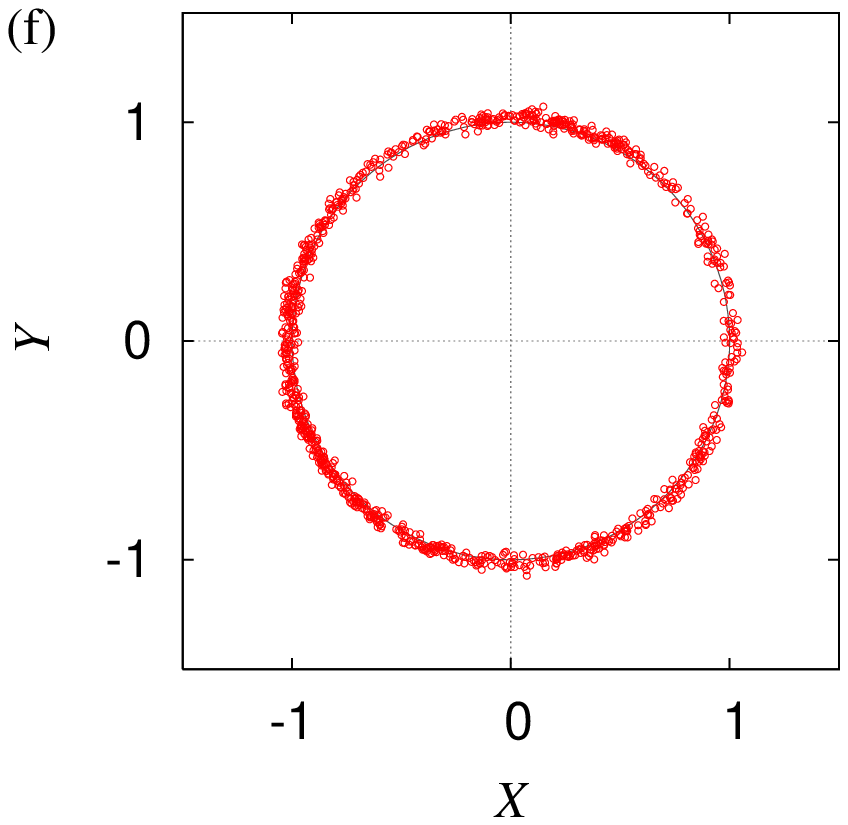}
\includegraphics[height=4cm]{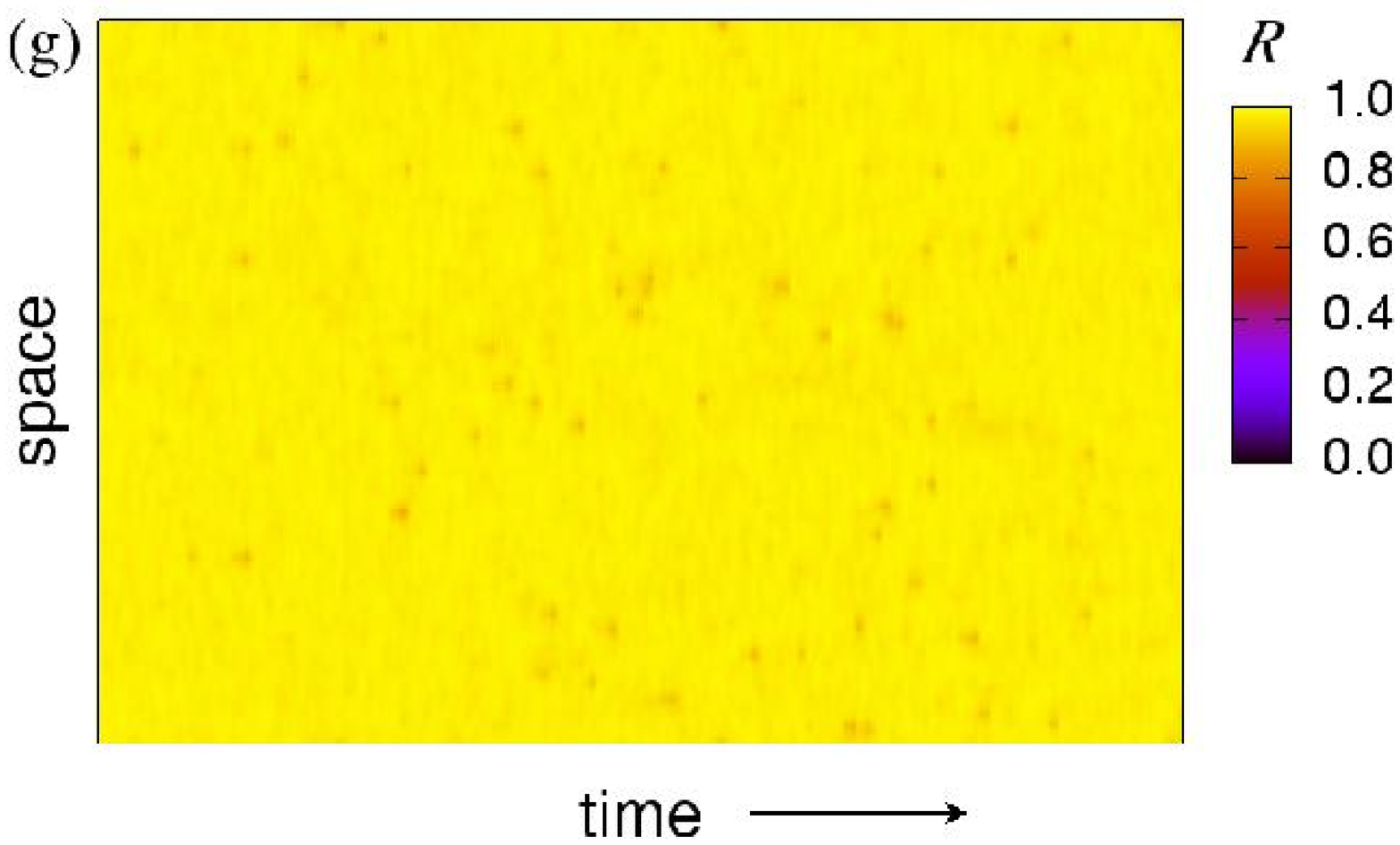}
\includegraphics[height=4cm]{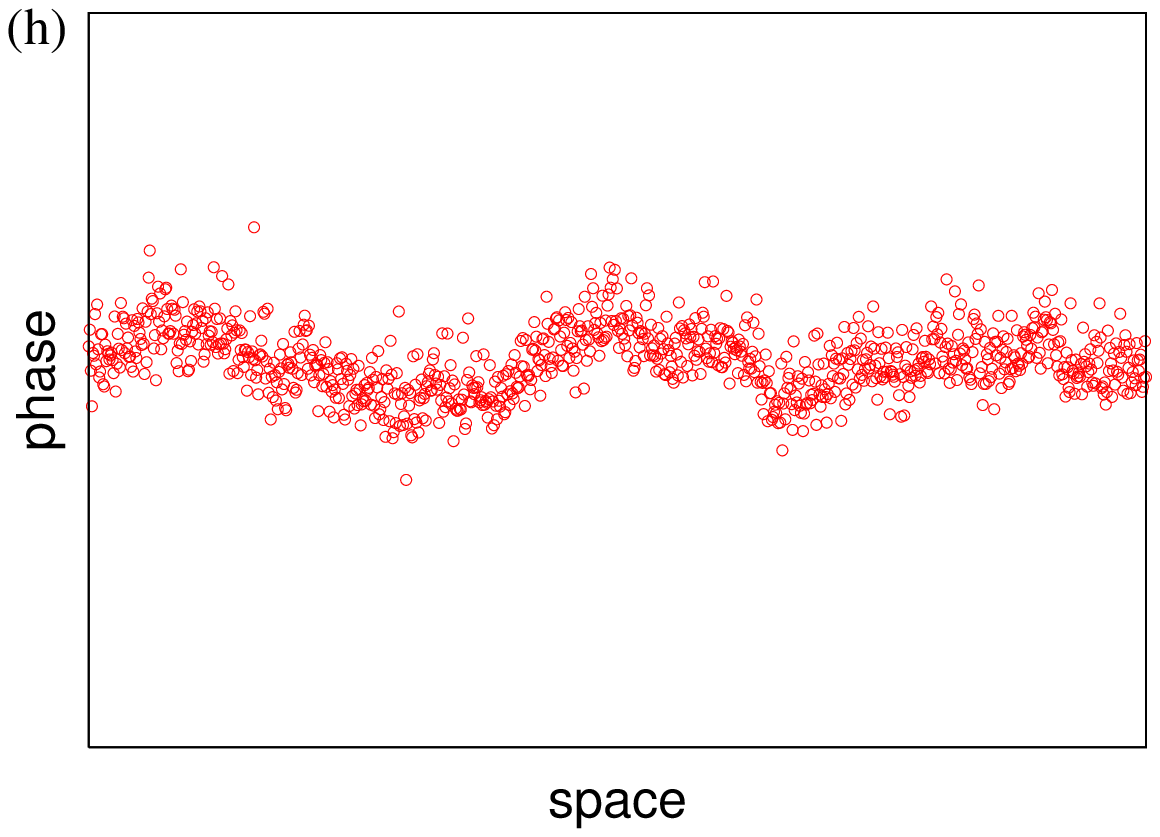}
\includegraphics[height=4cm]{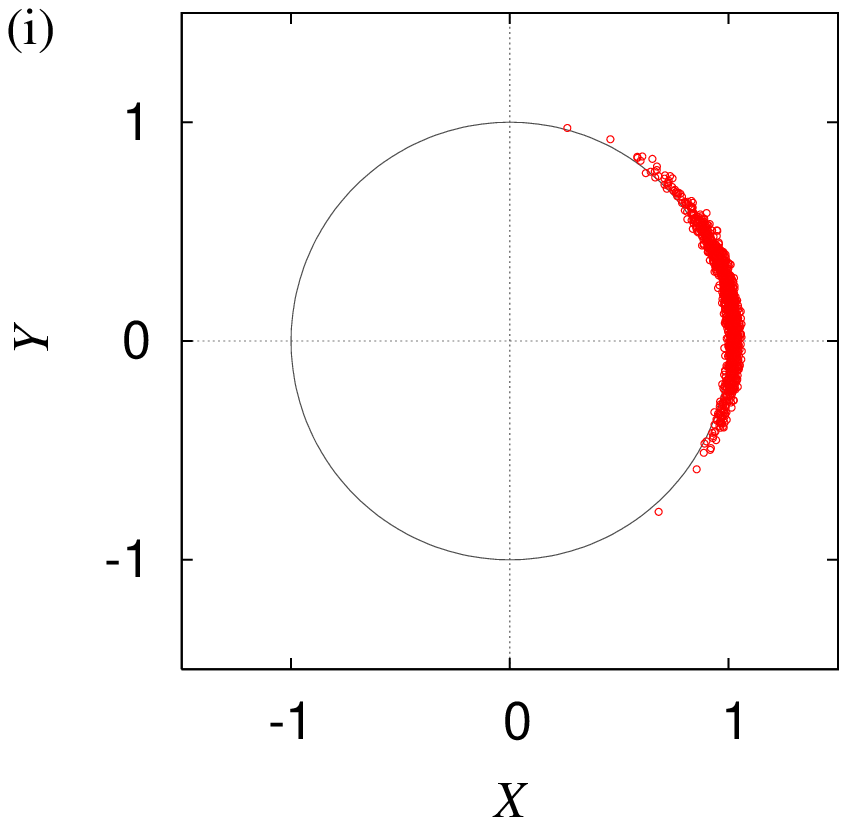}
\caption{(Color online) Numerical simulations of nonlocally coupled
  noisy Stuart-Landau oscillators for three representative cases.
  (a), (b), and (c): Case SL-I ($\beta=2.6$ and $\sigma=0.0001$);
  (d), (e), and (f): Case SL-II ($\beta=2.6$ and $\sigma=0.0006$);
  (g), (h), and (i): Case SL-III ($\beta=1.6$ and $\sigma=0.0006$).
  Other parameters are given by $\omega_0=\beta+1.0$ and $K=0.05$.
  The left panels (a), (d), and (g) show the spatiotemporal patterns
  of the order parameter modulus; the middle panels (b), (e), and (h)
  show the instantaneous spatial profile of the local oscillator
  phase; the right panels (c), (f), and (i) show the corresponding
  phase portraits in the $X$-$Y$ plane.
  The number of oscillators is $N=2^{10}$ and the separation between
  neighboring oscillators is $\varDelta x=0.1$.}
\label{fig:4}
\end{figure}
Figure~\ref{fig:4} summarizes the numerical results, where three
representative cases SL-I, SL-II, and SL-III are compared.
As previous, the parameter $\beta$ is the same for SL-I and SL-II, but
the noise intensity in SL-II is six times larger than that in SL-I.
In the case SL-I, the amplitude of the phase fluctuations is quite
small, and the modulus $R(x,t)$ of the order parameter is almost
spatially uniform and temporally constant.
In the case SL-II, the amplitude of the phase fluctuations becomes
quite large, and the modulus $R(x,t)$ of the order parameter exhibits
complex spatiotemporal dynamics.
Thus, the strong turbulent fluctuation arises from the spatially uniform
oscillation as the noise intensity is increased.
However, when the parameter $\beta$ is slightly changed, SL-III, the
noise with the same intensity cannot induce such turbulent
fluctuations.

From the same argument as the previous FitzHugh-Nagumo case, we
conclude that the strong fluctuation seen in the case SL-II represents
a genuine turbulence of the dynamical origin induced by the external noise.

\section{Reduction to Nonlocally coupled noisy phase oscillators}
\label{sec:3}

We have observed that nonlocally coupled limit-cycle oscillators can
exhibit noise-induced turbulent states.
In the present and the subsequent sections, we will develop a theory
that explains consistently the above numerical results.
In this section, we present our first step, namely, the derivation of
a Langevin phase equation from the original dynamical equation for the
nonlocally coupled limit cycles by means of the phase reduction
method.
The resulting equation describes a system of nonlocally coupled phase
oscillators, whose validity is demonstrated numerically for the
Stuart-Landau oscillators.
We also derive an equivalent nonlinear Fokker-Planck equation through
the mean-field theory, which will be the starting point for further
analysis.

\subsection{Phase reduction}

We apply the standard phase reduction method to our system of
nonlocally coupled noisy limit-cycle oscillators, which derives an
approximate equation consisting of only the phase variable from the
original dynamical equations in multiple variables~\cite{ref:kuramoto84}.
The phase reduction is allowed when the individual local oscillators
are perturbed only slightly.
Thus, the coupling strength and the external noise intensity should be
sufficiently small.
As shown in the phase portraits in Figs.~{\ref{fig:2}} and
{\ref{fig:4}} (right panels), the oscillators are always in
the near vicinity of the unperturbed limit-cycle orbits.
Therefore, the parameter values used in the previous numerical
analysis satisfy the above condition.

As we already mentioned, we use a specific definition of the phase,
determined on the phase space of the limit cycle in such a way that
$\dot{\phi}=\omega$ holds identically, where $\omega$ is the natural
frequency of the local oscillator. This can always be done by an
appropriate nonlinear transformation of the phase space variables of
the oscillator~\cite{ref:winfree80,ref:kuramoto84,ref:pikovsky01}.

Details of the derivation of a phase equation from Eq.~(\ref{eq:1})
are given in Appendix~\ref{sec:A}.
The resulting phase equation takes the form
\begin{equation}
  \partial_t\phi\left(x,t\right)=\omega+\int^{\infty}_{-\infty}
  dx'\,G\left(x-x'\right)\Gamma\bigl(\phi(x,t)-\phi(x',t)\bigr)
  +\sqrt{D}\,\xi\left(x,t\right),
\label{eq:14}
\end{equation}
where $\Gamma(\phi)$ represents the phase coupling function between
the oscillators, $D$ the effective noise intensity that inherits the
effect of the noise $\bd{\eta}(x,t)$ in the original equation, and
$\xi(x,t)$ a real scalar spatiotemporally white Gaussian noise
satisfying
\begin{equation}
\left\langle\xi\left(x,t\right)\right\rangle=0,\quad
\left\langle\xi\left(x,t\right)\xi\left(x',t'\right)\right\rangle
=2\delta\left(x-x'\right)\delta\left(t-t'\right).
\label{eq:15}
\end{equation}
The phase coupling function $\Gamma(\phi)$ can be calculated from the
dynamical equations of the coupled limit cycles.
The effective noise intensity $D$ can also be calculated once the
parameters of the original dynamical equations are given.
Note that the above equation describes a system of nonlocally coupled
noisy phase oscillators.

\subsection{FitzHugh-Nagumo oscillators}

We first consider the case of the FitzHugh-Nagumo oscillators.
By applying the formula developed in Appendix~\ref{sec:A},
the phase coupling function can be expressed as
\begin{equation}
  \Gamma\left(\phi-\phi'\right)=\frac{1}{2\pi}\int_0^{2\pi}d\lambda\;
  \Bigl[K_X Z_X\left(\lambda+\phi\right)+K_Y Z_Y\left(\lambda+\phi\right)\Bigr]
  X_0\left(\lambda+\phi'\right),
\label{eq:16}
\end{equation}
where $X_0(\phi)$ is the $X$-component of the unperturbed limit cycle,
and $Z_X(\phi)$ and $Z_Y(\phi)$ are the phase sensitivity functions of
the FitzHugh-Nagumo oscillator.
Though the limit-cycle solution of the FitzHugh-Nagumo model cannot be
obtained analytically, these quantities can be calculated numerically
with sufficient precision using standard methods~\cite{ref:izhikevich07}.
\begin{figure}
\centering
\includegraphics[height=6cm]{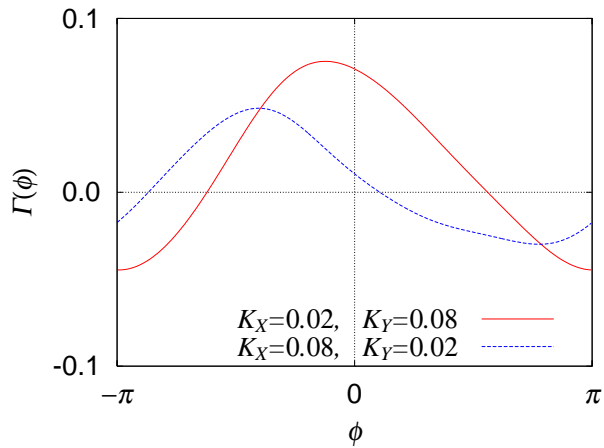}
\caption{(Color online) Phase coupling functions $\Gamma(\phi)$
  of the FitzHugh-Nagumo oscillators.
  The solid and dashed curves correspond to the Cases FN-I
  and FN-II ($K_X=0.02$ and $K_Y=0.08$) and the Case FN-III
  ($K_X=0.08$ and $K_Y=0.02$), respectively.
  The in-phase coupling condition is satisfied in both cases, i.e.,
  $d\Gamma(\phi)/d\phi|_{\phi=0}<0$.}
\label{fig:5}
\end{figure}
Figure~\ref{fig:5} displays the phase coupling function $\Gamma(\phi)$
calculated for the two sets of the coupling constants $K_X$ and $K_Y$
used in section~\ref{sec:2}.
For both parameter conditions, the coupling functions are
the in-phase type, as is clear from the property
$d\Gamma(\phi)/d\phi|_{\phi=0}<0$~\cite{ref:kuramoto84}.

The effective noise intensity $D$ can be expressed by the original
noise intensity $\sigma$ and the phase sensitivity functions $Z_X$
and $Z_Y$ as
\begin{equation}
  D=\frac{1}{2\pi}\int_0^{2\pi}d\lambda\;
  \sigma  \Bigl[Z_X^2(\lambda)+Z_Y^2(\lambda)\Bigr].
\label{eq:17}
\end{equation}

\subsection{Stuart-Landau oscillators}

In the case of the Stuart-Landau oscillators, the limit-cycle solution
and the phase sensitivity function can be obtained analytically.
The natural frequency, the phase coupling function, and the effective
noise intensity are given by $\tilde{\omega} = \omega_0 - \beta$,
$\Gamma(\phi - \phi') = - K\sqrt{1 + \beta^2}\sin(\phi - \phi' + \alpha)$,
and $\tilde{D} = \sigma(1 + \beta^2)$, respectively.
Thus, Eq.~(\ref{eq:14}) takes the following explicit form
\begin{equation}
  \partial_t \phi(x,t) = \tilde{\omega}
  -K\sqrt{1+\beta^2}\int_{-\infty}^\infty dx'\,
  G\left(x-x'\right)\sin\bigl(\phi(x,t)-\phi(x',t)+\alpha\bigr)
  +\sqrt{\tilde{D}}\,\xi\left(x,t\right),
\label{eq:18}
\end{equation}
where the parameter $\alpha$ is given by
\begin{equation}
  \alpha=\arg\left(1+i\beta\right).
\label{eq:19}
\end{equation}
By changing the time scale, we can simplify the above equation as
\begin{equation}
  \partial_t \phi\left(x,t\right) = \omega - \int_{-\infty}^{\infty}dx'\,
  G\left(x-x'\right)\sin\bigl(\phi(x,t)-\phi(x',t)+\alpha\bigr)
  +\sqrt{D}\,\xi\left(x,t\right),
\label{eq:20}
\end{equation}
where the rescaled noise intensity is given by
\begin{equation}
  D = \frac{\sigma\sqrt{1+\beta^2}}{K}.
\label{eq:21}
\end{equation}

To see the validity of the phase reduction, we present here results of
direct numerical simulations of the Langevin phase equation
(\ref{eq:18}).
\begin{figure}
\centering
\includegraphics[height=5cm]{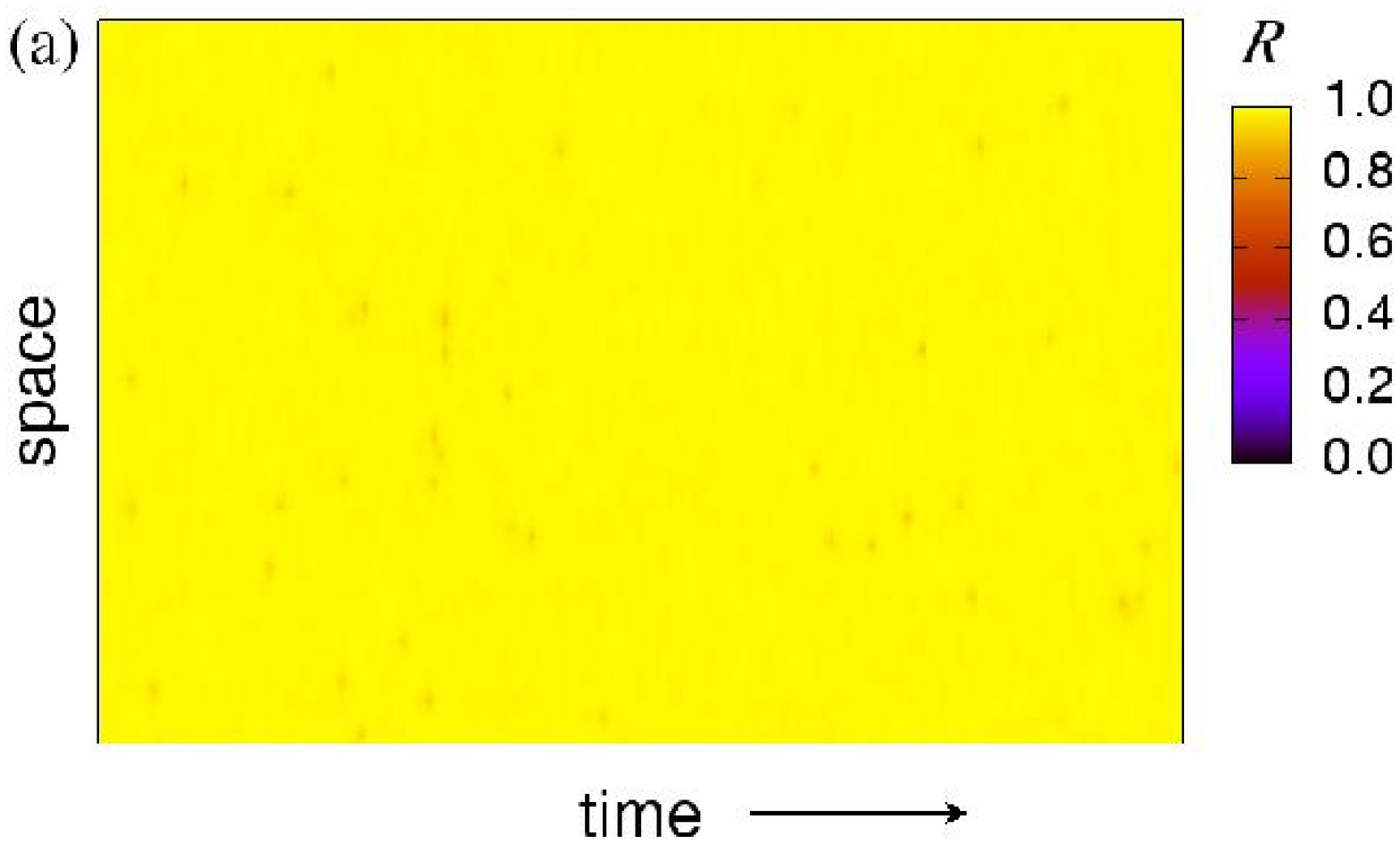}
\includegraphics[height=5cm]{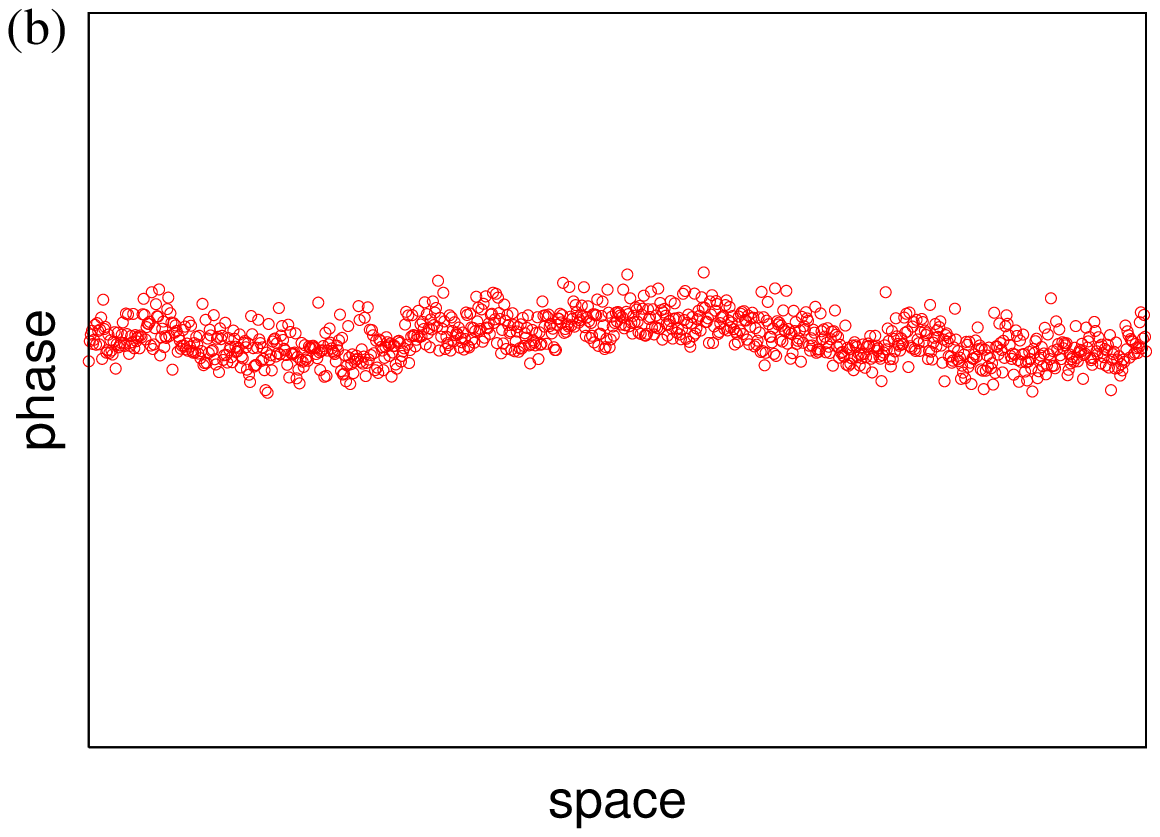}
\includegraphics[height=5cm]{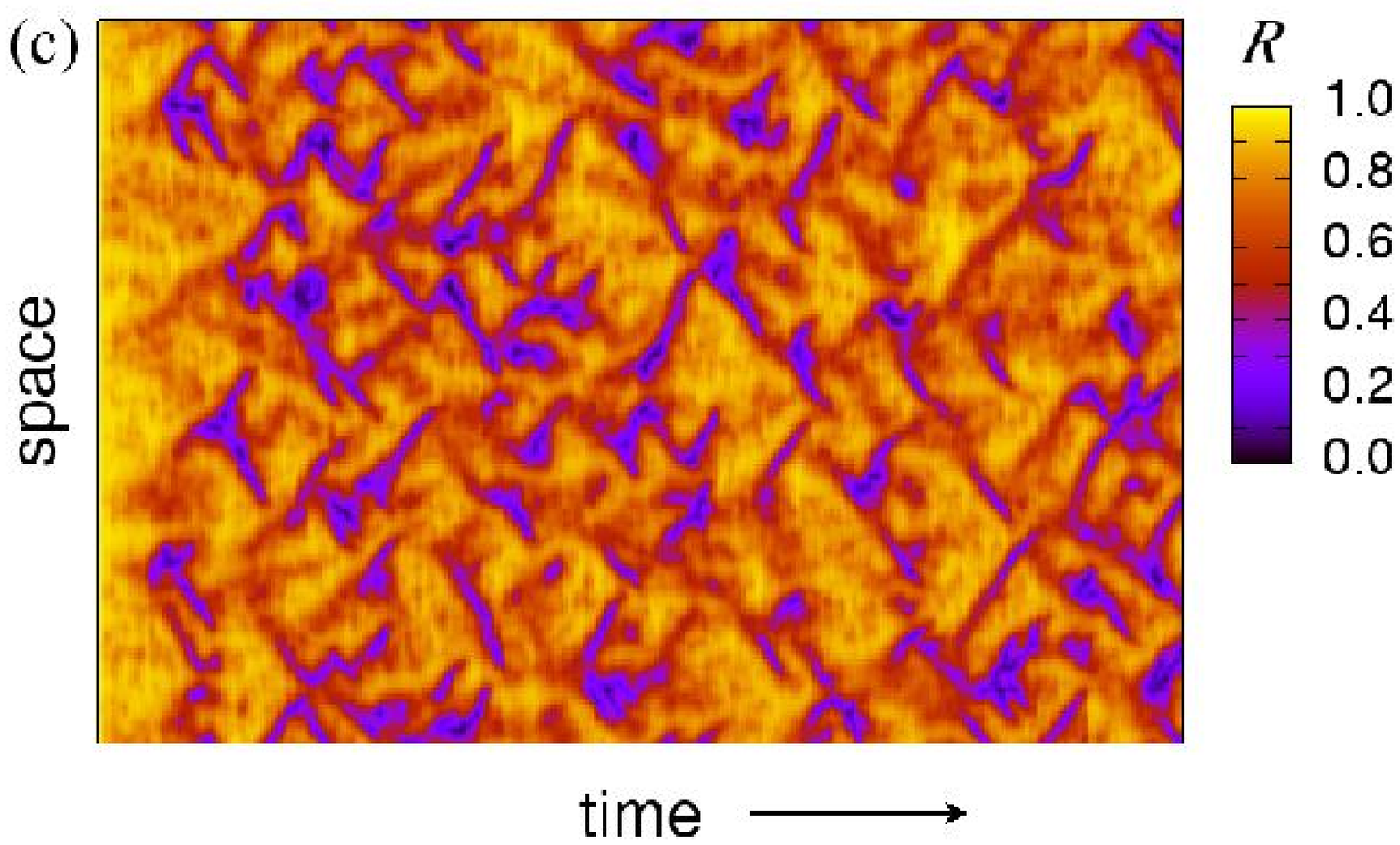}
\includegraphics[height=5cm]{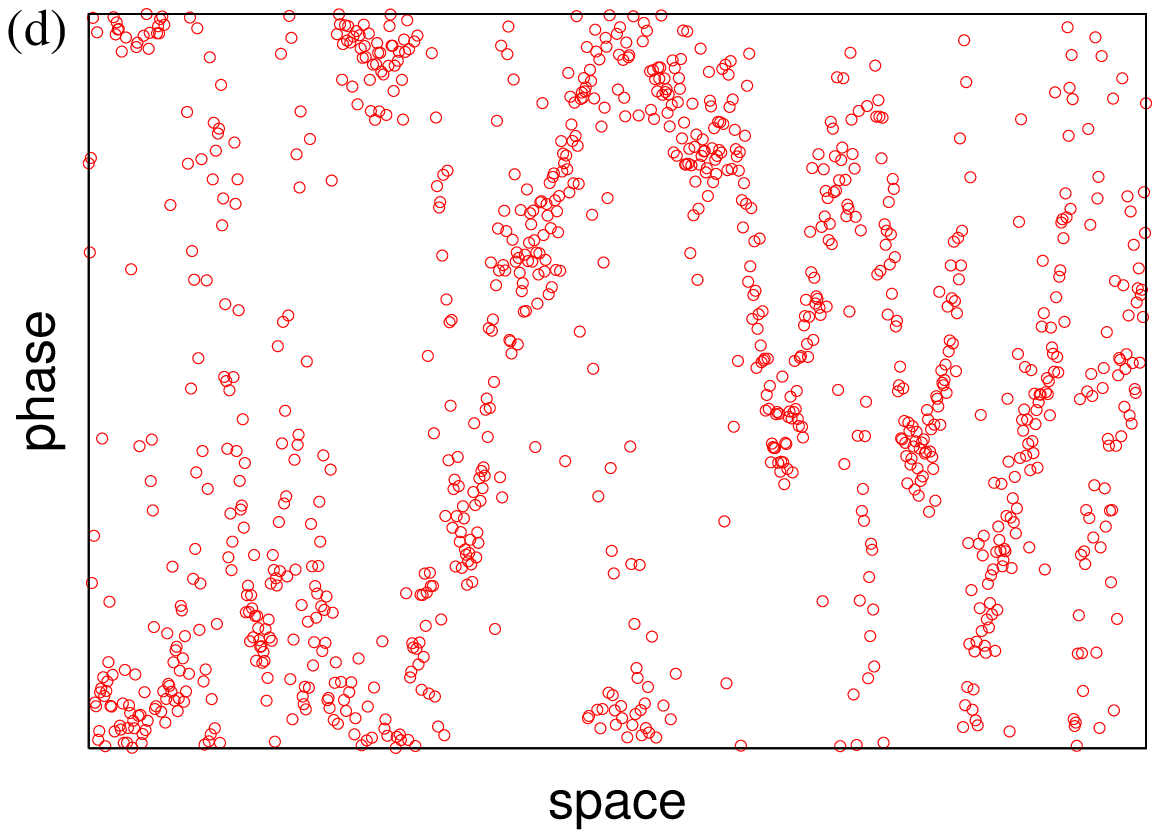}
\includegraphics[height=5cm]{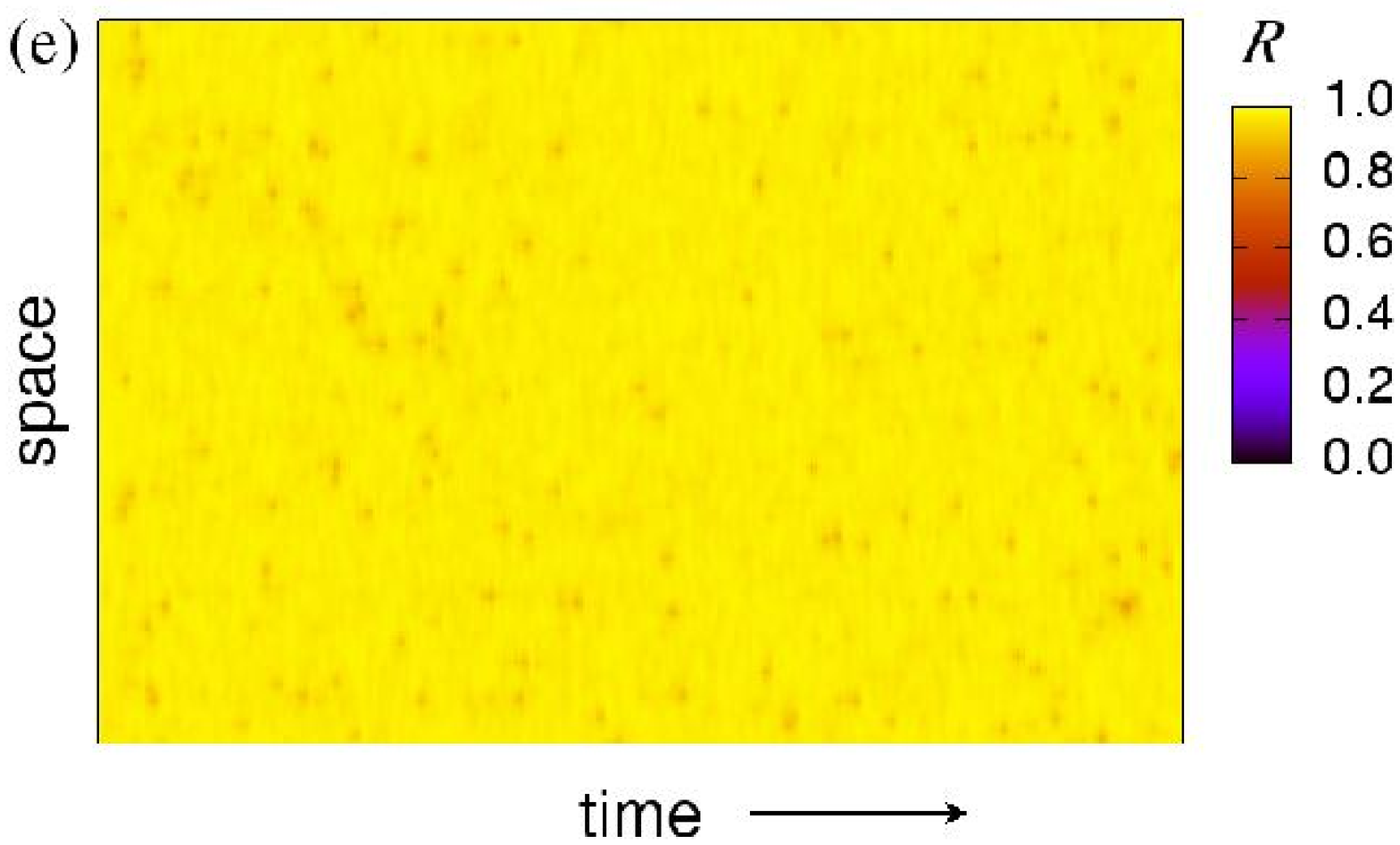}
\includegraphics[height=5cm]{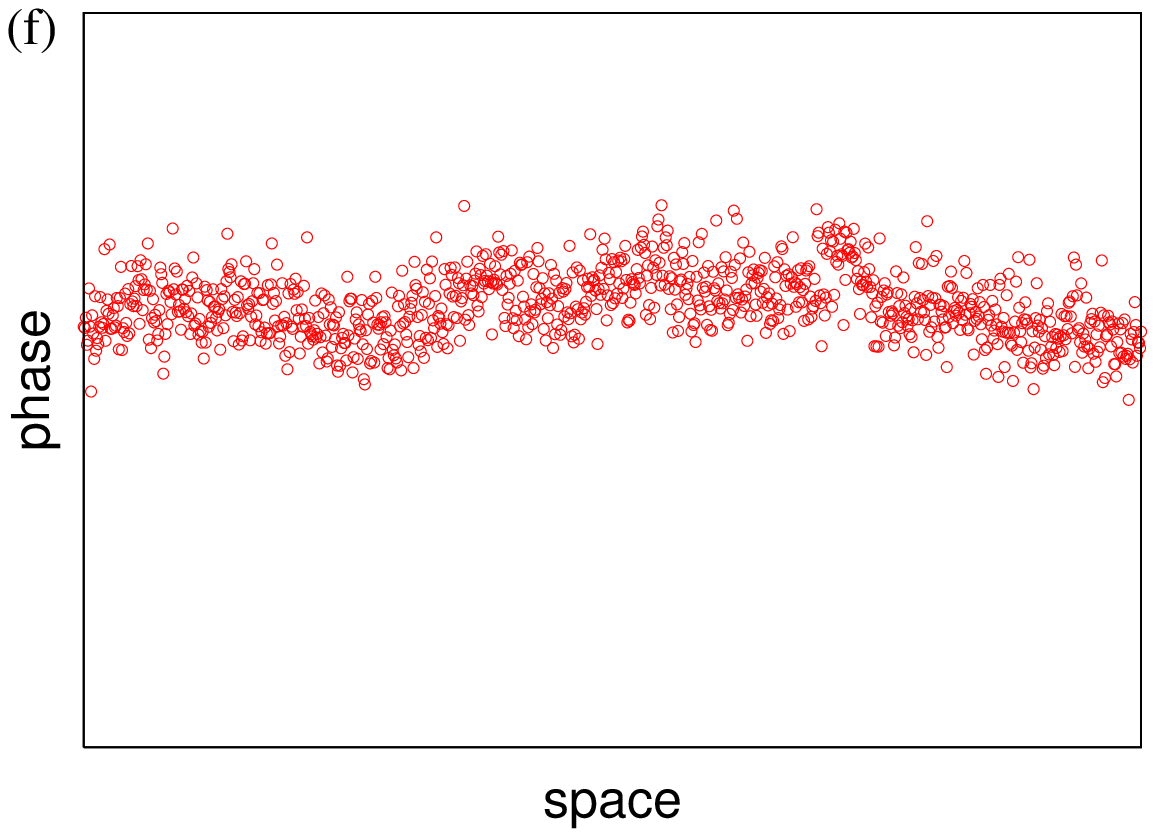}
\caption{(Color online) Numerical simulations of nonlocally coupled
  noisy phase oscillators corresponding to the nonlocally coupled
  noisy Stuart-Landau oscillators for Case SL-I [(a) and (b)], Case
  SL-II [(c) and (d)], and Case SL-III [(e) and (f)].
  (a), (c), and (e): Spatiotemporal patterns of the order parameter
  modulus.
  (b), (d), and (f): Instantaneous spatial profile of the local
  oscillator phase.
  The number of oscillators is $N=2^{10}$ and the separation between
  neighboring oscillators is $\varDelta x=0.1$.}
\label{fig:6}
\end{figure}
Figure~\ref{fig:6} displays the numerical results, which correspond
to the three cases, SL-I, SL-II, and SL-III, treated in
Section~\ref{sec:2}.
By comparing Fig.~\ref{fig:6} with Fig.~\ref{fig:4}, we can confirm
that the reduced Langevin phase equation nicely reproduces the
behavior of the original nonlocally coupled limit-cycle oscillators.

\subsection{Nonlinear Fokker-Planck equation}

We now transform the Langevin phase equation (\ref{eq:14})
derived above to an equivalent nonlinear Fokker-Planck equation, which
makes the following analysis far easier.
To do this, we note that the Langevin phase equation may be
viewed as describing the dynamics of a single local oscillator driven
by the nonlocal mean field of infinitely many other oscillators
sitting within the coupling range.
This means that the net coupling force experienced by this local
oscillator is a macro-variable, and therefore its statistical
fluctuation can be completely negligible.
For this reason, the nonlocal coupling term in the phase equation can
safely be replaced with its statistical average.

Let us denote by $f(\phi, x, t)$ the probability density function of
the phase $\phi$ of a single oscillator at a space-time point $(x,t)$.
Following the above argument, we average the coupling term in
Eq.~(\ref{eq:14}) by the single-oscillator phase distribution
$f(\phi,x,t)$.
The equation then takes the form of a single-oscillator Langevin
equation driven by a space-time dependent force $V(\phi,x,t)$,
that is,
\begin{equation}
  \partial_t\phi\left(x,t\right)=V\left(\phi,x,t\right)
  +\sqrt{D}\,\xi\left(x,t\right),
\label{eq:22}
\end{equation}
where
\begin{equation}
  V\left(\phi,x,t\right)=\omega+\int_{-\infty}^{\infty}dx'\,G\left(x-x'\right)
  \int_0^{2\pi}d\phi'\,\Gamma\left(\phi-\phi'\right)f\left(\phi',x',t\right).
\label{eq:23}
\end{equation}
Note that this equation now involves only one dynamical variable,
$\phi(x,t)$, as a result of statistical averaging.

The above single-oscillator Langevin equation can be easily
transformed to a single-oscillator Fokker-Planck equation
in the form~\cite{ref:risken89,ref:gardiner97}
\begin{equation}
  \frac{\partial f\left(\phi,x,t\right)}{\partial t}=-\frac{\partial}{\partial\phi}
  \left[\left\{\omega+\int_{-\infty}^{\infty}dx'\,G\left(x-x'\right)\int_0^{2\pi}d\phi'\,
      \Gamma\left(\phi-\phi'\right)f\left(\phi',x',t\right)\right\}f\left(\phi,x,t\right)\right]
  +D\frac{\partial^2 f\left(\phi,x,t\right)}{\partial\phi^2}.
\label{eq:24}
\end{equation}
Since the drift velocity itself involves the distribution
$f(\phi,x,t)$, this Fokker-Planck equation is nonlinear.
This equation is the starting point for further analysis.

\section{Amplitude equation near the onset of collective oscillation}
\label{sec:4}

In the following sections, we will further reduce the nonlinear
Fokker-Planck equation~(\ref{eq:24}) to analyze its dynamics
near the destabilization points.
In this section, we first apply the center-manifold reduction to
Eq.~(\ref{eq:24}) near the onset of collective oscillations.
Following Ref.~\cite{ref:shiogai03}, we derive the complex
Ginzburg-Landau equation, and then present results of numerical
simulations that confirm the theoretical conjecture on the
noise-induced turbulence.

\subsection{Hopf bifurcation and the complex Ginzburg-Landau equation}

The nonlinear Fokker-Planck equation~(\ref{eq:24}) has a trivial
constant solution, $f(\phi, x, t) \equiv 1 / 2\pi$, corresponding to
the completely desynchronized state of the oscillators.
When the noise intensity $D$ is sufficiently large, this constant
solution is stable. As $D$ is decreased, the constant solution is
destabilized via a Hopf bifurcation.

In the previous study~\cite{ref:shiogai03}, it was shown that the
nonlinear Fokker-Planck equation~(\ref{eq:24}) has a series of
critical noise intensities, below which the system can sustain
traveling waves of various wavenumbers.
Among them, the spatially uniform oscillating solution of the phase
distribution with wavenumber zero has the largest critical noise
intensity.
Thus, when we decrease $D$ from above, the trivial constant solution
gives way to a spatially uniform oscillating solution at a certain
critical value $D = D_c$.

In the vicinity of this Hopf bifurcation point, we can derive an
amplitude equation describing the slow dynamics of the destabilized
mode.
We introduce a complex amplitude $A(x,t)$ describing the deviation of
$f(\phi,x,t)$ from the constant solution $1/2\pi$ as
\begin{equation}
  f\left(\phi,x,t\right)=\frac{1}{2\pi}+\frac{1}{2\pi}
  \Bigl(A\left(x,t\right)e^{i\lambda\phi+i\Omega_c t}
  +A^{\ast}\left(x,t\right)e^{-i\lambda\phi-i\Omega_c t}\Bigr).
\label{eq:25}
\end{equation}
As explained in Ref.~\cite{ref:shiogai03}, we can derive the complex
Ginzburg-Landau equation for this complex amplitude
\begin{equation}
  \partial_t A\left(x,t\right)
  = \lambda^2 \left(D_c-D\right) A
  + d \partial_x^2 A - g \left|A\right|^2 A,
\label{eq:26}
\end{equation}
from the nonlinear Fokker-Planck equation by the center-manifold
reduction method~\cite{ref:kuramoto84}.
The parameters of the above complex Ginzburg-Landau equation can be
expressed in terms of the Fourier components $\Gamma_l$ of the
phase coupling function
\begin{equation}
  \Gamma(\phi) = \sum_{l=-\infty}^{\infty} \Gamma_l\,e^{il\phi}.
\label{eq:27}
\end{equation}
As derived in Ref.~\cite{ref:shiogai03}, they are given by
\begin{equation}
  \lambda = \arg \max_l \frac{\Im \Gamma_l}{l}, \quad
  D_c = \frac{\Im\Gamma_\lambda}{\lambda}, \quad
  \Omega_c = -\lambda\left(\omega+\Re\Gamma_\lambda+\Gamma_0\right),
\label{eq:28}
\end{equation}
\begin{equation}
  d = -i\lambda\Gamma_\lambda, \quad
  g = \frac{\lambda\Gamma_\lambda
    \left(\Gamma_{2\lambda}+\Gamma_{-\lambda}\right)}
  {2\Im\Gamma_\lambda-i\Re\Gamma_\lambda+i\Gamma_{2\lambda}},
\label{eq:29}
\end{equation}
where $\Re$ and $\Im$ denote the real part and the imaginary part,
respectively.
In what follows, we will restrict ourselves to the case of $\lambda =
1$ and $\Re g > 0$, i.e., the case that the first Fourier component of
the phase distribution undergoes a supercritical Hopf bifurcation.
This is actually the case with our two representative models of
limit-cycle oscillators.
When the noise intensity $D$ is decreased below $D_c$,
Eq.~(\ref{eq:26}) starts to exhibit oscillatory behavior.
In this region, by appropriately rescaling the variables, we
obtain the standard form of the complex Ginzburg-Landau equation
\cite{ref:kuramoto84,ref:pikovsky01,ref:manrubia03,
  ref:cross93,ref:bohr98,ref:aranson02}
\begin{equation}
  \partial_t A(x,t) = A + (1+ic_1)\partial_x^2 A - (1+ic_2)|A|^2 A,
\label{eq:30}
\end{equation}
where the real parameters $c_1$ and $c_2$ are given by
\begin{equation}
  c_1 = \frac{\Im d}{\Re d},\quad
  c_2 = \frac{\Im g}{\Re g}.
\label{eq:31}
\end{equation}
We use this standard form in the following discussion.

It is well known that the spatially uniform oscillating solution
of the complex Ginzburg-Landau equation becomes unstable and
spatiotemporal chaos develops when the Benjamin-Feir instability
condition $1 + c_1 c_2 < 0$ is satisfied.
Furthermore, it is also known that in the near vicinity of the
Benjamin-Feir line, the modulus of the complex Ginzburg-Landau
equation tends to be uniform, and the phase component
dominates the dynamics of the system.
In such a situation, the complex Ginzburg-Landau equation
can be further reduced to the Kuramoto-Sivashinsky equation
\cite{ref:kuramoto84}
\begin{equation}
  \partial_t \Theta(x,t) = -c_2 + (1+c_1 c_2)\partial_x^2\Theta
  +(c_2-c_1)(\partial_x\Theta)^2-\frac{c_1^2(1+c_2^2)}{2}\partial_x^4\Theta,
\label{eq:32}
\end{equation}
where $\Theta(x,t)$ is the phase of the complex amplitude $A(x,t)$,
and is essentially the same as the phase of the order parameter
defined in Eq.~(\ref{eq:10}) except for the sign, as we see later.
When the Benjamin-Feir instability condition is slightly exceeded, the
phase diffusion coefficient $1 + c_1 c_2$ of this equation becomes
slightly negative, leading to turbulent behavior. In such a case, by an
appropriate rescaling of the variables, we obtain the standard form of
the Kuramoto-Sivashinsky equation~\cite{ref:kuramoto84}
\begin{equation}
  \partial_t \Theta(x,t) = -\partial_x^2\Theta
  + (\partial_x\Theta)^2 - \partial_x^4\Theta.
\label{eq:33}
\end{equation}

\subsection{Stuart-Landau oscillators}

For the sake of simplicity, we treat the nonlocally coupled noisy
Stuart-Landau oscillators hereafter.
In this case, as seen from Eq.~(\ref{eq:20}) and Eq.~(\ref{eq:21}),
the phase coupling function is given by the simple sine function
\begin{equation}
  \Gamma\left(\phi\right) = -\sin\left(\phi+\alpha\right), \quad
  \left|\alpha\right| < \pi/2,
\label{eq:34}
\end{equation}
which is the in-phase type coupling.
The parameters of the reduced complex Ginzburg-Landau equation
can be calculated as
\begin{equation}
  D_c = \frac{\cos\alpha}{2},\quad
  c_1 = \frac{\Im d}{\Re d} = \tan\alpha,\quad
  c_2 = \frac{\Im g}{\Re g} = -\frac{\tan\alpha}{2}.
\label{eq:35}
\end{equation}
By choosing the parameter $\alpha$ appropriately, the Benjamin-Feir
instability condition can be satisfied.
Thus, it was conjectured in Refs.~\cite{ref:shiogai03,ref:kuramoto06}
that, since the reduced complex Ginzburg-Landau equation can exhibit
turbulent behavior, the corresponding nonlinear Fokker-Planck
equation, the Langevin phase equation, and the original nonlocally
coupled noisy limit-cycle oscillators, could also exhibit turbulent
behavior under suitable conditions.


To examine this conjecture, we conduct systematic numerical
simulations near the Hopf bifurcation curve.
\begin{figure}
\centering
\includegraphics[height=6cm]{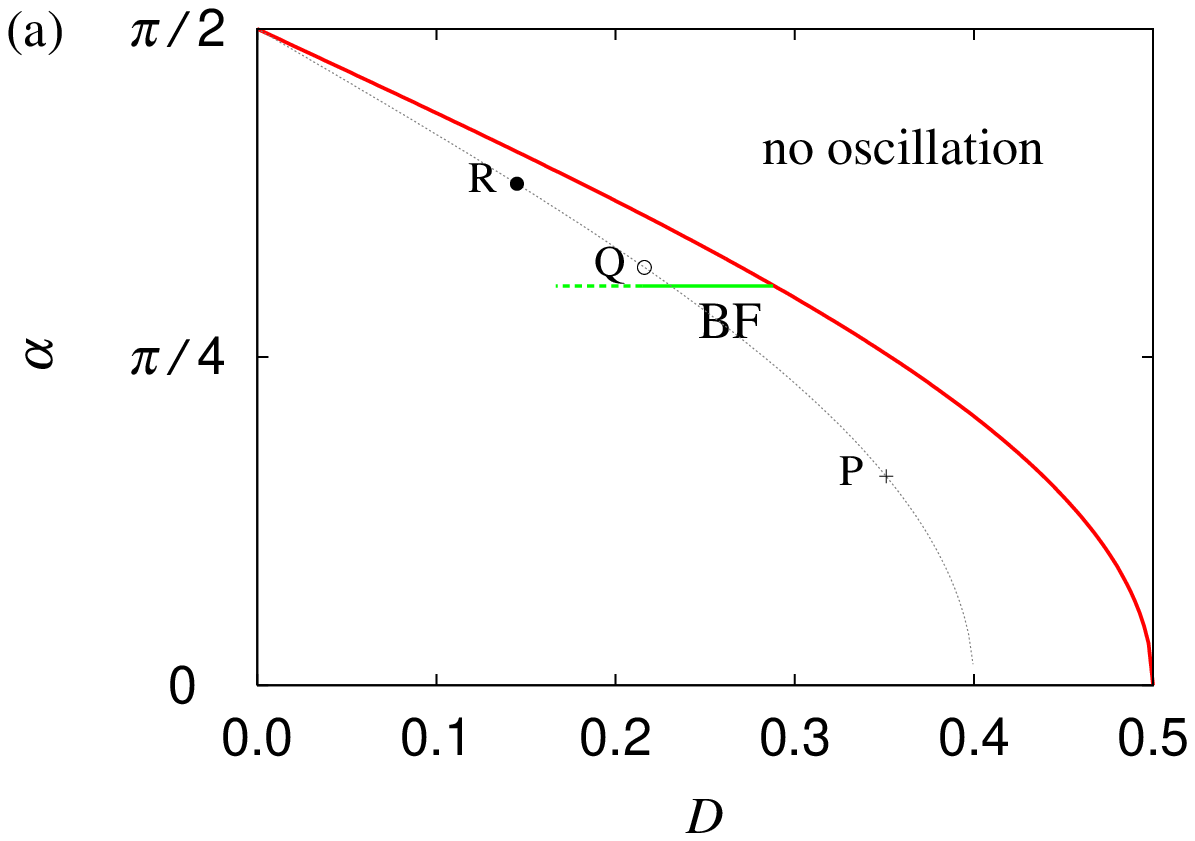}
\includegraphics[height=6cm]{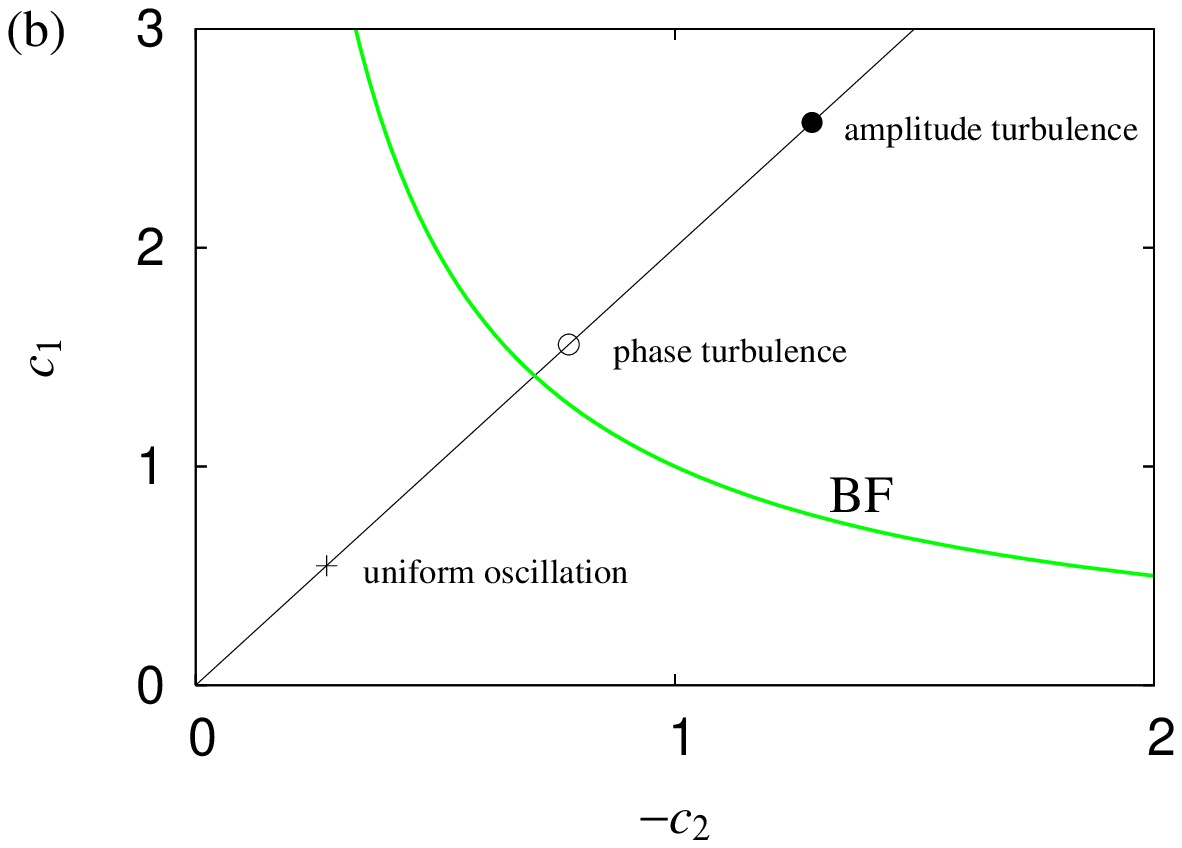}
\caption{(Color online) (a): Phase diagram plotted as a function of
  the noise intensity $D$ and the phase shift $\alpha$.
  The red solid curve represents the Hopf bifurcation line
  ($D=D_c=\cos(\alpha)/2$).
  For numerical simulations near the bifurcation, the noise intensity
  is chosen such that $D/D_c=0.8$ indicated by the dotted curve.
  The Benjamin-Feir (BF) critical line ($\alpha=\arctan\sqrt{2}$) is
  also indicated.
  (b): Phase diagram plotted as a function of the parameters $c_1=\Im
  d/\Re d=\tan(\alpha)$ and $c_2=\Im g/\Re g=-\tan(\alpha)/2$.
  Plus (Case P: $\alpha=0.5, D/D_c=0.8$),
  open circle (Case Q: $\alpha=1.0, D/D_c=0.8$), and
  filled circle (Case R: $\alpha=1.2, D/D_c=0.8$)
  correspond to spatially uniform oscillation, phase turbulence,
  and amplitude turbulence, respectively.
  The Benjamin-Feir (BF) line ($1+c_1 c_2=0$) is also indicated
  by the green curve.}
\label{fig:7}
\end{figure}
Theoretical bifurcation diagrams of the complex Ginzburg-Landau
equation are plotted as a function of the parameters $\alpha$ and $D$
in Fig.~\ref{fig:7}(a), and as a function of the parameters $c_1$ and
$c_2$ in Fig.~\ref{fig:7}(b).
The red solid line in Fig.~\ref{fig:7}(a) represents the Hopf
bifurcation curve, below which the complex amplitude $A(x,t)$
starts to oscillate.
In the numerical simulations, we fix the noise intensity at
$D/D_c = 0.8$, which corresponds to the black line.
We choose three representative points on this line, indicated by
P, Q, and R.
The green line represents the Benjamin-Feir instability curve $1+c_1 c_2 = 0$,
which can be expressed as $\alpha = \arctan\sqrt{2}$ from Eq.~(\ref{eq:35}).

Note that the parameters $c_1$ and $c_2$ cannot change independently
in the present model.
From Eq.~(\ref{eq:35}), only pairs of $(c_1, c_2)$ satisfying
$c_1 = -2c_2$ drawn as the straight line in Fig.~\ref{fig:7}(b)
can be realized.
The three representative cases P, Q, and R correspond to three
different dynamical states of the complex Ginzburg-Landau equation,
namely, spatially uniform oscillation, phase turbulence, and amplitude
turbulence, respectively~\cite{ref:shraiman92}.
On these points, we numerically simulate the Langevin phase equation
and the corresponding nonlinear Fokker-Planck equation (and, in some
cases, also the complex Ginzburg-Landau equation and the
Kuramoto-Sivashinsky equation).
Numerical methods used in the simulations are summarized in
Appendix~\ref{sec:C}.
In drawing the figures presented hereafter, numerical results obtained
from the different equations are appropriately rescaled to accord with
the standard form of the complex Ginzburg-Landau
equation~(\ref{eq:30}) or the Kuramoto-Sivashinsky
equation~(\ref{eq:33}).


\begin{figure}
\centering
\begin{tabular}{c c c}
 & (a) & (b) \\[1mm]
\rotatebox{90}{\hspace{2cm}{\Large time $\rightarrow$}} &
\includegraphics[height=5cm]{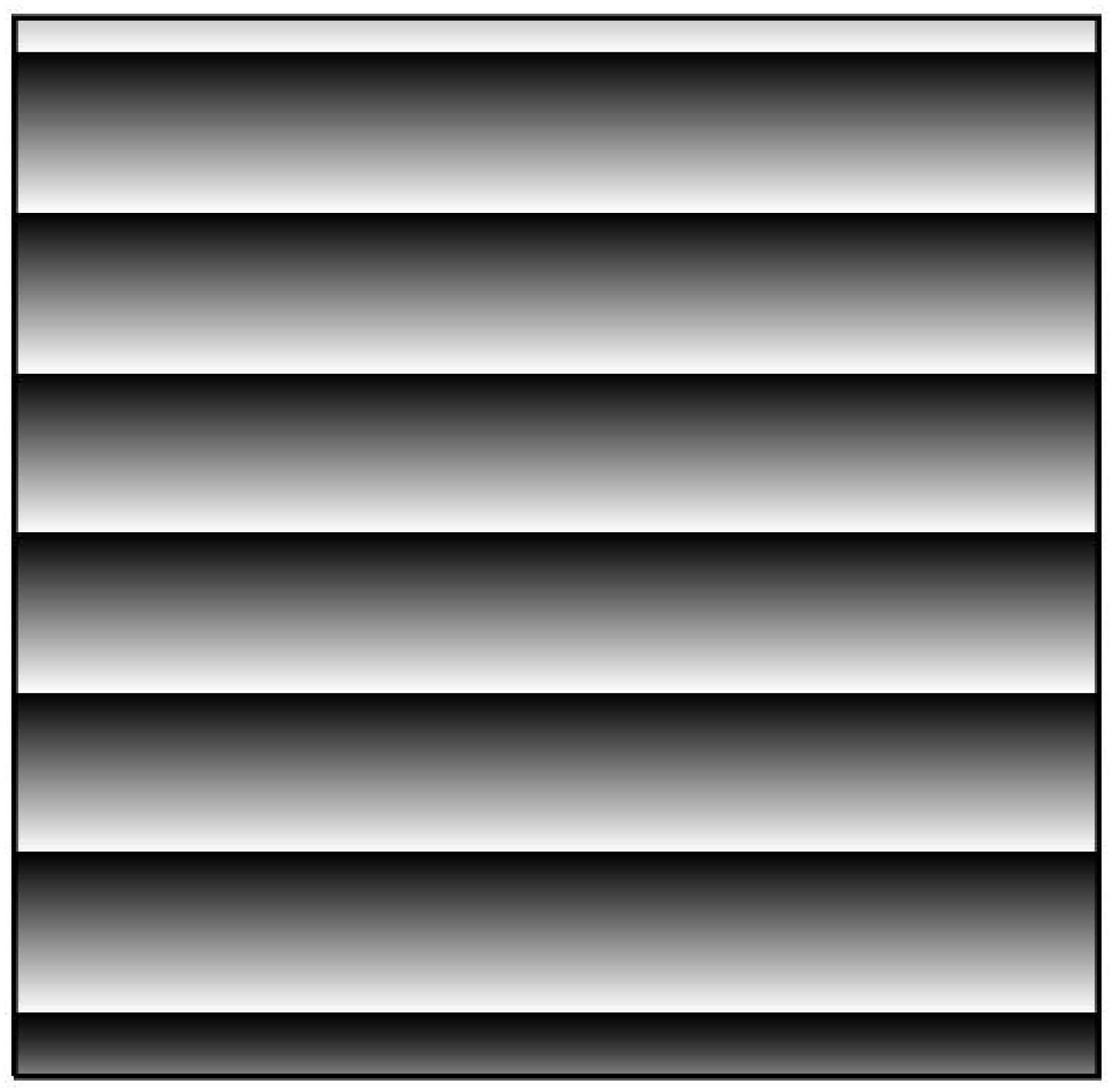} &
\includegraphics[height=5cm]{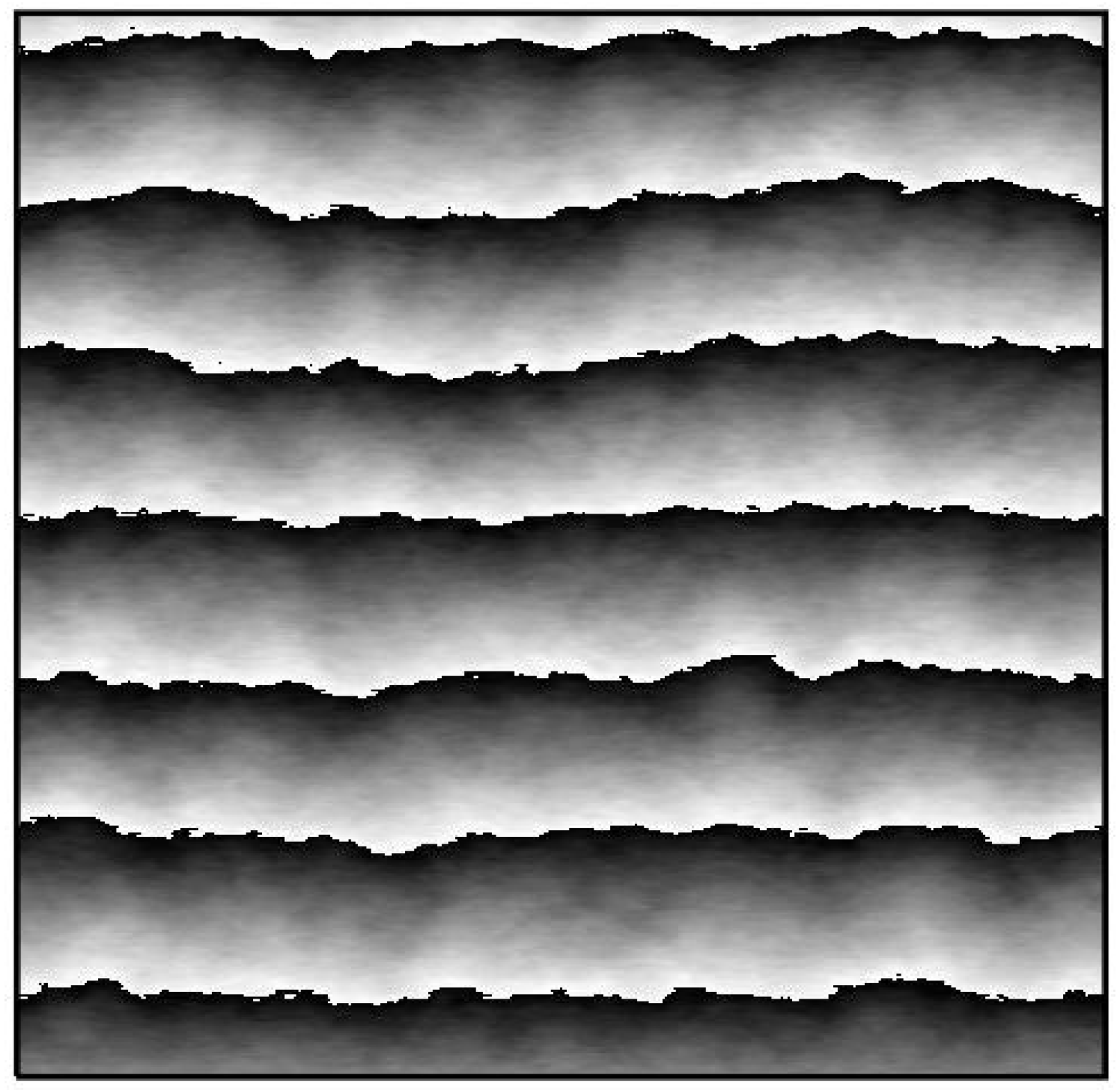} \\
\end{tabular}
\caption{Uniform oscillation.
  Parameter values are $\alpha=0.5$ and $D/D_c=0.8$ (Case P).
  Space (horizontal)-time (vertical) plot of order parameter
  phase $\Theta(x,t)$ for Fokker-Planck simulation (a)
  and Langevin simulation using $N=2^{15}$ oscillators (b).}
\label{fig:8}
\end{figure}
\begin{figure}
\includegraphics[height=5cm]{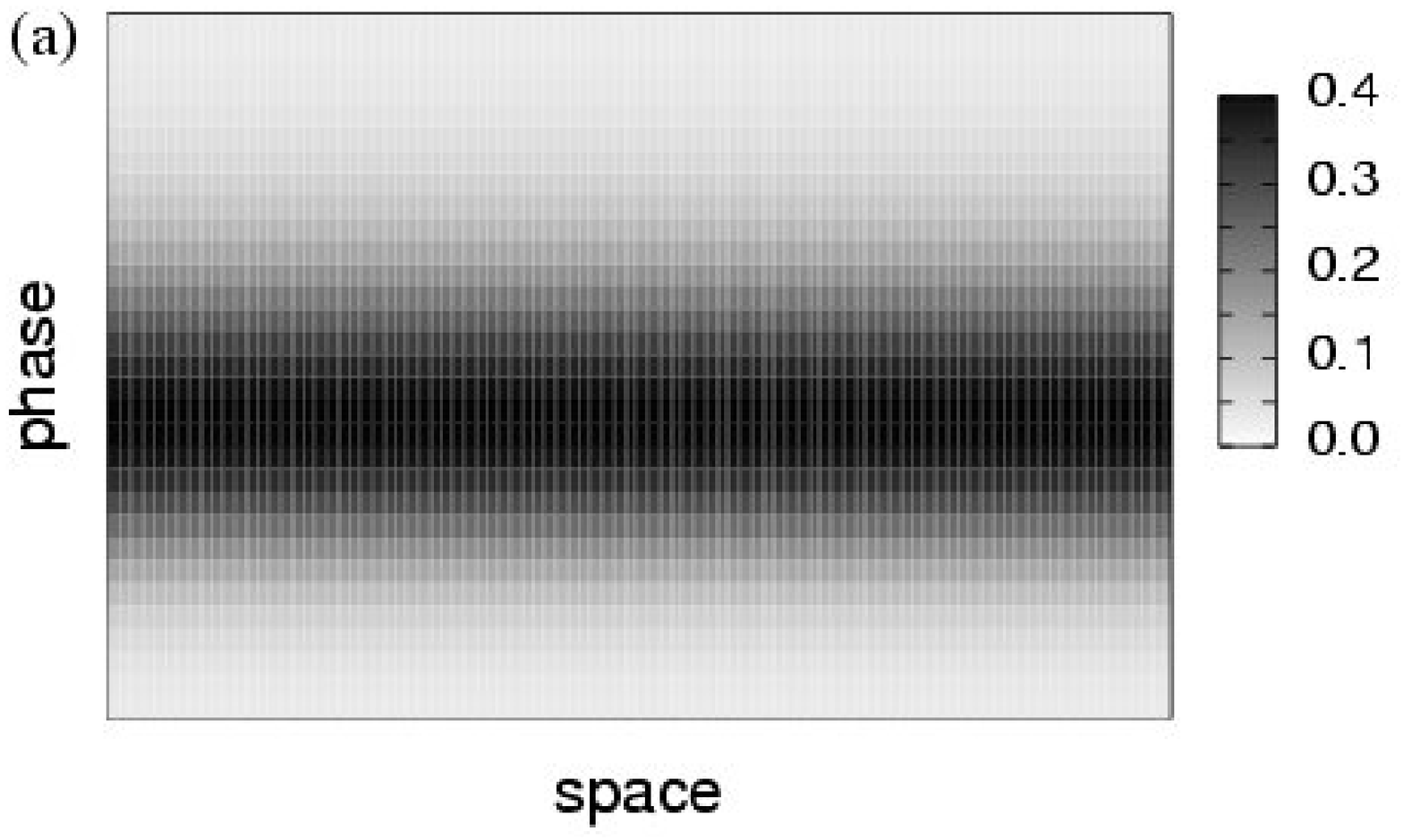}
\includegraphics[height=5cm]{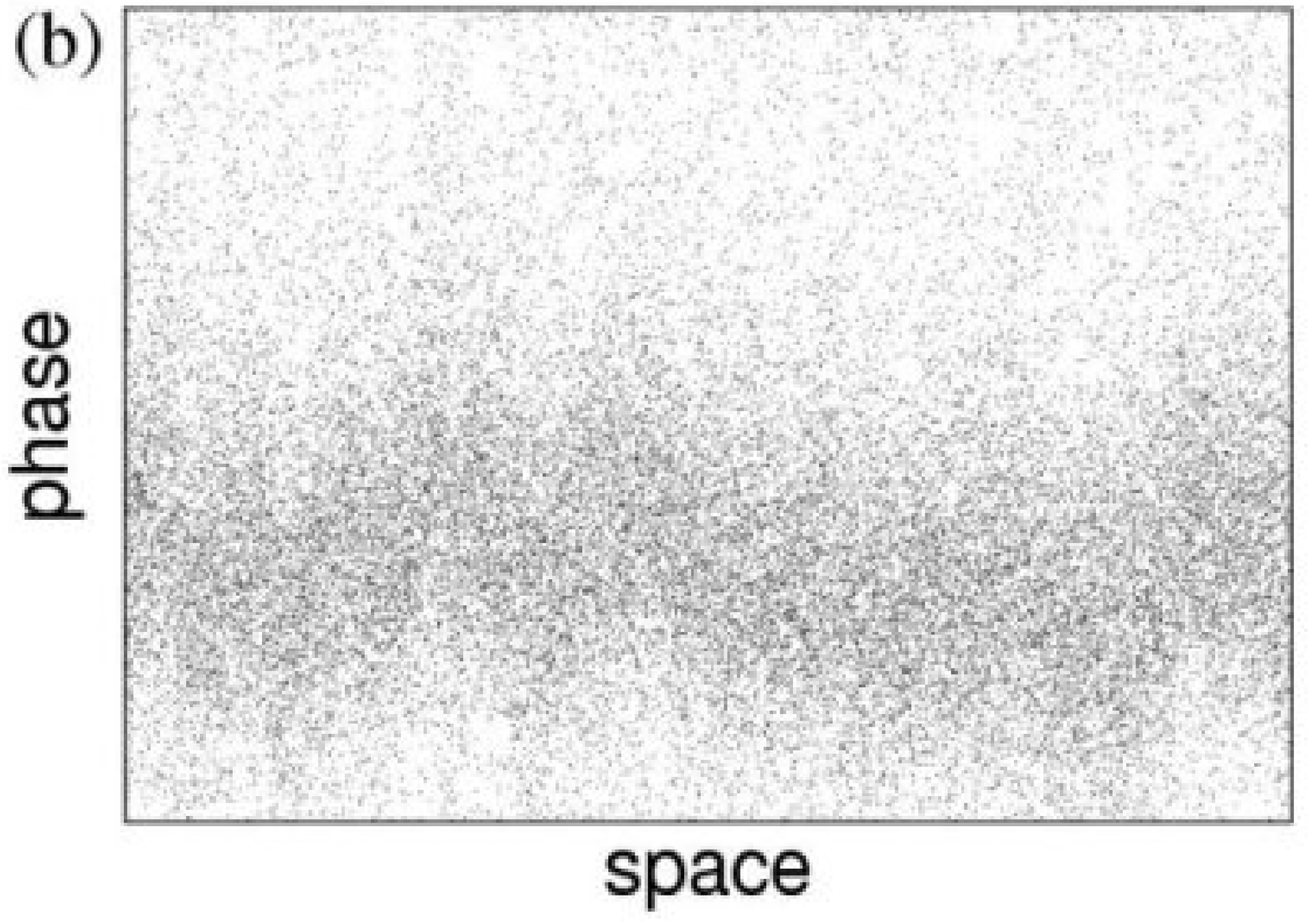}
\caption{Uniform oscillation.
  Parameter values are $\alpha=0.5$ and $D/D_c=0.8$ (Case P).
  (a): Instantaneous spatial profile of phase distribution function
  obtained from Fokker-Planck simulation.
  (b): Instantaneous spatial profile of local oscillator phase
  obtained from Langevin simulation using $N=2^{15}$ oscillators.}
\label{fig:9}
\end{figure}
Figures~\ref{fig:8} and \ref{fig:9} display the results at the
parameter point P, where we expect stable uniform oscillation obtained
by direct numerical simulations of the nonlinear Fokker-Planck
equation (\ref{eq:24}) and the Langevin phase equation (\ref{eq:14}).
Figures~\ref{fig:8}(a) and \ref{fig:8}(b) plot the temporal evolution
of the order parameter phase $\Theta(x, t)$ defined in
Eq.~(\ref{eq:10}).
As expected from the phase diagram of the reduced complex
Ginzburg-Landau equation, both equations exhibit spatially uniform
oscillating solutions.
Small non-uniformity seen in the Langevin simulation,
Fig.~\ref{fig:8}(b), is due to trivial statistical fluctuations.
Figures~\ref{fig:9}(a) and Fig.~\ref{fig:9}(b) display the snapshots
of the instantaneous spatial distribution of the phases.
We can confirm that the phase distributions are spatially uniform.

\begin{figure}
\begin{tabular}{c c c c}
 & (a) & (b) & (c) \\[1mm]
\rotatebox{90}{\hspace{2cm}{\Large time $\rightarrow$}} &
\includegraphics[height=5cm]{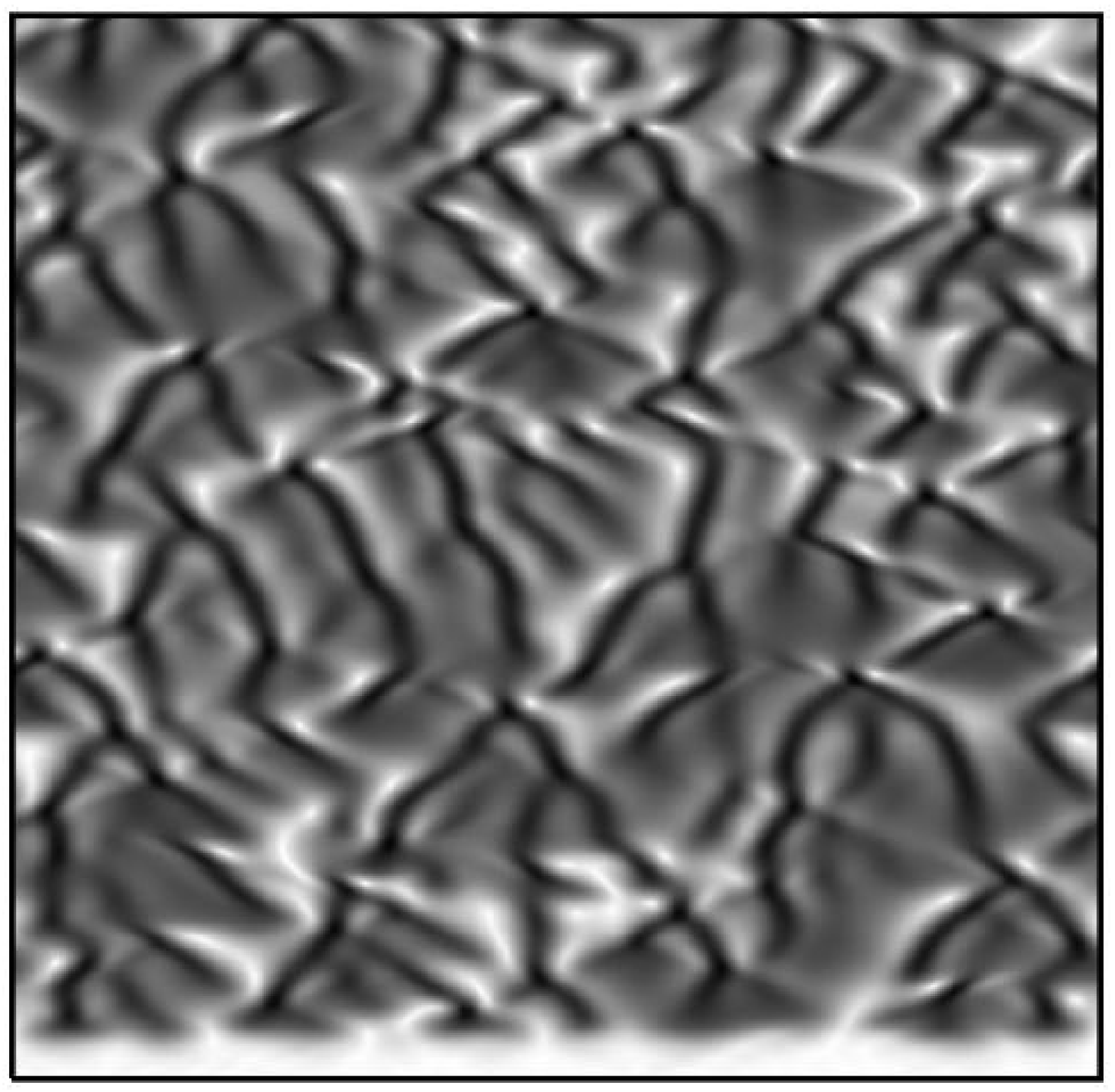} &
\includegraphics[height=5cm]{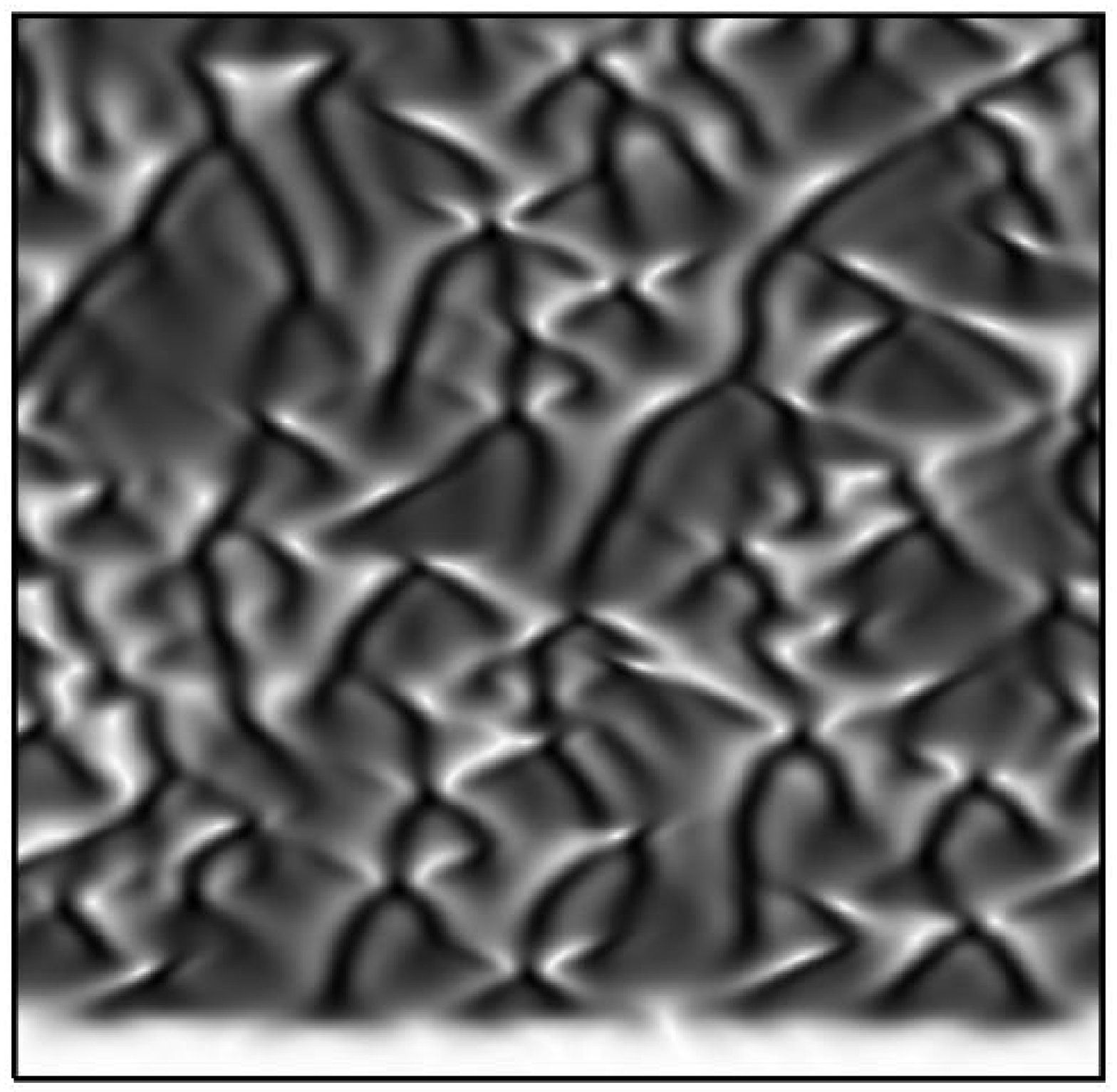} &
\includegraphics[height=5cm]{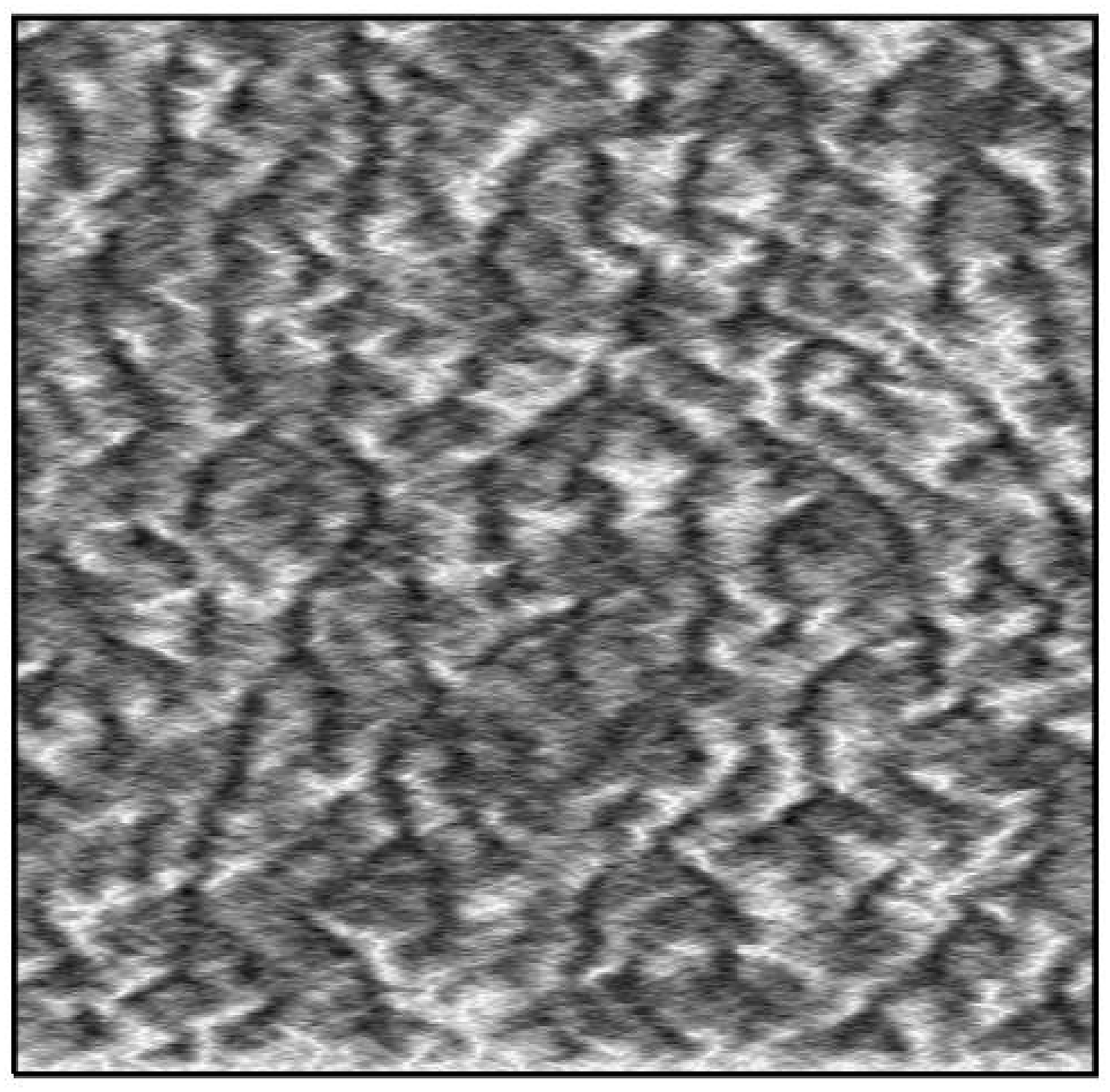} \\
\end{tabular}
\caption{Amplitude turbulence.
  Parameter values are $\alpha=1.2$ and $D/D_c=0.8$ (Case R).
  Space (horizontal) - time (vertical) plot of
  the order parameter modulus $R(x,t)$ is shown
  in rescaled units with which the complex Ginzburg-Landau
  equation takes the standard form, Eq.~(\ref{eq:30}),
  whose system size is $L\simeq 114$.
  Numerical data obtained form the complex Ginzburg-Landau
  equation (a), the nonlinear Fokker-Planck equation (b),
  and the Langevin phase equation using $N=2^{15}$
  oscillators (c) are compared.
}
\label{fig:10}
\end{figure}
\begin{figure}
\centering
\includegraphics[height=4cm]{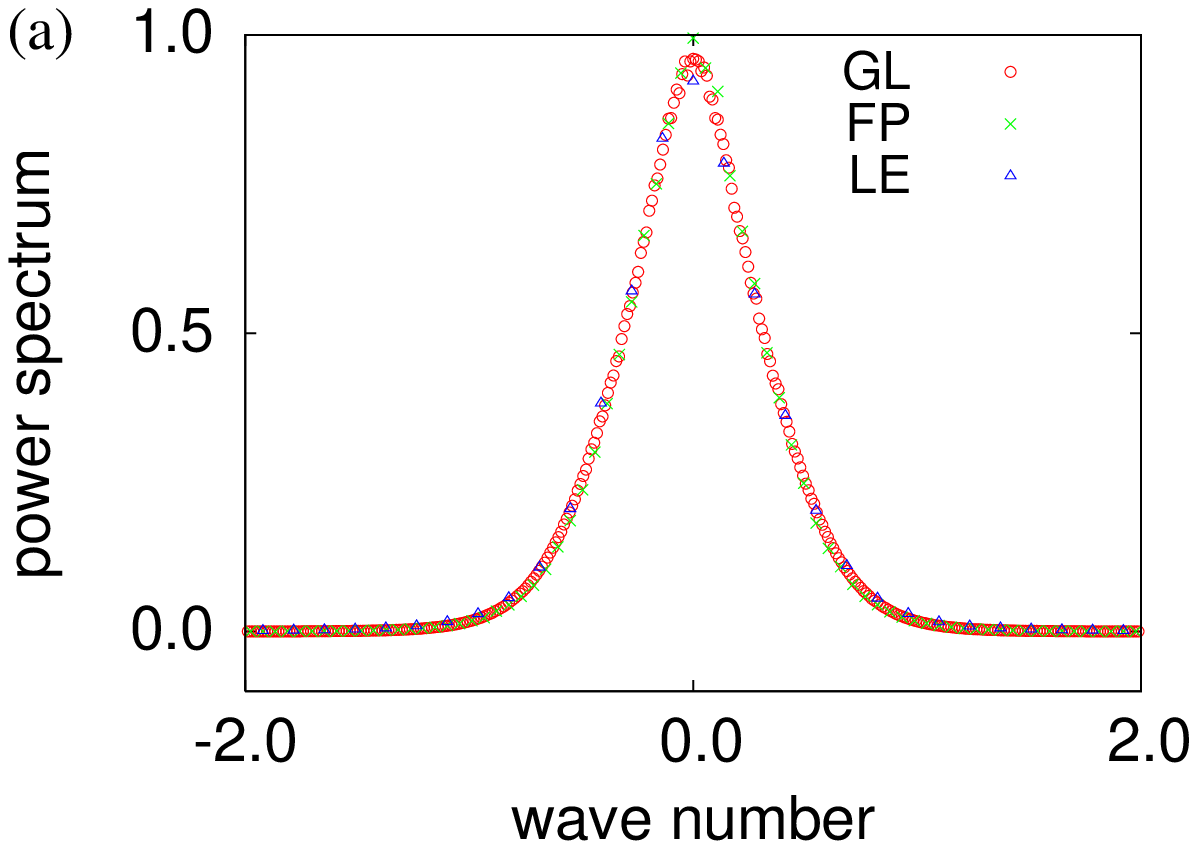}
\includegraphics[height=4cm]{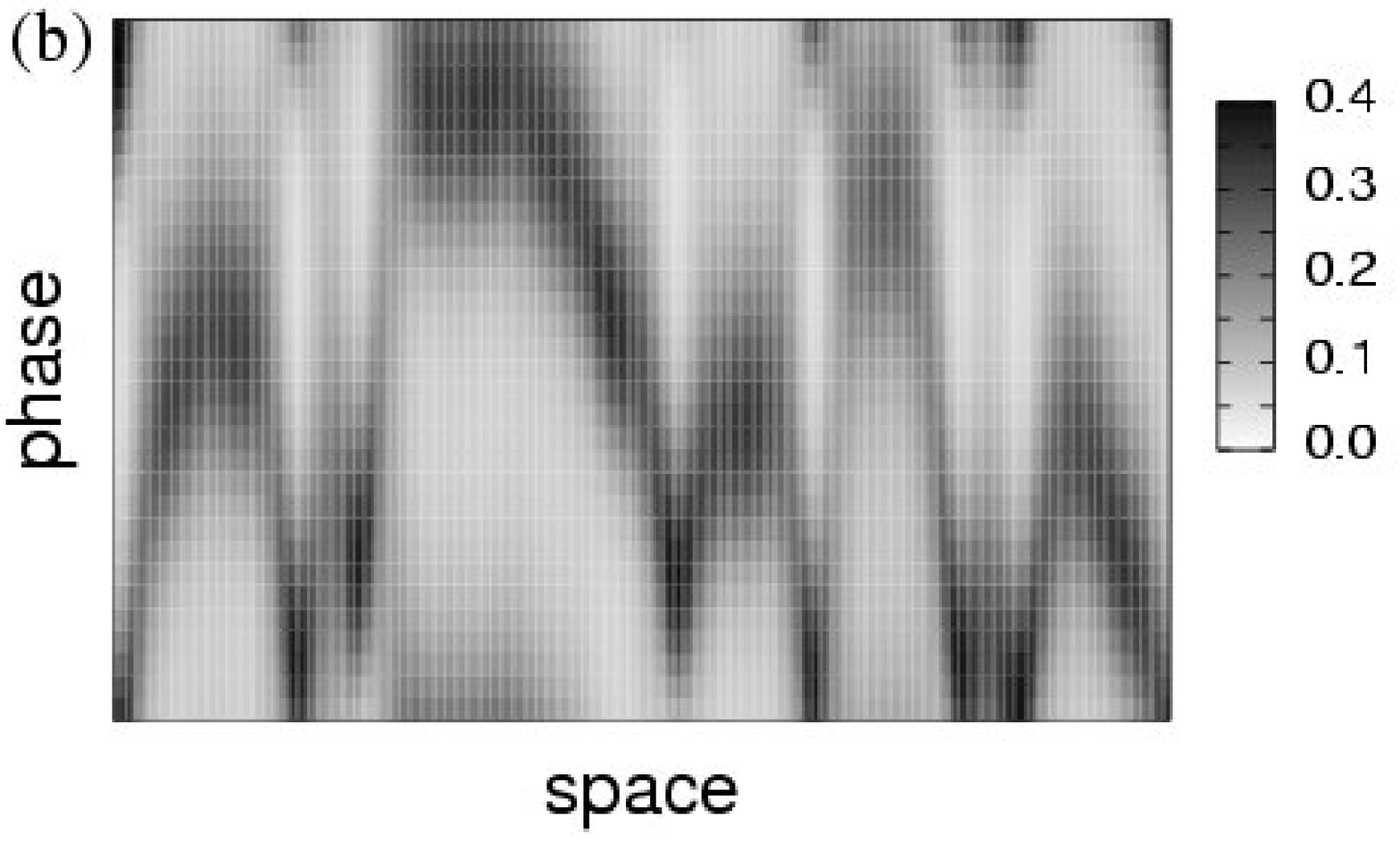}
\includegraphics[height=4cm]{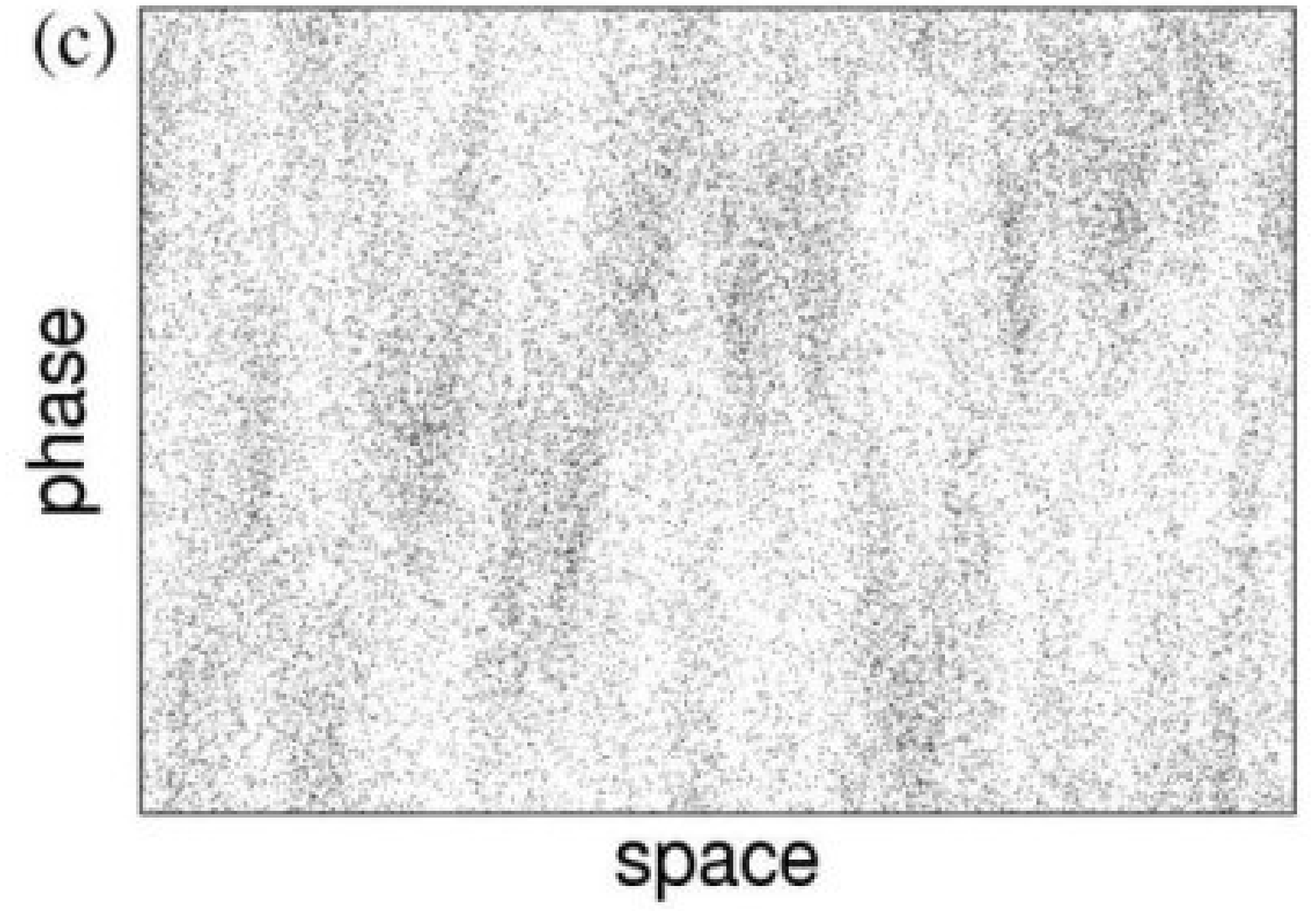}
\caption{(Color online) Amplitude turbulence.
  Parameter values are $\alpha=1.2$ and $D/D_c=0.8$ (Case R).
  (a): Spatial power spectrum of the order parameter
  as a function of the wavenumber in rescaled units with which
  the complex Ginzburg-Landau equation takes the standard form, Eq.~(\ref{eq:30}).
  Numerical data obtained from the complex Ginzburg-Landau equation (GL),
  the nonlinear Fokker-Planck equation (FP), and the Langevin phase equation (LE)
  are compared.
  (b): Instantaneous spatial profile of the phase distribution function
  obtained from the Fokker-Planck simulation.
  (c): Instantaneous spatial profile of the local oscillator phase
  obtained from the Langevin simulation using $N=2^{15}$ oscillators.}
\label{fig:11}
\end{figure}
Numerical results obtained at the parameter point R, where the
amplitude turbulence is expected in the reduced complex
Ginzburg-Landau equation, are shown in Figs.~\ref{fig:10} and
\ref{fig:11}.
Figures~\ref{fig:10}(a), \ref{fig:10}(b), and \ref{fig:10}(c) show
spatiotemporal patterns of the modulus $R(x,t)$ of the order
parameter defined in Eq.~(\ref{eq:10}), obtained by numerical
simulations of the complex Ginzburg-Landau equation, the nonlinear
Fokker-Planck equation, and the Langevin phase equation, respectively.
As mentioned above, the order parameter defined in Eq.~(\ref{eq:10})
gives the first Fourier mode of the phase distribution, which
corresponds to the complex conjugation of the complex amplitude
for the reduced complex Ginzburg-Landau equation as
\begin{align}
  R\left(x,t\right)e^{i\Theta\left(x,t\right)}
  &\equiv\int^{\infty}_{-\infty}dx'\,G\left(x-x'\right)
  e^{i\phi\left(x',t\right)} \label{eq:36} \\
  &=\int^{\infty}_{-\infty}dx'\,G\left(x-x'\right)
  \int^{2\pi}_{0}d\phi'\,e^{i\phi'}f\left(\phi',x',t\right) \label{eq:37} \\
  &=\int^{\infty}_{-\infty}dx'\,G\left(x-x'\right)
  A^{\ast}\left(x',t\right)e^{-i\Omega_c t} \label{eq:38} \\
  &\simeq A^{\ast}\left(x,t\right)e^{-i\Omega_c t}, \label{eq:39}
\end{align}
where we used the fact that the characteristic wavelength becomes
sufficiently longer than the length of the the nonlocal coupling
near the critical point in the last approximation.
The three patterns are quite similar to each other.
Figure~\ref{fig:11}(a) shows the spatial power spectrum
of the order parameter.
They are also almost identical to each other.
Figures~\ref{fig:11}(b) and \ref{fig:11}(c) show the snapshots of the
phase distributions obtained from the nonlinear Fokker-Planck equation
and the Langevin phase equation, respectively.
In both figures, we can observe strongly nonuniform phase
distributions due to turbulent fluctuations, which again resemble with
each other.
These results clearly indicate that the turbulent fluctuations
exhibited by the three equations are generated by the same dynamical
instability.

\begin{figure}
\begin{tabular}{c c c c}
 & (a) & (b) & (c) \\[1mm]
\rotatebox{90}{\hspace{4cm}{\Large time $\rightarrow$}} &
\includegraphics[height=9cm]{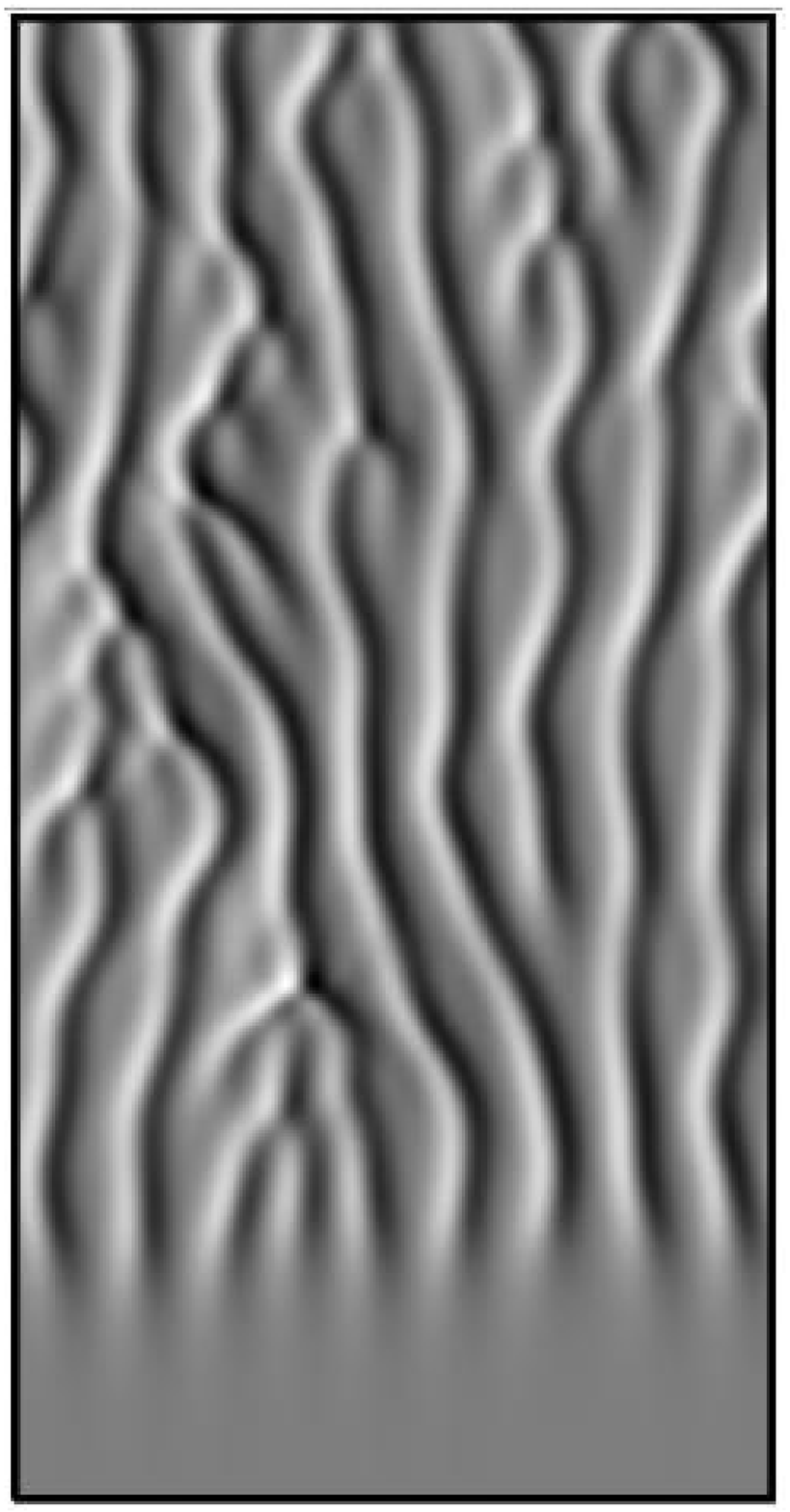} &
\includegraphics[height=9cm]{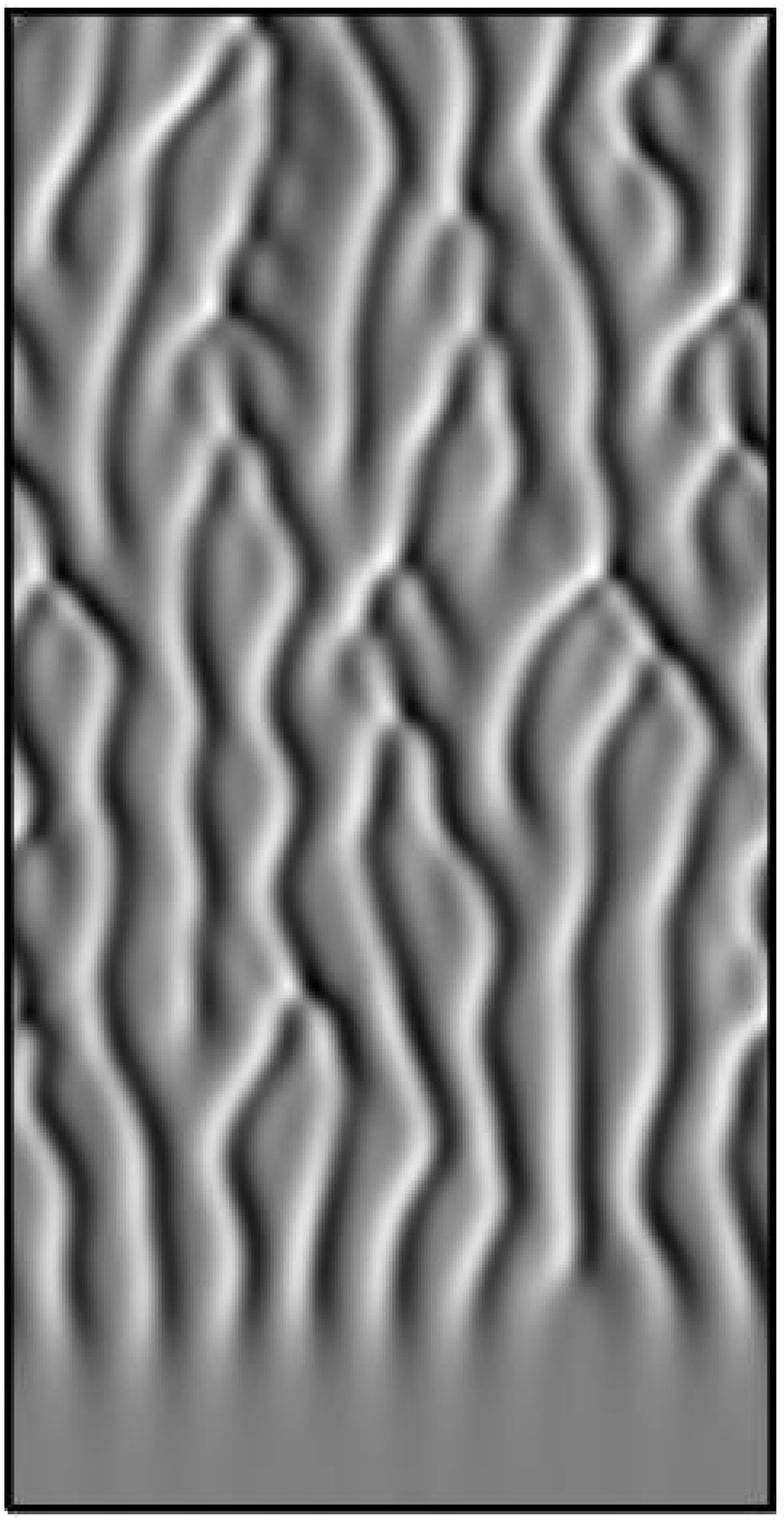} &
\includegraphics[height=9cm]{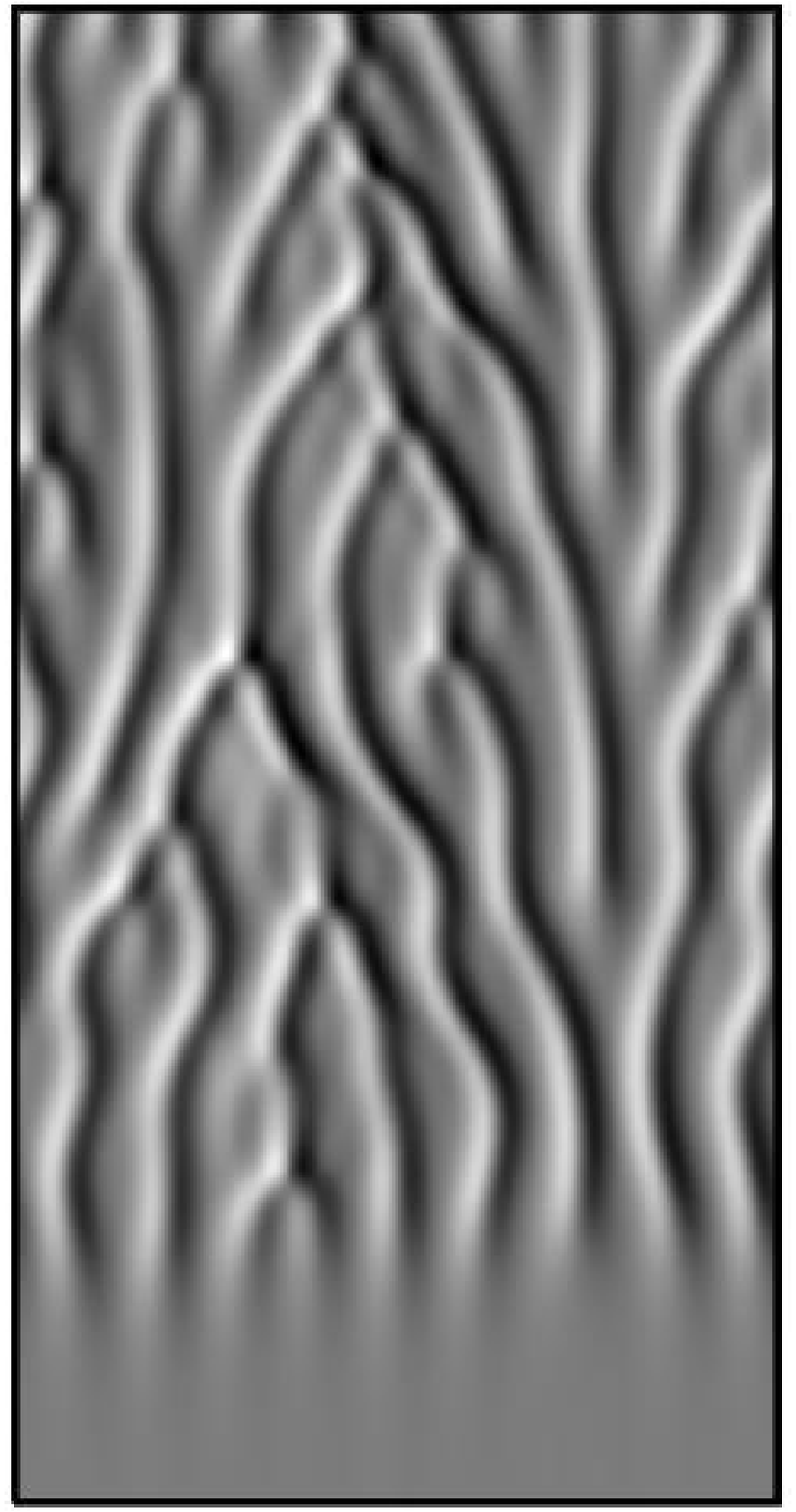} \\
\end{tabular}
\caption{Phase turbulence.  Parameter values are $\alpha=1.0$ and
  $D/D_c=0.8$ (Case Q).  Space (horizontal) - time (vertical) plot of
  the order parameter phase gradient $v(x,t) = 2 \partial_x
  \Theta(x,t)$ is shown in rescaled units with which the
  Kuramoto-Sivashinsky equation takes the standard form,
  Eq.~(\ref{eq:33}), whose system size is $L\simeq 76$.
  Numerical data obtained from the Kuramoto-Sivashinsky equation (a),
  the complex Ginzburg-Landau equation (b), and the nonlinear
  Fokker-Planck equation (c) are compared.}
\label{fig:12}
\end{figure}
\begin{figure}
\centering
\includegraphics[height=5cm]{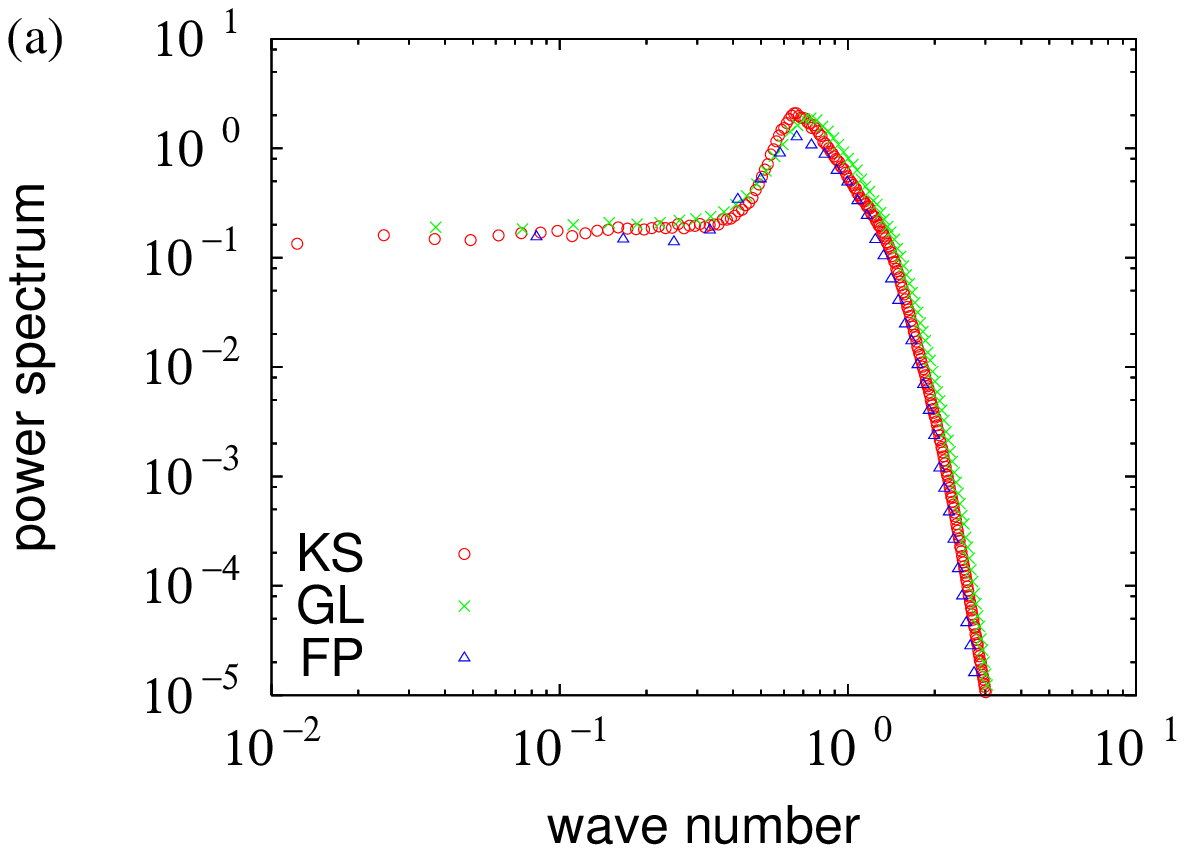}
\includegraphics[height=5cm]{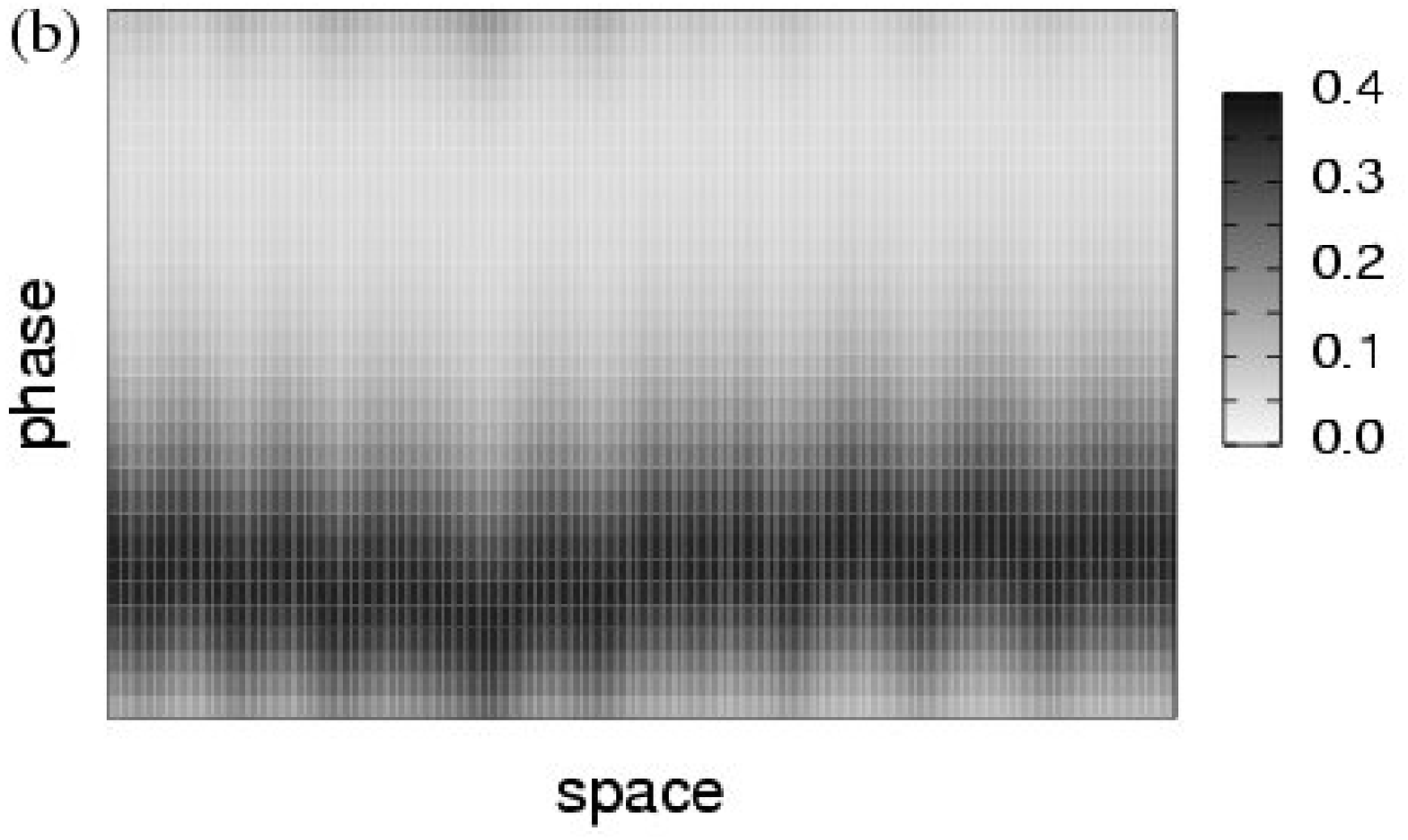}
\caption{(Color online) Phase turbulence.
  Parameter values are $\alpha=1.0$ and $D/D_c=0.8$ (Case Q).
  (a): Spatial power spectrum of the order parameter phase gradient
  $v(x,t) = 2 \partial_x \Theta(x,t)$
  is plotted as a function of the wavenumber in rescaled units with
  which the Kuramoto-Sivashinsky equation
  takes the standard form, Eq.~(\ref{eq:33}).
  Numerical data obtained from the Kuramoto-Sivashinsky equation (KS),
  the complex Ginzburg-Landau equation (GL), and the nonlinear
  Fokker-Planck equation (FP) are compared.
  (b): Instantaneous spatial profile of the phase distribution
  function obtained from the Fokker-Planck simulation.}
\label{fig:13}
\end{figure}
Finally, the numerical results for the parameter Q, for which phase
turbulence is expected, are shown in Figs.~\ref{fig:12} and
\ref{fig:13}.
Figure~\ref{fig:12} compares the results from the Kuramoto-Sivashinsky
equation, the complex Ginzburg-Landau equation, and the nonlinear
Fokker-Planck equation, where the spatiotemporal evolution of the
phase gradient $v(x,t) = 2 \partial_x \Theta(x,t)$ of the order
parameter is plotted for each equation.
They are remarkably similar to each other.
In Fig.~\ref{fig:13}(a), spatial power spectrum of the
order parameter phase gradient is plotted for each equation.
They also show excellent agreement with each other.
Figure \ref{fig:13}(b) shows a snapshot of the phase distribution
function obtained from the nonlinear Fokker-Planck equation.
As expected, long-wavelength phase fluctuations are observed.

These numerical results confirm the existence of the noise-induced
transition from the constant solution via the Hopf bifurcation and the
Benjamin-Feir instability.

\section{Phase equation near the destabilization point of the
  spatially uniform oscillation} \label{sec:5}

In the previous section, we have analyzed the nonlinear Fokker-Planck
equation near the Hopf bifurcation point of the constant solution
using the center-manifold reduction.
In this section, we investigate a different parameter region, where the
uniformly oscillating solution of the nonlinear Fokker-Planck equation
loses its stability against phase disturbances.
As a result, a transition line to the spatiotemporal chaos will be
determined, which completes the phase diagram of the noise-induced
turbulence.

\subsection{Lower transition line to the spatiotemporal chaos}

In the previous section, it was found that the constant solution of
the nonlinear Fokker-Planck equation, which is stable for large noise
intensity, loses its stability via the supercritical Hopf bifurcation
as the noise intensity is decreased, and the order parameter exhibits
spatiotemporal chaos near the Hopf bifurcation point under suitable
conditions.
However, the spatially uniform oscillating solution of all the
oscillators is stable in the absence of noise, since we assume
the in-phase coupling in the original nonlocally coupled
limit-cycle oscillators.
Thus, there should be a lower critical noise intensity, above which
the uniformly oscillating solution of the oscillators loses its
stability and gives way to the spatiotemporal chaos.
\begin{figure}
\centering
\includegraphics[height=6cm]{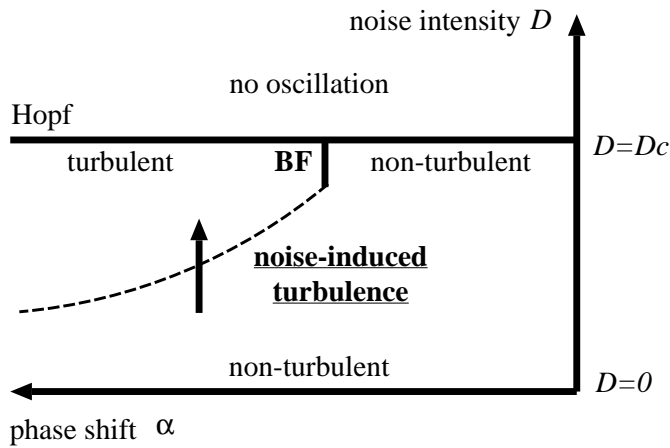}
\caption{Schematic phase diagram of the noise-induced turbulence as a
  function of the phase shift $\alpha$ and the noise intensity $D$.}
\label{fig:14}
\end{figure}
The schematic phase diagram which describes this situation is
illustrated in Fig.~\ref{fig:14}.
Below we analyze the situation where the order parameter becomes
turbulent as the noise intensity is increased from zero.
We use the method of phase reduction once again to derive a phase
equation describing the dynamics of slowly varying wavefronts.

\subsection{Slow phase modulation of the spatially uniform oscillation}

We here give the outline of the analysis.
Details of the analytical calculations are given in
Appendix~\ref{sec:B}.
First, we focus on the spatially uniform oscillating solution
$f_0(\theta-\Theta_0)$ of the nonlinear Fokker-Planck equation,
where $\theta=\phi-\Omega t$.
Here, $\Omega$ is the frequency of the collective oscillation, and the
constant $\Theta_0$ is the initial phase that can be chosen arbitrary.
We then allow this phase constant $\Theta_0$ to be slowly space-time
dependent, and denote it as $\Theta(x,t)$.
\begin{figure}
\centering
\includegraphics[height=6cm]{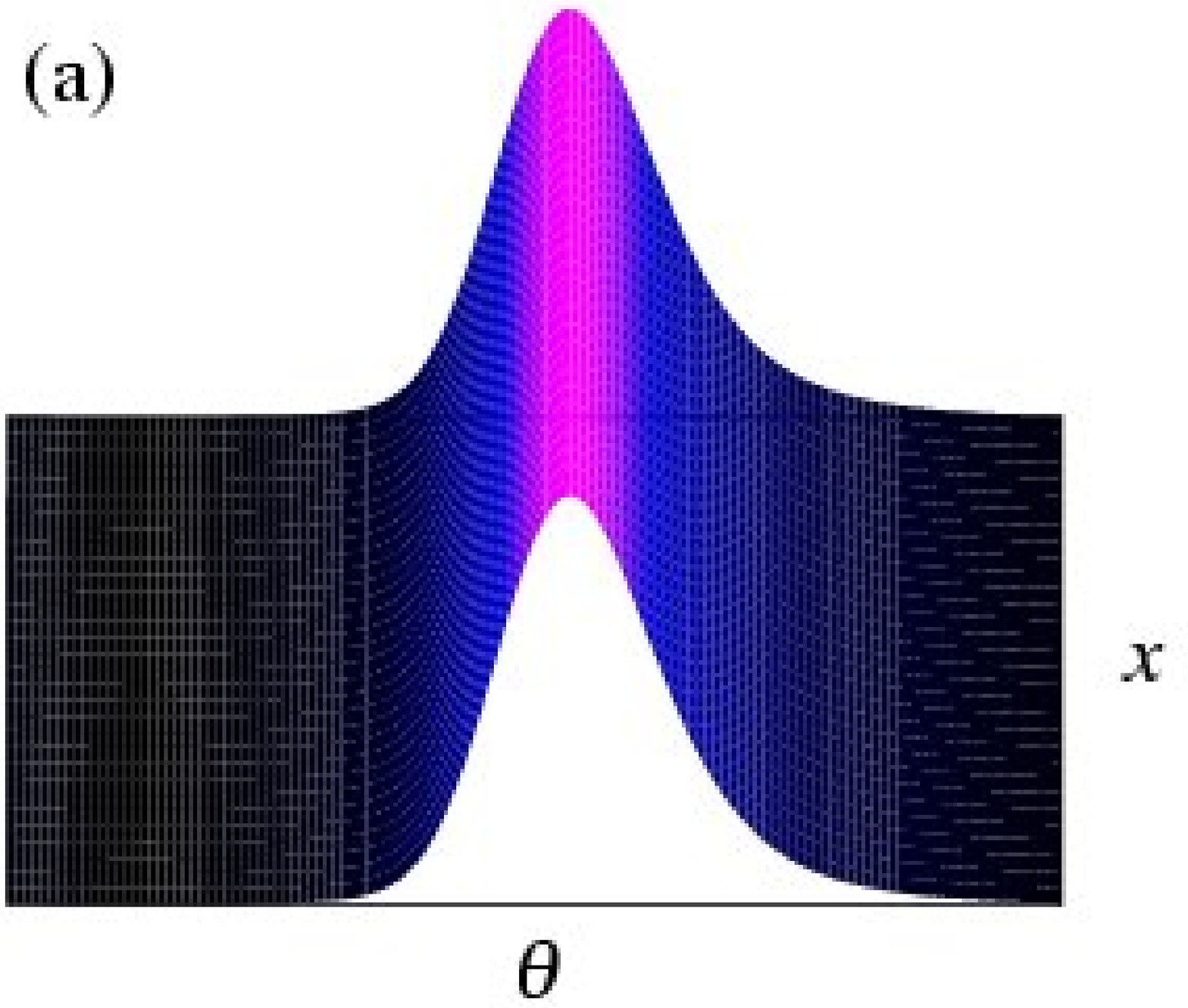}
\includegraphics[height=6cm]{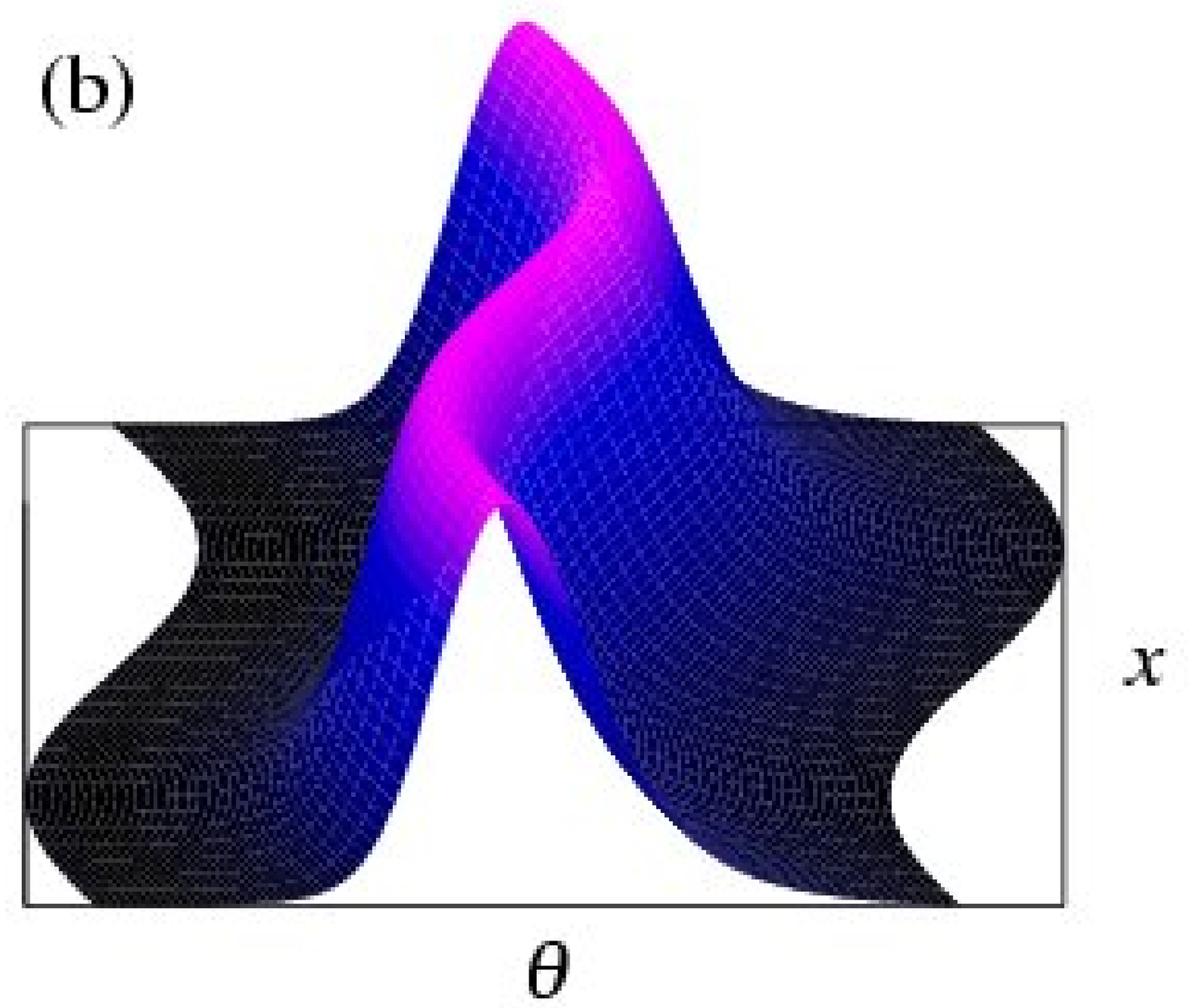}
\caption{(Color online) Schematic phase distribution function
  $f_0(\theta-\Theta)$ with a spatially uniform phase,
  $\Theta=\Theta_0$ (a), and with a spatially slowly varying phase,
  $\Theta=\Theta(x,t)$ (b).}
\label{fig:15}
\end{figure}
Figure~\ref{fig:15}(a) shows a schematic picture of a spatially
uniform oscillating solution, whose order parameter phase $\Theta_0$
is constant, and Fig.~\ref{fig:15}(b) the case that the order
parameter phase $\Theta(x,t)$ is slowly varying.
The phase of the slowly modulated oscillating solution defined here is
essentially the same as the phase of the order parameter defined
previously in Eq.~(\ref{eq:10}) except for the mean drift term $\Omega
t$, as long as the spatial scale of the variation is sufficiently
long compared to the coupling length.
Thus, we use the same notation $\Theta(x,t)$ for both phase variables.
As shown in Appendix~\ref{sec:B}, we can derive an equation for the
phase $\Theta(x,t)$ in a closed form using the analytical procedure
given in Ref.~\cite{ref:kuramoto84}.
If we truncate the series of gradient expansion retaining the first
few terms, the derived phase equation has the form
\begin{equation}
  \partial_t \Theta\left(x,t\right)
  = \bar{\nu} \partial_x^2 \Theta
  + \bar{\mu} \left(\partial_x \Theta\right)^2
  - \bar{\lambda} \partial_x^4 \Theta
  + \cdot \cdot \cdot,
\label{eq:40}
\end{equation}
where the parameters $\bar{\nu}$, $\bar{\mu}$, and $\bar{\lambda}$ can
be calculated using the phase coupling function $\Gamma(\phi)$, the
spatially uniform oscillating solution $f_0(\theta)$, and its
associated left- and right eigenfunctions.
Among these parameters, the phase diffusion coefficient $\bar{\nu}$ is
the most important quantity.
When $\bar{\nu}$ changes its sign from positive to negative, the spatially
uniform oscillating solution loses its stability, and spatiotemporal
chaos sets in.

\subsection{Phase diagram for the Stuart-Landau case}

For the case of the Stuart-Landau oscillators with the phase coupling
function given in Eq.~(\ref{eq:34}), we numerically calculated the
phase diffusion coefficient $\bar{\nu}$ at various values of the phase
shift $\alpha$ and the noise intensity $D$, and determined the
$\bar{\nu}=0$ curve on the $\alpha$-$D$ plane.
\begin{figure}
\centering
\includegraphics[height=4cm]{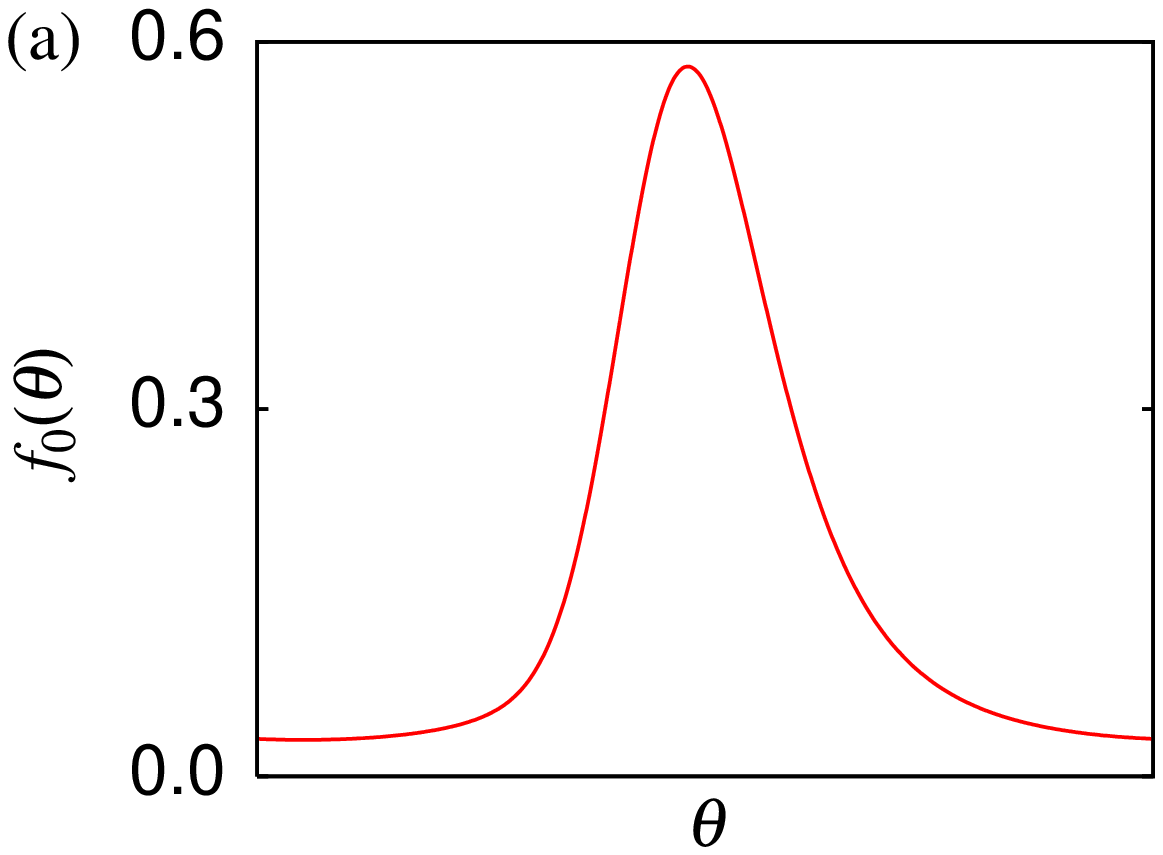}
\includegraphics[height=4cm]{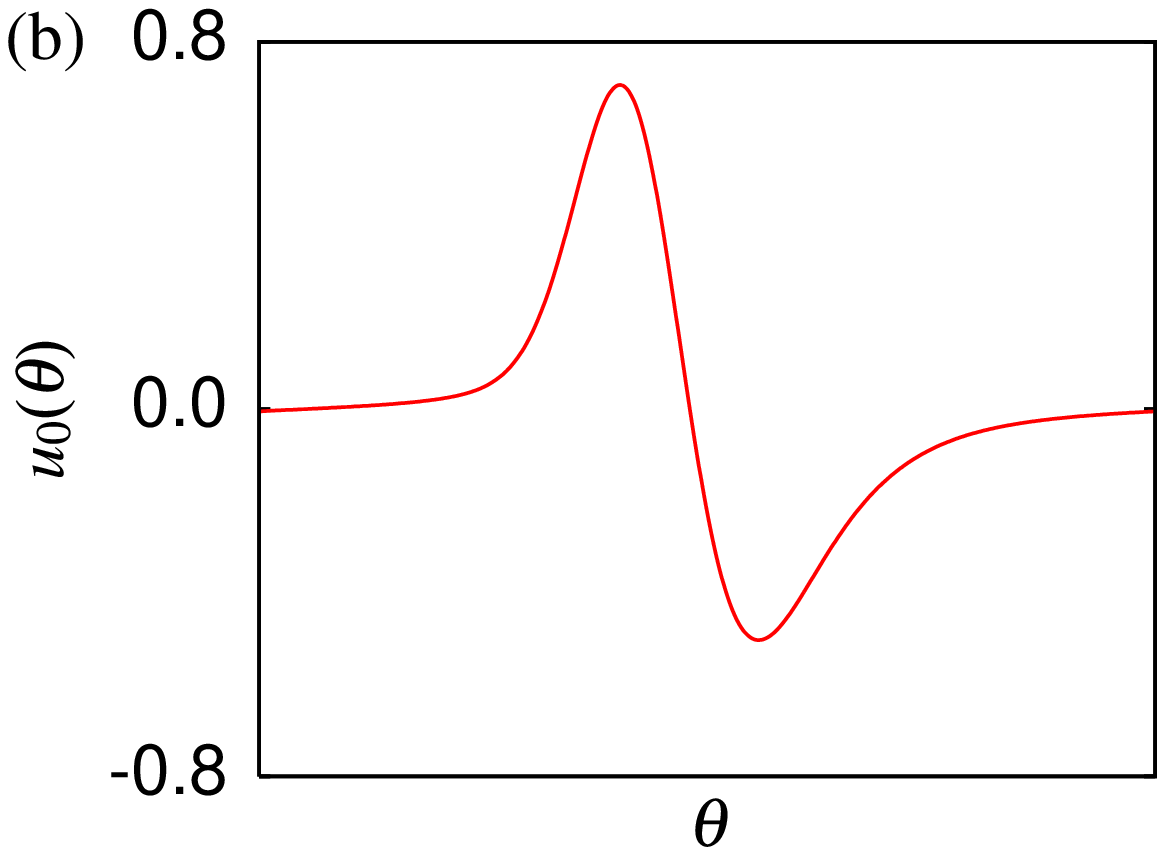}
\includegraphics[height=4cm]{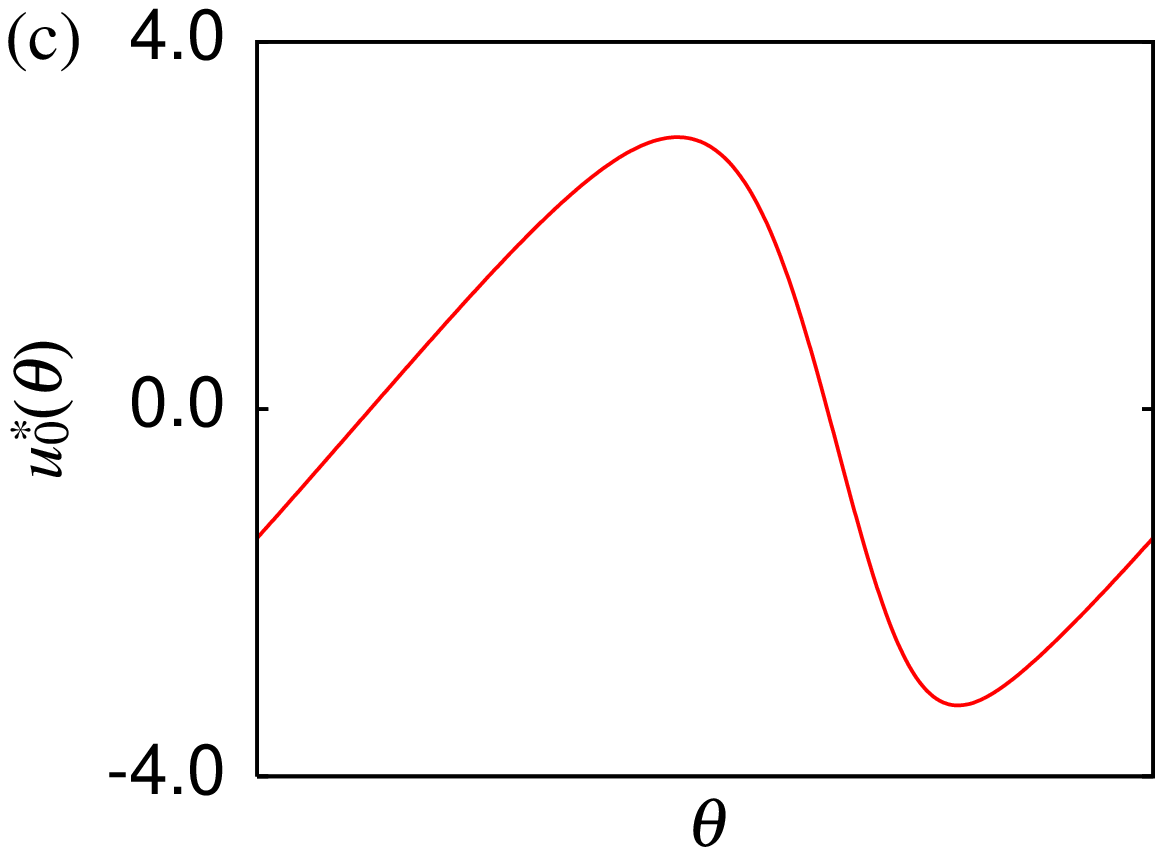}
\caption{(Color online) Numerically obtained spatially uniform
  oscillating solution (a), right zero eigenfunction (b), and
  left zero eigenfunction (c).
  The parameter values are $\alpha=1.2$ and $D/D_c=0.5$,
  which give $\bar{\nu}\simeq -0.371652.$}
\label{fig:16}
\end{figure}
Typical functional shapes of $f_0(\theta)$, $u_0(\theta)$, and
$u_0^{\ast}(\theta)$ obtained from the numerical calculations are
illustrated in Fig.~\ref{fig:16}.

\begin{figure}
\centering
\includegraphics[height=5.7cm]{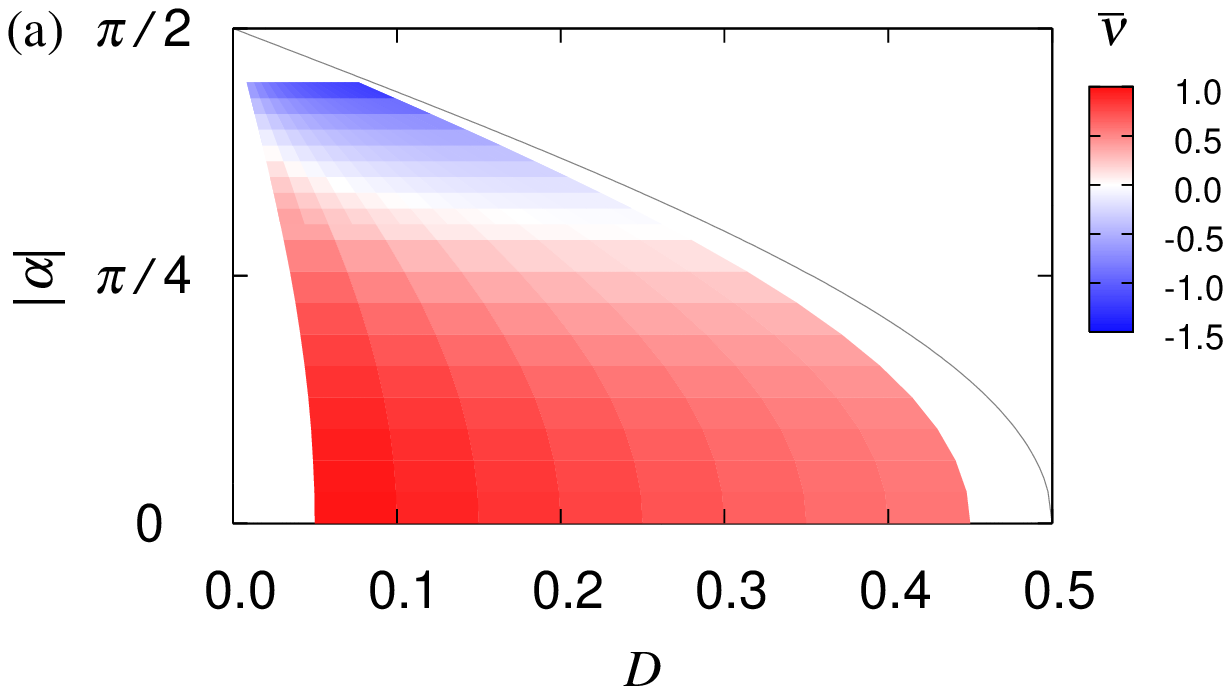}
\includegraphics[height=5.7cm]{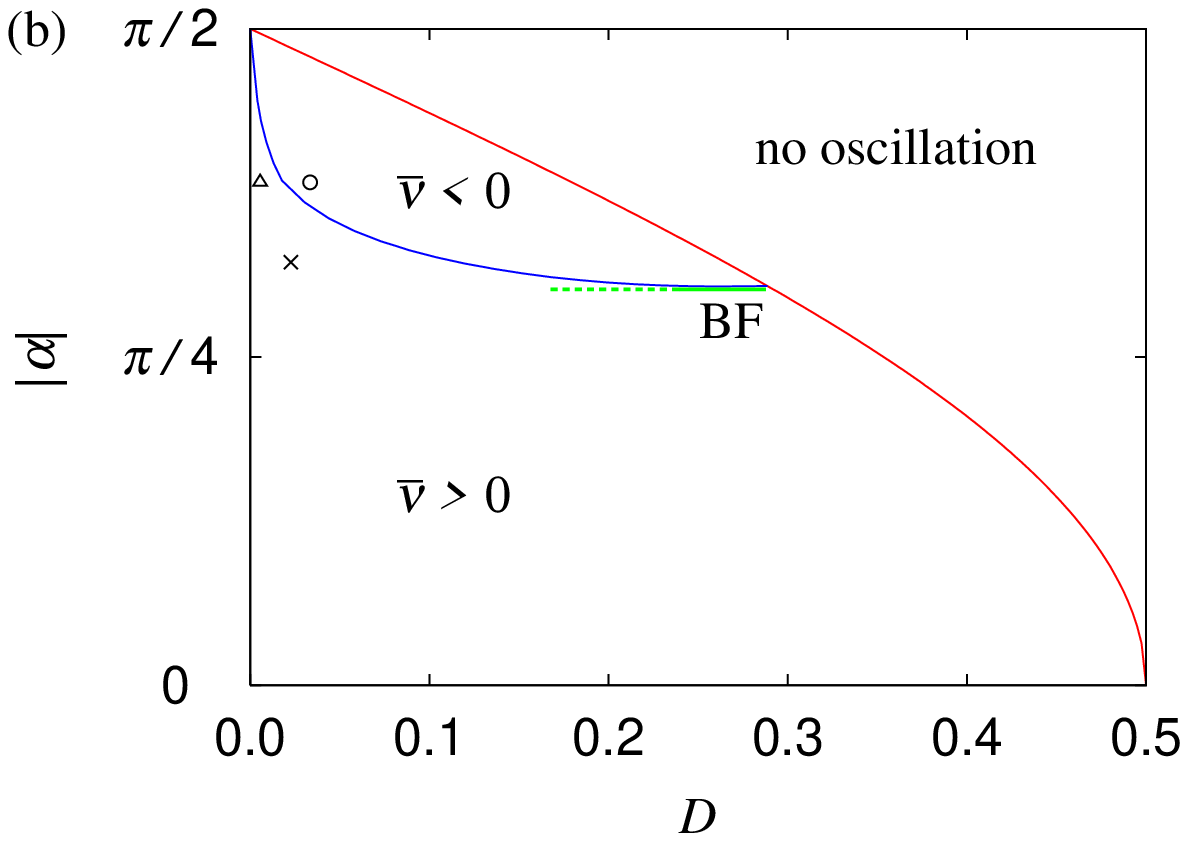}
\caption{(Color online) Dependence of the phase diffusion coefficient
  $\bar{\nu}$ on the phase shift $\alpha$ and the noise
  intensity $D$.
  (a):
  $\bar{\nu}$ was numerically evaluated in the parameter region
  with $\alpha \in [0.0, 1.4]$ and $D/D_c \in [0.1, 0.9]$.
  $\bar{\nu}$ is negative in the blue region,
  and positive in the red region.
  (b): The red and blue curves represent the Hopf bifurcation line ($D=D_c$)
  and the noise-induced transition line to turbulent state ($\bar{\nu}=0$).
  The Benjamin-Feir (BF) line is also indicated by the green line.
  Cases SL-I, SL-II, and SL-III are indicated by the triangle, the open
  circle, and the cross, respectively.}
\label{fig:17}
\end{figure}
In Fig.~\ref{fig:17}(a), dependence of the phase diffusion coefficient
$\bar{\nu}$ on $\alpha$ and $D$ is illustrated.
$\bar{\nu}$ is positive in the red region, and negative in the blue
region.
Figure \ref{fig:17}(b) represents the whole phase diagram.
The blue curve represents $\bar{\nu}=0$.
The red line represents the Hopf bifurcation curve, and the green
line the Benjamin-Feir criticality near the Hopf bifurcation.
In the parameter region surrounded by the red curve and the blue
curve, the phase diffusion coefficient $\bar{\nu}$ is negative,
indicating that spatiotemporal chaos appears in this region.
The Benjamin-Feir criticality line, which we obtained in the previous
section by the center-manifold reduction, is also consistent with the
results of the present phase reduction analysis.

Let us now look back on the numerical results of the original
nonlocally coupled noisy Stuart-Landau oscillators (Fig.~\ref{fig:4})
and also its reduced phase model (Fig.~\ref{fig:6}).
Three sets of the parameters used in Fig.~\ref{fig:4} and \ref{fig:6},
SL-I, SL-II, and SL-III, are also plotted on the phase diagram in
Fig.~\ref{fig:17}(b).
As can be seen, only the case SL-II is in the region of noise-induced
turbulent states. The other two cases, SL-I and SL-III, are in the
parameter region where the spatially uniform oscillation is stable.
Thus, our theory consistently explains the occurrence of noise-induced
turbulence.

\section{Concluding Remarks} \label{sec:6}

We studied a system of nonlocally coupled limit-cycle oscillators
subject to spatiotemporal white Gaussian noise, which exhibits a
remarkable phenomenon called noise-induced turbulence.
We considered the case that the coupling and the noise are
sufficiently weak, and reduced the system to nonlocally coupled
noisy phase oscillators by using the phase reduction method.
We then derived an equivalent nonlinear Fokker-Planck equation from
the Langevin phase equation, utilizing the fact that the mean-field
theory exactly holds owing to the nonlocal coupling.
The center-manifold reduction was applied to the situation where the
trivial constant solution of the nonlinear Fokker-Planck equation
becomes unstable and starts to oscillate, and we derived the complex
Ginzburg-Landau equation.
Furthermore, we also derived the Kuramoto-Sivashinsky-type equation
from the nonlinear Fokker-Planck equation by applying the phase
reduction method once again.
\begin{table}
\centering
\includegraphics[height=4.5cm]{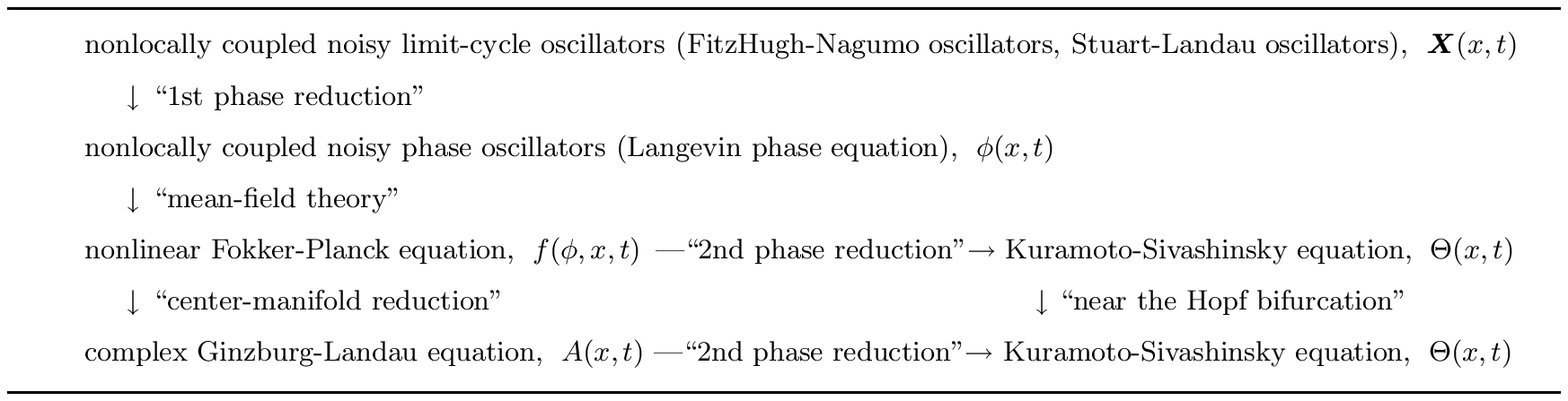}
\caption{Hierarchy of equations for noise-induced turbulence.}
\label{table:1}
\end{table}
The hierarchy of our equations derived for the noise-induced
turbulence is summarized in Table.~\ref{table:1}.
Our theoretical analysis and numerical simulations thus provide strong
evidence for the existence of noise-induced turbulence in systems of
nonlocally coupled noisy limit-cycle oscillators.

Finally, we briefly discuss how to distinguish the noise-induced
effective deterministic dynamics of the system from the noisy patterns
when the precise dynamical equation describing the system is not available.
In principle, we will be able to detect the bifurcations of the effective
deterministic dynamics, if appropriately filtered spatial patterns
exhibit qualitative changes as the external noise intensity is varied.
For our nonlocally coupled noisy oscillators, we introduced the
space-time dependent complex order parameter given by Eq.~(\ref{eq:10})
to observe the effective deterministic dynamics of the system,
where the nonlocal coupling function $G(x)$ was used as the
kernel function that filters the noisy patterns to eliminate
nonessential statistical fluctuations.
For experimental systems, however, there would be cases that the appropriate
kernel function is not known in advance and should be determined empirically.
The precise form of the kernel function is not important, so that
the essential parameter is the width of the kernel function, namely,
the coarse-graining scale of the noisy patterns.
One possible method to estimate the appropriate width of the kernel function
is to utilize a spatial correlation function $C(x)$ of the local oscillators
at a sufficiently strong noise intensity, where no collective oscillations
arises.
The correlation length of $C(x)$ calculated in such a regime will directly
reflect the coupling length of the system, which gives a reasonable
estimate of the appropriate kernel width.
We can then filter the noisy patterns using a suitable localized kernel
function, such as the Gaussian or the exponential kernel, whose width
is given by the obtained correlation length.
For our nonlocally coupled oscillators, the correlation length of $C(x)$
directly reflects the coupling length of $G(x)$.
Using the method explained above, we could actually observe spatiotemporal
patterns similar to those observed using the complex order parameter,
Eq.~(\ref{eq:10}), without assuming prior knowledge of the nonlocal
coupling function $G(x)$.

We believe that {\it noise-induced turbulence} can be experimentally
realized in the near future.

\begin{acknowledgments}
This work is supported by the Grant-in-Aid for the 21st Century COE
``Center for Diversity and Universality in Physics'' from the Ministry
of Education, Culture, Sports, Science and Technology (MEXT) of Japan.
\end{acknowledgments}


\appendix

\section{Phase reduction of coupled noisy limit-cycle oscillators}
\label{sec:A}

We consider the following Langevin equation describing an ensemble of
coupled identical limit-cycle oscillators subject to independent noises
\begin{equation}
  \partial_t\bd{X}_n\left(t\right) = \bd{F}\left(\bd{X}_n\right)
  +\sum_{n'}\bd{V}_{n,n'}\left(\bd{X}_n,\bd{X}_{n'}\right)
  +\sqrt{\sigma}\,\bd{\eta}_n\left(t\right),
\label{eq:a1}
\end{equation}
where $\bd{X}_n(t)$ represents the state of the $n$-th oscillator at
time $t$, $\bd{F}(\bd{X}_n)$ the individual dynamics,
$\bd{V}_{n,n'}(\bd{X}_n,\bd{X}_{n'})$ the interaction between $n$- and
$n'$-th oscillators, $\sigma$ the noise intensity, and
$\bd{\eta}_n(t)$ the external noise added independently to each
oscillator.  The noise $\bd{\eta}_n(t)$ is assumed to be white
Gaussian, whose statistics are given by
\begin{equation}
  \left\langle\eta^j_n\left(t\right)\right\rangle=0,\quad
  \left\langle\eta^j_n\left(t\right)\eta^{j'}_{n'}\left(t'\right)\right\rangle
  =2\delta_{j,j'}\delta_{n,n'}\delta\left(t-t'\right),
\label{eq:a2}
\end{equation}
where the superscripts $j$ and $j'$ denote the vector component.  When
the coupling term and the noise term are sufficiently small, the phase
reduction method is applicable. Applying the phase reduction
method~\cite{ref:kuramoto84}, we obtain the following Langevin
equation for the phase variables:
\begin{equation}
  \partial_t\phi_n\left(t\right)=\omega+\sum_{n'}
  \bd{Z}\left(\phi_n\right)\cdot\bd{V}_{n,n'}\left(\phi_n,\phi_{n'}\right)
  +\sqrt{\sigma}\bd{Z}\left(\phi_n\right)\cdot\bd{\eta}_n\left(t\right).
\label{eq:a3}
\end{equation}
Here, $\omega$ is the natural frequency of the oscillators,
$\bd{Z}(\phi_n)$ is the phase sensitivity function~\cite{ref:kuramoto84},
and $\bd{V}_{n,n'}(\phi_n,\phi_{n'})$ is the abbreviation of
$\bd{V}_{n,n'}(\bd{X}_0(\phi_n),\bd{X}_0(\phi_{n'}))$, where
$\bd{X}_0$ is the unperturbed limit-cycle solution.  By virtue of the
independent Gaussian statistics, we can rewrite the above equation
(\ref{eq:a3}) in the following form
\begin{equation}
  \partial_t\phi_n\left(t\right)=\omega+\sum_{n'}
  \bd{Z}\left(\phi_n\right)\cdot\bd{V}_{n,n'}\left(\phi_n,\phi_{n'}\right)
  +\sqrt{\sigma}\left\|\bd{Z}\left(\phi_n\right)\right\|
  \xi_n\left(t\right),
\label{eq:a4}
\end{equation}
where $\|\bd{Z}(\phi_n)\|=\sqrt{\bd{Z}(\phi_n)\cdot\bd{Z}(\phi_n)}$
and the statistics of the white Gaussian noise $\xi_n$ is given by
\begin{equation}
  \left\langle\xi_n\left(t\right)\right\rangle=0,\quad
  \left\langle\xi_n\left(t\right)\xi_{n'}\left(t'\right)\right\rangle
  =2\delta_{n,n'}\delta\left(t-t'\right).
\label{eq:a5}
\end{equation}
We introduce the new slow phase variables $\bar{\phi}_n$ as
\begin{equation}
  \phi_n = \omega t+\bar{\phi}_n,
\label{eq:a6}
\end{equation}
and rewrite the above Langevin equation (\ref{eq:a4}) as
\begin{equation}
  \partial_t\bar{\phi}_n\left(t\right)=\sum_{n'}\bd{Z}\left(\omega t+\bar{\phi}_n\right)
  \cdot\bd{V}_{n,n'}\left(\omega t+\bar{\phi}_n, \omega t+\bar{\phi}_{n'}\right)
  +\sqrt{\sigma}\left\|\bd{Z}\left(\omega t+\bar{\phi}_n\right)\right\|
  \xi_n\left(t\right).
\label{eq:a7}
\end{equation}
The corresponding Fokker-Planck equation describing the evolution of the
probability density function $P(\{\bar{\phi}_n\},t)$ of the phase variables
is given by
\begin{equation}
  \frac{\partial}{\partial t}P\left(\{\bar{\phi}_n\}, t\right)
  =\sum_n\left(-\frac{\partial}{\partial\bar{\phi}_n}\right)
  J_n\left(\{\bar{\phi}_n\}, t\right),
\label{eq:a8}
\end{equation}
where
\begin{align}
  J_n\left(\{\bar{\phi}_n\}, t\right)
  &=\left[\sum_{n'}\bd{Z}\left(\omega t+\bar{\phi}_n\right)\cdot
    \bd{V}_{n,n'}\left(\omega t+\bar{\phi}_n, \omega t+\bar{\phi}_{n'}\right)
    +\frac{1}{2}\frac{\partial}{\partial\bar{\phi}_n}
    \Bigl(\sigma\left\|\bd{Z}(\omega t+\bar{\phi}_n)\right\|^2\Bigr)\right]
  P\left(\{\bar{\phi}_n\}, t\right) \nonumber \\
  &-\frac{\partial}{\partial\bar{\phi}_n}
  \Biggl[\sigma\left\|\bd{Z}(\omega t+\bar{\phi}_n)\right\|^2
    P\left(\{\bar{\phi}_n\}, t\right)\Biggr].
\label{eq:a9}
\end{align}
Rapidly oscillating quantities in $J_n(\{\bar{\phi}_n\},t)$
are now time averaged over one period of the oscillator.
The resulting averaged Fokker-Planck equation has the following form
\begin{equation}
  \frac{\partial}{\partial t}P\left(\{\bar{\phi}_n\}, t\right)
  =\sum_n\left(-\frac{\partial}{\partial\bar{\phi}_n}\right)
  \left(\sum_{n'}\Gamma_{n,n'}\left(\bar{\phi}_n-\bar{\phi}_{n'}\right)
  P\left(\{\bar{\phi}_n\},t\right)-D\frac{\partial}{\partial\bar{\phi}_n}
  P\left(\{\bar{\phi}_n\},t\right)\right),
\label{eq:a10}
\end{equation}
where the phase coupling function $\Gamma_{n,n'}(\bar{\phi}_n-\bar{\phi}_{n'})$
is given by
\begin{equation}
  \Gamma_{n,n'}\left(\bar{\phi}_n-\bar{\phi}_{n'}\right)=
  \frac{1}{2\pi}\int_0^{2\pi}d\lambda\;
  \bd{Z}\left(\lambda+\bar{\phi}_n\right)\cdot
  \bd{V}_{n,n'}\left(\lambda+\bar{\phi}_n, \lambda+\bar{\phi}_{n'}\right),
\label{eq:a11}
\end{equation}
and also the effective noise intensity $D$ is given by
\begin{equation}
  D = \frac{1}{2\pi}\int_0^{2\pi}d\lambda\;
  \sigma\bd{Z}\left(\lambda+\bar{\phi}_n\right)
  \cdot\bd{Z}\left(\lambda+\bar{\phi}_n\right).
\label{eq:a12}
\end{equation}
The Langevin equation corresponding to the averaged Fokker-Planck
equation (\ref{eq:a10}) is expressed as
\begin{equation}
  \partial_t\bar{\phi}_n\left(t\right)
  =\sum_{n'}\Gamma_{n,n'}\left(\bar{\phi}_n-\bar{\phi}_{n'}\right)
  +\sqrt{D}\,\xi_n\left(t\right).
\label{eq:a13}
\end{equation}
Finally, we obtain the following Langevin equation for the phases:
\begin{equation}
  \partial_t\phi_n\left(t\right) = \omega+\sum_{n'}
  \Gamma_{n,n'}\left(\phi_n-\phi_{n'}\right)+\sqrt{D}\,\xi_n\left(t\right).
\label{eq:a14}
\end{equation}
Now let us consider the nonlocally coupled oscillator system.
The above formulae can be written in the form
\begin{align}
  \sum_{n'}\bd{V}_{n,n'}\left(\bd{X}_n,\bd{X}_{n'}\right)&\to
  \int dx'\,G\left(x-x'\right)\bd{V}\left(\bd{X},\bd{X}'\right), \label{eq:a15} \\
  \sum_{n'}\Gamma_{n,n'}\left(\phi_n-\phi_{n'}\right)&\to
  \int dx'\,G\left(x-x'\right)\Gamma\left(\phi-\phi'\right), \label{eq:a16}
\end{align}
where
\begin{equation}
  \Gamma\left(\phi-\phi'\right) =
  \frac{1}{2\pi}\int_0^{2\pi}d\lambda\;
  \bd{Z}\left(\lambda+\phi\right)\cdot
  \bd{V}\left(\lambda+\phi, \lambda+\phi'\right).
\label{eq:a17}
\end{equation}
Here, $\bd{X}'$ and $\phi'$ are the abbreviations
of $\bd{X}(x',t)$ and $\phi(x',t)$, respectively.

\section{Second phase reduction of the nonlinear Fokker-Planck equation}
\label{sec:B}

In this appendix, we present analytical procedures to derive a phase
equation describing the slow phase modulation of the spatially uniform
oscillating solution.
We assume long-wavelength modulation to the spatially uniform
oscillation, so that the spatial derivatives of the function
$f(\phi,x,t)$ are small quantities.
We thus expand the nonlocal coupling term of the Fokker-Planck
equation as
\begin{equation}
  \int_{-\infty}^{\infty}dx'\,G\left(x-x'\right)f\left(\phi,x',t\right)
  = \sum_{m=0}^{\infty}G_{2m}\,\partial_x^{2m}f\left(\phi,x,t\right),
\label{eq:b1}
\end{equation}
where $G_{2m}$ is the $2m$-th moment of $G(x)$.
It is given by $G_{2m} = \int_{-\infty}^{\infty} dx\, G(x) x^{2m} /
(2m)!$, where $G_0=1$ holds from the normalization condition.
Due to the spatial reflection symmetry, only even moments $G_{2m}$
remain.
For the coupling function used in this paper, $G(x)=\exp(-|x|)/2$, we
obtain $G_{2m}=1$ for all $m$.
The nonlinear Fokker-Planck equation is expanded as
\begin{align}
  \frac{\partial f(\phi,x,t)}{\partial t}=
  &-\frac{\partial}{\partial\phi}
  \left[\left\{\omega+\int_0^{2\pi}
    d\phi'\,\Gamma\left(\phi-\phi'\right)
    f\left(\phi',x,t\right)\right\} f\left(\phi,x,t\right)\right]
  +D\frac{\partial^2 f(\phi,x,t)}{\partial\phi^2}
  \nonumber \\
  &-\frac{\partial}{\partial\phi} \left[\int_0^{2\pi}d\phi'\,
    \Gamma\left(\phi-\phi'\right) \Bigl\{G_2\partial_x^2
    f\left(\phi',x,t\right)\Bigr\} f\left(\phi,x,t\right)\right]
  \nonumber \\
  &-\frac{\partial}{\partial\phi} \left[\int_0^{2\pi}d\phi'\,
    \Gamma\left(\phi-\phi'\right) \Bigl\{G_4\partial_x^4
    f\left(\phi',x,t\right)\Bigr\} f\left(\phi,x,t\right)\right]
  \nonumber \\
  &-\cdot\cdot\cdot.
\label{eq:b2}
\end{align}

Let us denote by $f(\phi, x, t) = f_0(\theta) = f_0(\phi - \Omega t)$
the spatially uniform oscillating solution of the nonlinear
Fokker-Planck equation, where $\Omega$ is the collective frequency.
Inserting this expression in the nonlinear Fokker-Planck equation,
we find that $f_0(\theta)$ satisfies the following equation
\begin{equation}
  D\frac{d^2}{d\theta^2}f_0\left(\theta\right)+\left(\Omega-\omega\right)
  \frac{d}{d\theta}f_0\left(\theta\right)-\frac{d}{d\theta}
  \Bigl[g_0\left(\theta\right)f_0\left(\theta\right)\Bigr]=0,
\label{eq:b3}
\end{equation}
where
\begin{equation}
  g_0\left(\theta\right) = \int_0^{2\pi}d\theta'\,
  \Gamma\left(\theta-\theta'\right)f_0\left(\theta'\right).
\label{eq:b4}
\end{equation}
Let $u(\theta, t)$ represent disturbance to the spatially uniform
oscillating solution defined by $f(\phi, x, t) = f_0(\theta) + u(\theta, t)$.
Eq.~(\ref{eq:b2}) is linearized in $u(\theta, t)$, i.e., $\partial _t
u = \hat{L} u$, where the linearized operator $\hat{L}$ is defined as
\begin{equation}
  \hat{L}u = D\frac{d^{\, 2}}{d\theta^2}u\left(\theta\right)
  +\left(\Omega-\omega\right)\frac{d}{d\theta}u\left(\theta\right)
  -\frac{d}{d\theta}\Bigl[g_0\left(\theta\right)u\left(\theta\right)\Bigr]
  -\frac{d}{d\theta}\left[f_0\left(\theta\right)\int_0^{2\pi}d\theta'\,
    \Gamma\left(\theta-\theta'\right)u\left(\theta'\right)\right],
\label{eq:b5}
\end{equation}
whose adjoint operator $\hat{L}^{\ast}$, defined by
$\int_0^{2\pi}d\theta\, u^{\ast}(\theta) \hat{L} u(\theta) =
\int_0^{2\pi}d\theta\, u(\theta) \hat{L}^{\ast} u^{\ast}(\theta)$,
is expressed as
\begin{equation}
  \hat{L}^{\ast}u^{\ast} = D\frac{d^{\, 2}}{d\theta^2}u^{\ast}\left(\theta\right)
  -\left(\Omega-\omega\right)\frac{d}{d\theta}u^{\ast}\left(\theta\right)
  +g_0\left(\theta\right)\frac{d}{d\theta}u^{\ast}\left(\theta\right)
  +\int_0^{2\pi}d\theta'\,\Gamma\left(\theta'-\theta\right)f_0\left(\theta'\right)
  \frac{d}{d\theta'}u^{\ast}\left(\theta'\right).
\label{eq:b6}
\end{equation}
The left- and the right eigenfunctions and their eigenvalues of these linear
operators are determined from
\begin{equation}
  \hat{L}u_l=\lambda_l u_l,\quad
  \hat{L}^{\ast}u_l^{\ast}=\lambda_l u_l^{\ast} \quad
  \left(l=0,1,2,\cdot\cdot\cdot\right).
\label{eq:b7}
\end{equation}
The eigenfunctions are assumed to be orthonormalized as
\begin{equation}
  \int_0^{2\pi}d\theta\,u_l^\ast\left(\theta\right)
  u_m\left(\theta\right) = \delta_{lm}.
\label{eq:b8}
\end{equation}
Here we should note that the right zero eigenfunction can
be chosen as
\begin{equation}
  \hat{L} u_0 = 0, \quad
  u_0\left(\theta\right) = 
  \frac{d}{d\theta} f_0\left(\theta\right), \quad
  \lambda_0 = 0,
\label{eq:b9}
\end{equation}
which follows from the differentiation of
Eq.~(\ref{eq:b3}) with respect to $\theta$.

We now apply the second-order phase reduction to
the nonlinear Fokker-Planck equation~(\ref{eq:b2})
by treating spatial derivatives as perturbations.
Following the procedure developed in Ref.~\cite{ref:kuramoto84},
the Kuramoto-Sivashinsky-type phase equation describing the
slowly varying phase modulation can be derived in the form
\begin{equation}
  \partial_t \Theta\left(x,t\right)
  = \bar{\nu} \partial_x^2 \Theta
  + \bar{\mu} \left(\partial_x\Theta\right)^2
  - \bar{\lambda} \partial_x^4 \Theta
  + \cdot \cdot \cdot,
\label{eq:b10}
\end{equation}
where
\begin{equation}
  \bar{\nu} = -G_2\int_0^{2\pi}d\varphi\,u_0^\ast\left(\varphi\right)
  \frac{d}{d\varphi}\Bigl[a_0\left(\varphi\right)f_0\left(\varphi\right)\Bigr],
\label{eq:b11}
\end{equation}
\begin{equation}
  \bar{\mu} = G_2\int_0^{2\pi}d\varphi\,u_0^\ast\left(\varphi\right)
  \frac{d}{d\varphi}\Bigl[b_0\left(\varphi\right)f_0\left(\varphi\right)\Bigr],
\label{eq:b12}
\end{equation}
and
\begin{align}
  \bar{\lambda}&=G_4\int_0^{2\pi}d\varphi\,u_0^\ast\left(\varphi\right)
  \frac{d}{d\varphi}\Bigl[a_0\left(\varphi\right)f_0\left(\varphi\right)\Bigr]
  \nonumber \\
  &+G_2\sum_{l\ne 0}\lambda_l^{-1}
  \left\{\int_0^{2\pi}d\varphi\,u_0^\ast\left(\varphi\right)\frac{d}{d\varphi}
  \Bigl[a_l\left(\varphi\right)f_0\left(\varphi\right)\Bigr]\right\}
  \left\{\int_0^{2\pi}d\varphi\,u_l^\ast\left(\varphi\right)\frac{d}{d\varphi}
  \Bigl[a_0\left(\varphi\right)f_0\left(\varphi\right)\Bigr]\right\}.
\label{eq:b13}
\end{align}
Here, the quantities $a_l(\varphi)$ and $b_l(\varphi)$ are defined by
\begin{align}
  a_l\left(\varphi\right) &=\int_0^{2\pi}d\varphi'\,
  \Gamma\left(\varphi-\varphi'\right)
  u_l\left(\varphi'\right), \label{eq:b14} \\
  b_l\left(\varphi\right) &=\int_0^{2\pi}d\varphi'\,
  \Gamma\left(\varphi-\varphi'\right)
  \frac{d}{d\varphi'}u_l\left(\varphi'\right). \label{eq:b15}
\end{align}
In general, the spatially uniform oscillating solution $f_0(\theta)$
cannot be obtained analytically.
Thus, $f_0(\theta)$ and the associated zero eigenfunctions should
be calculated numerically in order to evaluate the phase diffusion
coefficient $\bar{\nu}$ (\ref{eq:b11}).
This can be done with sufficient precision by applying the
numerical relaxation method using $M=2^9$ modes for the phase.
See Appendix~\ref{sec:C} for the details of the numerical methods.

\section{Numerical methods} \label{sec:C}

We used spatially periodic boundary conditions in all the numerical
simulations throughout this paper.
We confirmed our numerical simulation results are not changed even if
we further increase the number of grid points $N$ for the space or the
number of modes $M$ for the phase.

\subsection{Algorithm for the nonlocally coupled noisy limit-cycle
  oscillators}

We used an explicit Euler scheme with a time step $\varDelta t = 0.01$
for the equation
\begin{equation}
  \partial_t\bd{X}\left(x,t\right) = \bd{F}\left(\bd{X}\left(x,t\right)\right)
  +\hat{K}\int_{-\infty}^{\infty}dx'\,G\left(x-x'\right)
  \bd{X}\left(x',t\right) + \sqrt{\sigma}\,\bd{\eta}\left(x,t\right).
\label{eq:c1}
\end{equation}
The system is discretized using $N=2^{10}$ grid points.
The nonlocal coupling term can be efficiently calculated by using the
fast Fourier transform (FFT) technique because it is simply a
convolution form with the kernel given by Eq.~(\ref{eq:2}).

\subsection{Algorithm for the nonlocally coupled noisy phase
  oscillators}

The system is discretized using $N=2^{10}$ grid points for Fig.~\ref{fig:6}
and $N=2^{15}$ grid points for Figs.~\ref{fig:8}-\ref{fig:11}.
We used an explicit Euler scheme with a time step $\varDelta t = 0.01$
for the equation in the following form
\begin{equation}
  \partial _t \phi\left(x,t\right)= \omega - R\left(x,t\right)
  \sin\bigl(\phi\left(x,t\right) - \Theta\left(x,t\right) + \alpha\bigr)
  + \sqrt{D}\,\xi\left(x,t\right),
\label{eq:c2}
\end{equation}
\begin{equation}
  R\left(x,t\right)\exp\Bigl(i\Theta\left(x,t\right)\Bigr)
  = \int_{-\infty}^{\infty}dx'\,G\left(x-x'\right)
  \exp\Bigl(i\phi\left(x',t\right)\Bigr),
\label{eq:c3}
\end{equation}
where the FFT is used for the calculation of the spatial convolution.

\subsection{Algorithm for the nonlinear Fokker-Planck equations}

We used a pseudo-spectral method with $M=2^5$ modes for the phase
component, and spatial discretization with $N=2^{10}$ grid points for
the equation
\begin{equation}
  \frac{\partial f\left(\phi,x,t\right)}{\partial t}=-\frac{\partial}{\partial\phi}
  \Bigl[\Bigl\{\omega-R\left(x,t\right) \sin\bigl(\phi -\Theta\left(x,t\right) 
    + \alpha\bigr)\Bigr\}f\left(\phi,x,t\right)\Bigr]
  +D\frac{\partial^2 f\left(\phi,x,t\right)}{\partial\phi^2},
\label{eq:c4}
\end{equation}
where
\begin{equation}
  R\left(x,t\right)\exp\Bigl(i\Theta\left(x,t\right)\Bigr)
  =\int^{\infty}_{-\infty}dx'\,G\left(x-x'\right)
  \int^{2\pi}_{0}d\phi'\,e^{i\phi'}f\left(\phi',x',t\right).
\label{eq:c5}
\end{equation}
A modified Euler scheme with a time step $\varDelta t=0.01$ is used
for the temporal integration. FFT is used for the calculation of the
spatial convolution.

\subsection{Algorithm for the complex Ginzburg-Landau equation}

We used a pseudo-spectral method for the complex amplitude using
$N=2^{10}$ modes for the equation
\begin{equation}
  \partial_t A(x,t) = A + (1+ic_1) \partial_x^2 A - (1+ic_2) |A|^2 A,
\label{eq:c6}
\end{equation}
where a modified fourth-order Runge-Kutta scheme with a time step
$\varDelta t = 0.01$ is used for the temporal integration.

\subsection{Algorithm for the Kuramoto-Sivashinsky equation}

We used a pseudo-spectral method using $N=2^{10}$ modes for the
equation in the following form
\begin{equation}
  \partial_t v(x,t) = -\partial_x^2 v
  + v \,\partial_x v - \partial_x^4 v,
\label{eq:c7}
\end{equation}
where $v(x,t) = 2 \partial_x \Theta(x,t)$.
Temporal integration was done by a modified fourth-order Runge-Kutta
scheme with a time step $\varDelta t = 0.01$.

\subsection{Algorithm for the numerical relaxation method}

To determine the spatially uniform oscillating solution
$f_0(\theta)$, we numerically evolve the following
nonlinear Fokker-Planck equation
\begin{equation}
  \frac{\partial f_0\left(\phi,t\right)}{\partial t}=
  -\frac{\partial}{\partial\phi}\left[\left\{\omega
    +\int_0^{2\pi}d\phi'\,\Gamma\left(\phi-\phi'\right)
    f_0\left(\phi',t\right)\right\}f_0\left(\phi,t\right)\right]
  +D\frac{\partial^2 f_0\left(\phi,t\right)}{\partial\phi^2},
\label{eq:c8}
\end{equation}
from an appropriate initial condition, which converges to a steadily
rotating wave packet with the collective drift velocity $\Omega$,
$f_0\left(\theta=\phi - \Omega t\right)$.
The right eigenfunction $u_0(\theta)$ is simply obtained by
differentiating $f_0\left(\theta\right)$ by $\theta$.
To obtain the left eigenfunction $u^{\ast}_0(\theta)$,
we evolve the following equation for $u^{\ast}_0(\theta, t)$,
\begin{equation}
  \frac{\partial\, u_0^{\ast}(\theta,t)}{\partial t}
  =D\frac{\partial^2\,u_0^{\ast}\left(\theta,t\right)}{\partial\theta^2}
  -(\Omega - \omega) \frac{\partial\,u_0^{\ast}\left(\theta,t\right)}{\partial\theta}
  +g_0\left(\theta\right)\frac{\partial\, u_0^{\ast}\left(\theta,t\right)}{\partial\theta}
  +\int_0^{2\pi}d\theta'\,\Gamma\left(\theta'-\theta\right)f_0\left(\theta'\right)
  \frac{\partial\, u_0^{\ast}\left(\theta',t\right)}{\partial\theta'},
\label{eq:c9}
\end{equation}
from an appropriate initial condition using the $f_0(\theta)$
and the $\Omega$ obtained above. We constantly rescale
$u^{\ast}_0(\theta, t)$ throughout the numerical evolution,
so that the normalization condition
$\int_0^{2\pi}d\theta\,u_0^{\ast}(\theta, t)u_0(\theta)=1$
is always satisfied.
After sufficient relaxation, $u^{\ast}_0(\theta, t)$
converges to the desired left eigenfunction $u^{\ast}_0(\theta)$.
For the numerical calculations, we used a pseudo-spectral method using
$M=2^9$ modes, and a modified Euler scheme with a time step
$\varDelta t = 0.01$ for the temporal integration.


\end{document}